# Simulating Alzheimer's amyloid-like pathology on a fly central nervous system connectome

Taeho Jo[1,2*]

[1] Department of Radiology and Imaging Sciences, Indiana University School of Medicine, Indianapolis, Indiana, United States of America

[2] Indiana Alzheimer's Disease Research Center, Indiana University School of Medicine, Indianapolis, Indiana, United States of America

* Corresponding author

E-mail: tjo@iu.edu (TJ)

Short title: Simulating Alzheimer's amyloid-like pathology on a fly connectome

## Abstract

In Alzheimer's disease, the amount of pathology in the brain is an imperfect predictor of how much function is lost. Pathology leads to that loss through neural circuits, but these cannot be observed neuron by neuron in patients. This study therefore simulated the central nervous system of the adult male *Drosophila melanogaster*: 166,700 integrate-and-fire neurons wired by the MaleCNS connectome. The simulator reproduced a published fly brain model (rate correlation 0.9997). Three kinds of damage reported in amyloid-β flies, a lowered spike threshold, synapse loss and neuron loss, were imposed with assumed magnitudes, and circuit responses were measured in 1,424 simulated flies. The number of neurons lost was a poor guide to the deficit, which depended partly on how many synaptic contacts went with them and whether they included input neurons or one command neuron. With the same 12% of neurons removed, odour-evoked whole-brain activity was 93%, 47% or 31% of the healthy value when damage fell on poorly connected, random or highly connected neurons; these flies lost 20%, 35% and 48% of contacts. When the contacts lost were equalised, the three groups differed less; whether the remaining difference reflects where the damage fell could not be established. In 48 flies, restoring only the contacts that surviving neurons had lost recovered 62% of the whole-brain deficit, and restoring the removed neurons 39%. In flies that had lost neurons delivering the odour, these accounted for about half the deficit due to neuron loss. One neuron could decide an outcome: all 66 flies that lost the giant fibre on the stimulated side, which commands escape, had a giant fibre rate below 100 Hz, against 1 of 254 that kept it. Whole-brain activity revealed damage at milder severity than the learning index, because it changed much relative to its variation between trials.

## Author summary

In Alzheimer's disease, brain damage is an imperfect guide to impairment, and the circuits behind the mismatch cannot be watched neuron by neuron in people. This study simulated the fully mapped central nervous system of a male fruit fly, 166,700 neurons, and tested it with an odour, a threat and odour training. It was damaged increasingly in three ways seen in flies making a human Alzheimer's protein:

neurons become easier to excite, lose synapses and die. Losing the same number of highly or poorly connected neurons left whole-brain activity at a third of normal or nearly normal, partly because highly connected cells take more synapses with them. Putting lost synapses or dead neurons back into damaged flies showed that lost synapses reduced whole-brain activity more. Much of the effect of dead neurons came from cells that deliver the odour or, where escape signals were weak, command cells such as the giant escape neurons. Raised excitability alone caused no deficit and partly offset the other two. Whole-brain activity revealed damage at milder levels than a learning score or simulated behaviour, because it changed much relative to its trial-to-trial variation. These are model predictions: where and how strongly damage strikes was assumed.

## Introduction

In Alzheimer's disease the amount of pathology is an imperfect predictor of the clinical deficit. In an autopsy series, neocortical synapse density tracked cognitive scores far better than counts of plaques or tangles did [1]. The biomarker frameworks of clinical research order markers by when they become abnormal [2] or group them by the pathological process each measures [3]; neither says how that pathology leads, through neural circuits, to symptoms. Network accounts place the disease in large-scale brain systems [4] and in their most connected hubs [5–7], and region-level models have reproduced aspects of its atrophy pattern and of hub vulnerability [8, 9]. Neuron by neuron, however, the circuits that link damage to deficit cannot be observed in people, so the relation between the two is known only as an association.

A nervous system whose wiring diagram is known at synaptic resolution allows a complementary approach. Pathology can be imposed on individual neurons by stated rules, such as removing a neuron with a given probability, and the whole network simulated. A deficit is then the consequence of those rules, and each rule can be switched off, strengthened, weakened or applied in the same amount to other neurons. The output is a prediction about the modelled system rather than a measurement of the disease, but one that experiments in the same species can test.

*Drosophila melanogaster* offers both transgenic models of amyloid-β toxicity and the wiring diagram. With human amyloid-β42 expressed in the nervous system, neurons reproducibly become more excitable, with loss of the A-type potassium channel Kv4, and are progressively lost, among them the Kenyon cells of the mushroom body, the fly's centre for olfactory learning; olfactory associative learning is impaired [10, 11]. Synapses, examined at the neuromuscular junction, have fewer release sites in flies that express amyloid-β [12] and fewer boutons in flies that generate it from co-expressed human amyloid precursor protein (APP) and the β-secretase BACE [13]. Separately, electron microscopy connectomes cover the adult fly brain, the ventral nerve cord and the complete central nervous system (CNS) [14–18]. A leaky integrate-and-fire model of a whole female brain, wired by the FlyWire connectome, agreed with experiment in 91% of 164 tested predictions about feeding and grooming circuits [19], and a connectome-constrained network predicted activity across the visual system [20].

No published study or preprint appears to have joined these two bodies of work by placing pathology specific to amyloid-β, tau or APP on an adult fly connectome and reporting a quantitative result. The nearest are a connectome description of the ageing fly's escape circuit [21] and a doctoral thesis on amyloid-β expression in a larval circuit, with connectome analysis and mathematical modelling [22]. Simulations based on fly connectomes have removed or silenced neurons and synapses [23–27], but none applied rules derived from a specific disease. In a *Caenorhabditis elegans* connectome model, a single level of simulated axonal injury produced a range of behavioural deficits [28], as one amount of damage does here.

This study simulates the whole CNS of an adult male fly, wired by the MaleCNS connectome [18], and imposes three rules modelled on amyloid-β flies, together called amyloid-like pathology: synapse loss, neuron loss and a lowered spike threshold, which stands in for raised excitability. One setting per fly, its severity, scales all three. Each simulated fly is the same connectome with its own random draw of lost synapses and neurons; healthy flies have none. Every fly receives the same stimuli, among them an odour and, on one side, a looming visual stimulus that triggers escape, both delivered through neurons that can themselves be damaged. The readouts measured include whole-brain activity, the spikes of all CNS neurons during the odour. The questions were how much each rule contributes to the deficit in these readouts; whether equal numbers of lost neurons, a common measure of degeneration in fly models, produce equal deficits wherever the loss falls, and how far lost synapses explain any difference; which part of a fly's damage accounts for its deficit; and which readouts separate damaged from healthy flies at the lowest severity.

Synapse loss and neuron loss produced the deficit, and the lowered spike threshold partly offset it. The number of neurons lost was a poor guide to the deficit: for the same number, whole-brain activity differed widely between flies whose damage fell on highly connected, random or poorly connected neurons, partly because unequal numbers of synapses were lost. When the synapses lost were equalised, by raising the severity of two groups, the three differed less; but this also lowered their thresholds further, and the groups differed in how much odour input they lost, so the remaining difference cannot be attributed to where the damage fell. When single damaged flies were simulated again, putting back the synapses that surviving neurons had lost recovered more of the whole-brain deficit than putting back the removed neurons. Where neuron loss accounted for much of a deficit, the neurons responsible were mostly input neurons, which deliver the odour, or, for the escape response, the giant fibres, a pair of neurons that command it: chance loss of the one on the stimulated side split flies of one severity into two groups, one with the pair firing at less than about half the healthy rate. Whole-brain activity and activity in the lateral horn, a second olfactory centre, separated damaged from healthy flies at the lowest severity, because they changed much relative to their variation between trials. The direction of each rule comes from amyloid-β flies, whereas its size and the choice of which neurons are damaged most are assumed.

## Results

### A whole-CNS model carrying graded amyloid-like pathology

The network was built from MaleCNS v1.0, the connectome of the central nervous system (CNS) of an adult male *Drosophila* [18]: 166,700 neurons joined by 24,469,412 excitatory or inhibitory connections, which comprise 120,260,398 synaptic contacts, each a single synapse (Fig 1A). Each neuron is a leaky integrate-and-fire neuron, which fires when synaptic input drives its membrane potential to a threshold, with the parameters of a published whole-brain model [19] (the base model). The simulator written for this study (the engine) was checked against the published code [29], run in Brian2 [30] on its original connectome, FlyWire v630, a female fly brain [15, 31]: under sugar stimulation, the 438 neurons active in either simulator fired at nearly identical rates (Pearson r = 0.9997; S1 Fig; S1 Text).

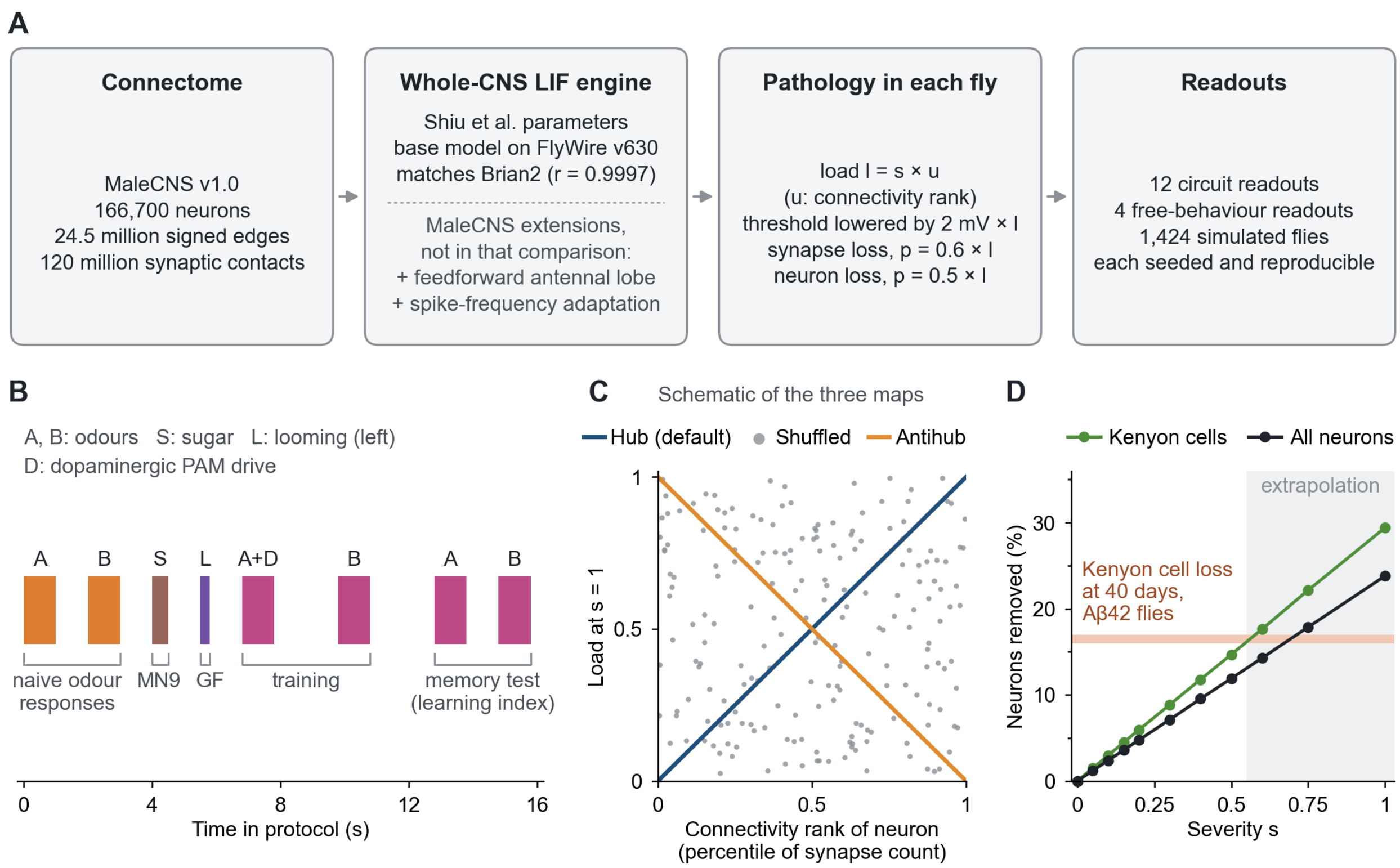

**Fig 1. A whole-CNS model carrying graded amyloid-like pathology.** (A) The model. Signed edges, excitatory or inhibitory connections; LIF, leaky integrate-and-fire; base model, the published whole-brain model [19]; MaleCNS extensions, the two extensions added to it; r, Pearson correlation of per-neuron firing rates between the simulator written for this study and Brian2 (S1 Fig); s, severity. The "12 circuit readouts" are eleven circuit readouts and the projection neuron rate; the "free-behaviour readouts" are the four arena readouts. (B) Circuit protocol: odours A and B, sugar (S), looming (L) and odour A paired with the teaching signal, activation of dopaminergic protocerebral anterior medial (PAM) neurons (A+D). Brackets mark groups of steps; sugar is measured at the MN9 feeding motor neurons and looming at the giant fibres (GF). (C) Idealised drawing, not model output: load at s = 1.0 against connectivity rank under the three vulnerability maps. (D) Mean percentage of all neurons and of Kenyon cells removed per fly (severity series, 32 flies per severity). Salmon band (16 to 17%), Kenyon cell loss at 40 days in flies expressing amyloid-β42 (Aβ42) in the mushroom body alone [11]; its width, around the published 16.5% (S1 Text), is not an error range. Grey, extrapolation ($s \geq 0.55$).

Every simulated fly received the same sequence of stimuli, the circuit protocol (Fig 1B), each delivered by driving a set of neurons with random spike trains: odours A and B; sugar, delivered through sugar receptor neurons and measured at the feeding motor neuron MN9 [32]; looming, the sight of an approaching object, delivered through the visual neurons LPLC2 and LC4 on the left and measured at the giant fibres, two descending neurons that command escape [33]; training, which paired odour A with 100-Hz activation of protocerebral anterior medial (PAM) dopaminergic neurons, the teaching signal in appetitive conditioning; and a memory test of both odours.

On MaleCNS the base model did not suit such a sequence: activity persisted after every stimulus, and an odour activated far more Kenyon cells of the mushroom body than it does in flies (S1 Text), so two extensions were made. Spike-frequency adaptation, a current that grows with each spike and slows firing, ended persistent activity after sugar and looming but not after odours, and lowered the MN9 and giant fibre responses by 42% and 17%. For odours, the antennal lobe, the first olfactory relay, was made feedforward: its projection neurons, which carry odours to the mushroom body and lateral horn, are driven directly, and chemical synapses among antennal lobe neurons are removed. Antennal lobe processing is thus outside the model, and the check against Brian2 covered neither extension.

With both extensions, odour A, delivered through 8 projection neurons, made a median of 16.6% of Kenyon cells fire in flies without pathology (healthy flies), and odour B, through 9 others, 8.5% (S1 Text).

Amyloid-like pathology followed three rules, each in the direction observed in flies whose neurons make human amyloid-β and of assumed size (Table 1): a spike threshold lowered by up to 2 mV (the threshold shift), modelling raised excitability [11]; loss of each outgoing synaptic contact [12, 13] with probability up to 0.6 (synapse loss); and removal of the neuron [10, 11] with probability up to 0.5 (neuron loss). Each rule acts on a neuron in proportion to its load: the severity s, one number per fly from 0 (healthy) to 1, times a value that a vulnerability map assigns to the neuron. In the default, the hub map, this value is the neuron's connectivity rank, its percentile among all neurons by number of synaptic contacts, an assumption that damages highly connected neurons most; two comparison maps assign the same values at random (the shuffled map) or in reverse order, damaging poorly connected neurons most (the antihub map) (Fig 1C). Pathology does not propagate between neurons.

Only sensory neurons are exempt from pathology, so it also damages the neurons that deliver every stimulus except sugar. Most of the stimulated projection, LPLC2 and LC4 neurons are highly connected: under the hub map about 22 to 25% of them are expected to be removed at s = 0.5, against 12 to 13% of the PAM neurons, about the fraction of all neurons removed (S1 Text). Part of every odour- and looming-evoked deficit is therefore loss of input; no simulation spared these neurons while damaging the rest. Each fly's lost contacts and neurons are drawn at random, so flies of one severity differ; healthy flies are one intact network and differ only in the random timing of their input spikes.

**Table 1. The three pathology rules and two further assumptions.**

| Rule | Observation | Implementation | What is assumed |
|---|---|---|---|
| Threshold shift | Amyloid-β42 flies: raised excitability after loss of the Kv4 potassium channel; cultured neurons fire at 46% less current [11] | Threshold lowered by 2 mV × load (at most 29% of the 7-mV rest-to-threshold gap) | Direction from flies; size assumed, smaller than measured |
| Synapse loss | Amyloid-β42 and amyloid-β40 flies: fewer active zones at the adult neuromuscular junction [12]; flies making amyloid-β from a human precursor protein: fewer boutons at the larval junction [13] | Each outgoing contact lost with probability 0.6 × load | Direction from flies, at the neuromuscular junction, not the central brain; size and uniformity across cell types assumed |
| Neuron loss | Amyloid-β42 flies: progressive neurodegeneration [10, 11]; about 17% fewer Kenyon cells at 40 days with mushroom-body-restricted expression [11] | Neuron removed (never fires) with probability 0.5 × load | Direction from flies; 0.5 assumed, not fitted |
| Vulnerability map | Humans: amyloid deposition concentrates in highly connected cortical regions [5] | Load (0 to 1) = s × connectivity rank; sensory neurons exempt | Assumed; untested in flies; compared with shuffled and antihub maps |
| Propagation between neurons | Flies: no trans-synaptic propagation of tau detected where tested [34]; mutant huntingtin aggregates, via glia [35], and seeded amyloid-β deposits [36] do spread | None; neurons damaged independently | Assumed |

One measurement ties severity to real flies: at 40 days, Kenyon cell loss was about 17% in flies expressing amyloid-β42 in the mushroom body alone and about 30% with expression in all neurons [11], a difference the model cannot show, because it damages each neuron independently of the others. The scale is placed by the lower figure, 17%, reached in the model between s = 0.5 (14.7%) and 0.6 (17.6%): severities up to 0.5 lie within the measured loss, and figures shade those from 0.55 upward as extrapolation (Fig 1D; a second limit, near s = 0.3, is given under Limitations). The measurement was not used to set the size of any rule (Table 1); severity measures damage, not age.

Table 2 defines the readouts, the quantities measured in each fly. Whole-brain activity counts the spikes of all CNS neurons during odour A. Valence, called naive valence before training, is the sum of mushroom body output neuron (MBON) rates, with glutamatergic MBONs, which promote avoidance [37], counted as negative: a higher valence stands for attraction. Training weakens synapses from Kenyon cells onto MBONs, and the learning index, the resulting change in valence for odour A minus that for odour B (a rise in healthy flies), reflects MBON activity, not behaviour.

The last four readouts of Table 2 come from a separate set of 176 flies, the arena cohort, which after the circuit protocol steered a model body through a virtual arena with food, odours and looming threats. Its two odour readouts are not used here: healthy flies did not prefer the trained odour (S1 Text). The simulations reported here comprise 1,424 flies in all (Fig 1A).

**Table 2. Readouts.**

| Readout | Definition | Measured during |
|---|---|---|
| Whole-brain activity | Spikes of all 166,700 neurons (ventral nerve cord and stimulated projection neurons included) | Odour A |
| Lateral horn rate | Lateral horn cell types (names beginning with "LH") | Odour A |
| Projection neuron rate | All 686 antennal lobe projection neurons; mostly the 8 stimulated ones (odour input delivered) | Odour A |
| Kenyon cell rate | All 4,064 Kenyon cells | Odour A |
| Kenyon cells recruited | Fraction of all Kenyon cells, removed ones included, that fired | Odour A |
| MBON rate | All 97 mushroom body output neurons | Odour A |
| Naive valence | Sum of MBON rates, glutamatergic MBONs negative | Odour A |

| Readout | Definition | Measured during |
|---|---|---|
| Learning index | Change in valence with training, odour A minus odour B | Before and after training |
| Learning index divided by MBON rate | Undefined if MBON rate is below 0.5 Hz | As above |
| MN9 feeding drive | Both MN9 motor neurons | Bilateral sugar, 0.5 s |
| Giant fibre rate | Both giant fibres (DNp01) | Left looming, 0.3 s |
| Giant fibre latency | First 1-ms bin with a giant fibre spike | Left looming, 0.3 s |
| PAM pairing rate | Mean of the 15 PAM cell-type rates; all 316 neurons driven at 100 Hz (teaching signal delivered) | Odour A with PAM activation, 1 s |
| Time feeding | Time on food with MN9 rate above 10 Hz | Arena, 120 s |
| Jumps per threat | Take-offs (at least two giant fibre spikes within 30 ms, spontaneous ones included) per looming threat met | Arena, 120 s |
| Time near trained odour | Time within 14 mm of the odour A source | Arena, 120 s |
| Odour preference index | $(t_A - t_B) / (t_A + t_B)$, time near each source | Arena, 120 s |

Rates (Hz) are means over all the neurons named, removed neurons counted as silent; odour A means its 1-s presentation before training. The first twelve rows were tested for a trend with severity; the thirteenth is used only in S1 Text; the last four are arena readouts (Materials and methods).

## Circuit readouts decline with severity

The severity series comprised 64 healthy flies and 32 flies at each of ten severities from 0.05 to 1.0, damaged under the default model: all three rules act on every non-sensory neuron (pan-neuronal pathology), most strongly on the most connected neurons (the hub map). All but one of the twelve readouts tested (Table 2) changed strongly with severity in the direction of impairment (Fig 2): ten fell and giant fibre latency rose (Spearman correlation with severity, $|\rho|$ from 0.50 to 0.95, each Holm-corrected $p < 10^{-22}$; S1 Text). The exception, the learning index divided by MBON rate, checks whether learning fell by more than mushroom body output (below).

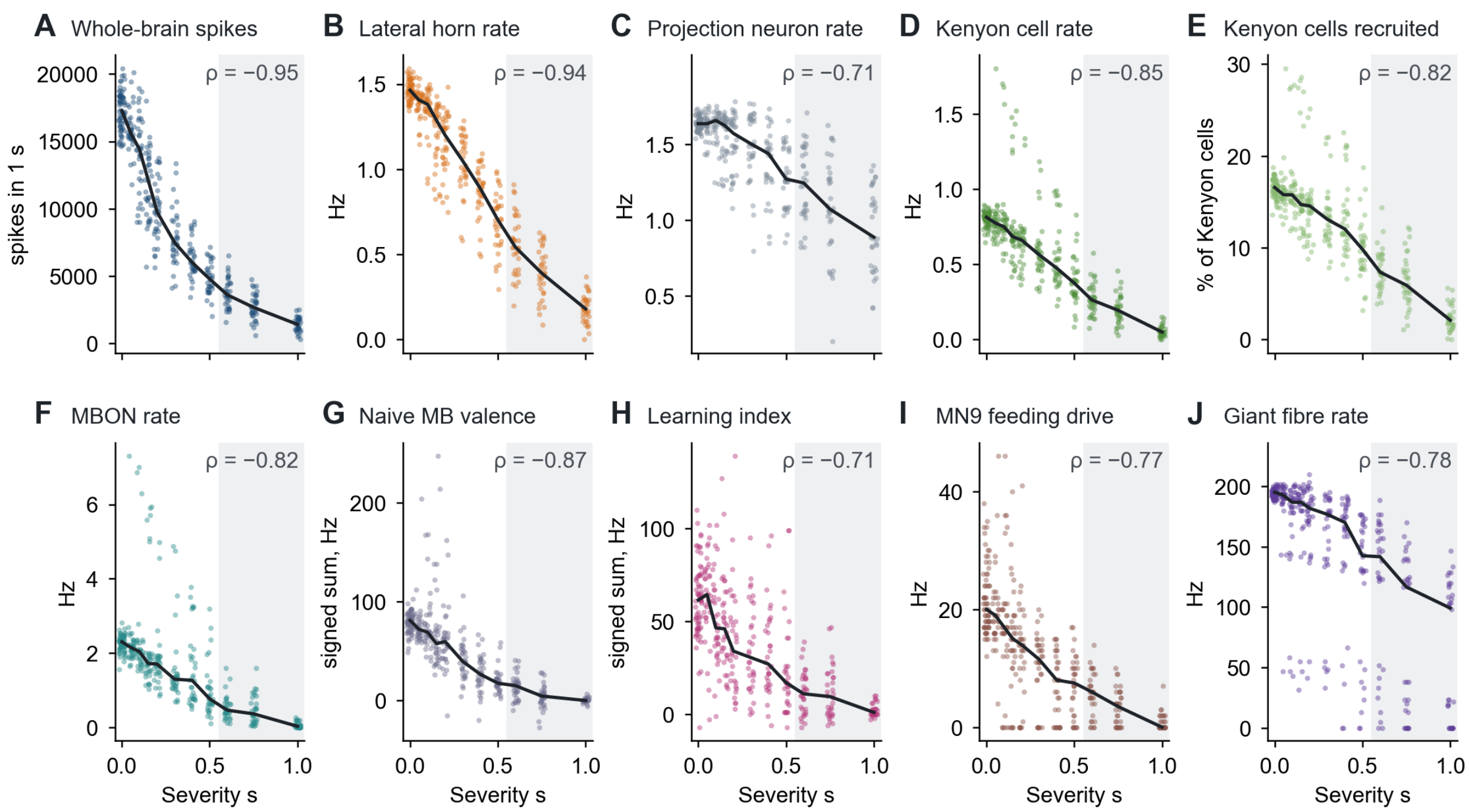

**Fig 2. Circuit readouts decline with severity.** Severity series, default model: 64 healthy flies (s = 0) and 32 flies at each of ten severities from 0.05 to 1.0. Points, flies (jittered); black line, per-severity medians; ρ, Spearman correlation with severity across all 384 flies; grey, s ≥ 0.55 (extrapolation; Fig 1D). (A) Whole-brain activity (printed as whole-brain spikes). (B) Lateral horn rate. (C) Projection neuron (PN) rate. (D) Kenyon cell rate. (E) Kenyon cells recruited. (F) Mushroom body output neuron (MBON) rate. (G) Naive valence (printed as naive MB valence; MB, mushroom body). (H) Learning index. (I) MN9 feeding drive. (J) Giant fibre rate. Definitions, Table 2.

At s = 0.5 readouts fell by different amounts (Fig 2A to 2J; medians as a percentage of the healthy median; S1 Text). Whole-brain activity fell to 28% (95% bootstrap interval 24 to 31%) and lateral horn rate to 48%, further than the odour input delivered, which the projection neuron (PN) rate measures (78%). MBON rate fell to 33% (27 to 46%) and the learning index to 28% (21 to 38%). MN9 feeding drive fell to 38% and giant fibre rate to 73%; both declined gradually and, in some flies, showed a discrete drop seen in no healthy fly (discrete phenotypes, below).

The PN rate is mostly the output of the 8 projection neurons that deliver odour A, and it fell with severity (ρ = −0.71) in step with the expected removal of these neurons (S1 Text). This loss of input shows in the odour responses: at s = 0.5, flies with a lower PN rate had lower whole-brain activity (ρ = 0.84; 0.83 without the projection neurons' own spikes).

The default model at s = 0.5 and 1.0 was simulated three times: in this series and again within the comparisons of rules and of maps that follow. These replicate groups of 32 flies differ only in their random draws and show how much a group median varies by chance (S2 Fig). At s = 0.5 their whole-brain medians were 28%, 24% and 31% of the healthy median, the highest 1.27 times the lowest (Kruskal-Wallis p = 0.03, although the condition is the same). Differences between single 32-fly groups smaller than this replicate variation, about 27% for whole-brain activity and 39 to 44% for MBON rate and the learning index, cannot be distinguished from chance (S1 Text).

Divided by each fly's own MBON rate, the learning index fell much less (to 84.9% of the healthy median at s = 0.5, in the 25 flies with a defined ratio, against 28%), so most of its fall reflects weaker mushroom body output. The ratio is not a pure measure of plasticity, because synapse loss alone raises it and neuron loss alone lowers it (S3 Fig; S1 Text).

### Synapse loss and neuron loss produce the deficit; the threshold shift partly offsets it

The contribution of each rule was examined by simulating each rule alone, all three together, and synapse loss with neuron loss but without the threshold shift (the rule ablation; 32 flies per condition at s = 0.5 and 1.0; Fig 3A). At s = 0.5 all three rules together (the second replicate group) reduced whole-brain activity to 24% of the healthy median (20 to 30%), MBON rate to 31% (27 to 43%) and the learning index to 34% (25 to 46%).

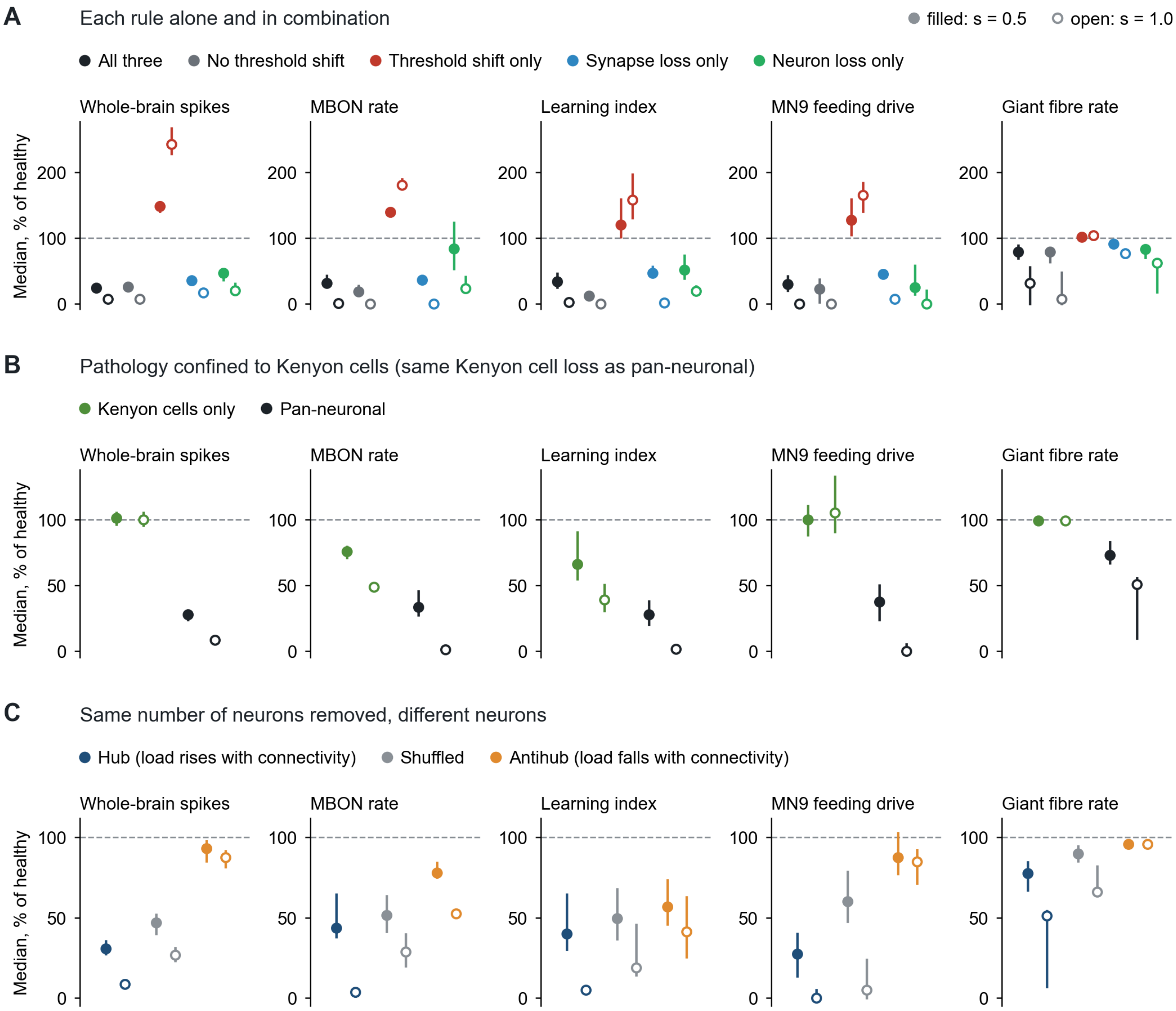

**Fig 3. Which pathology rule and which neurons produce the deficit.** Group medians as a percentage of the median of the 64 healthy flies of the severity series, with 95% percentile bootstrap intervals of that ratio. Filled symbols, s = 0.5; open symbols, s = 1.0; 32 flies per condition; dashed line, healthy median. (A) Rule ablation: all three rules, both loss rules without the threshold shift, and each rule alone. (B) Kenyon-cell-only pathology (green) against pan-neuronal pathology (black; groups of the severity series). (C) Map comparison: the three vulnerability maps. The default model appears in A (all three), B (pan-neuronal) and C (hub) as replicate groups with independent random draws of the damage (S2 Fig).

Each loss rule alone produced much of the whole-brain deficit: activity fell to 36% (34 to 37%) under synapse loss alone and to 47% (37 to 53%) under neuron loss alone. The MBON deficit came mainly from synapse loss, which alone lowered MBON rate to 36% (33 to 39%). Under neuron loss alone the median MBON rate was not detectably lowered (84%, 54 to 124%; Mann-Whitney p = 0.27 against healthy flies), with some flies below the healthy range and others above it (examined below).

At s = 0.5 the threshold shift alone raised whole-brain activity to 148% (140 to 154%) and MBON rate to 139%. With the loss rules present, the group simulated without it showed no detectable difference in whole-brain activity or four other rates (p from 0.18 to 0.96 against all three rules; S1 Text), but a lower MBON rate, 19% against 31% ($p = 0.006$), and learning index, 12% against 34% ($p < 0.001$): the shift had offset part of the mushroom body deficit and, as single flies show below, a smaller part of the others.

### The number of neurons lost is a poor measure of damage

Counts of lost neurons, for example Kenyon cells, are a common measure of neurodegeneration in fly models [11]. Whether such a count predicts the deficit was first tested by applying pathology to the Kenyon cells alone, each with the load it has under the default model, leaving all other neurons intact (Kenyon-cell-only pathology; 32 flies at each of s = 0.5 and 1.0; Fig 3B). This is analogous to amyloid-β42 expression confined to the mushroom body, if amyloid-β42 harms only the cells that express it. At s = 0.5 Kenyon-cell-only pathology removed 14.7% of Kenyon cells, as pan-neuronal pathology does, but only 0.36% of all neurons, against 11.9%. At the same Kenyon cell loss, the mushroom body deficit was much smaller than under pan-neuronal pathology: at s = 0.5 the learning index was 66% of the healthy median (54 to 90%) against 28% ($p = 2 \times 10^{-5}$) and MBON rate 76% (71 to 79%) against 33% ($p = 2 \times 10^{-6}$); readouts outside the mushroom body did not differ detectably from healthy flies (S1 Text). The same Kenyon cell loss thus goes with a small or a large mushroom body deficit, depending on whether the rest of the network is damaged.

Does equal neuron loss produce equal deficits wherever it falls? The hub, shuffled and antihub maps assign the same loads to different neurons (Fig 1C). With 32 flies per map at s = 0.5 and 1.0 (the map comparison), the fraction of all neurons removed did not differ between maps, as designed (11.9% at s = 0.5, Kruskal-Wallis $p = 0.11$).

Which neurons were damaged changed most readouts (Fig 3C). At s = 0.5, whole-brain activity was 31% (27 to 35%), 47% (40 to 52%) and 93% (85 to 98%) of the healthy median under the hub, shuffled and antihub maps ($p = 3 \times 10^{-16}$), a threefold antihub/hub ratio of medians (3.04, 2.57 to 3.36). Lateral horn rate, Kenyon cell rate, MN9 feeding drive and giant fibre rate followed the same order (S1 Text).

The mushroom body readouts distinguished the hub map from the shuffled map less clearly. MBON rate was 43% (38 to 59%), 52% (42 to 63%) and 78% (75 to 84%) of the healthy median under the three maps ($p = 6 \times 10^{-5}$), but the hub and shuffled maps did not differ (Mann-Whitney $p = 0.36$), and the learning index did not differ between maps at $s = 0.5$ ($p = 0.18$). The shuffled/hub ratio exceeded replicate variation (S2 Fig) for whole-brain activity (1.53 against 1.27) but not for MBON rate or the learning index (S1 Text).

Part of the hub/antihub difference in the readouts taken during the odour is loss of input: the PN rate was 90%, 89% and 100% of the healthy median under the three maps at $s = 0.5$ (hub against antihub $p = 0.002$). These hub-map and shuffled-map groups lost input alike, so loss of input is unlikely to explain most of their difference; other hub-map groups lost more (78% in the severity series), the rate resting on eight neurons (S1 Text).

The three maps removed equal numbers of neurons but unequal numbers of synaptic contacts (Fig 4A). Counting contacts taken by the synapse-loss rule and all contacts of removed neurons (contacts lost), the hub, shuffled and antihub maps removed 48%, 35% and 20% of all contacts at $s = 0.5$ (S1 Text), because the hub map puts the largest loads on the neurons with the most contacts. The number of neurons removed is therefore a poor measure of damage in this model, but this comparison cannot say whether it matters where in the network the damage falls, beyond the neurons that deliver the stimulus, or only how many contacts are lost.

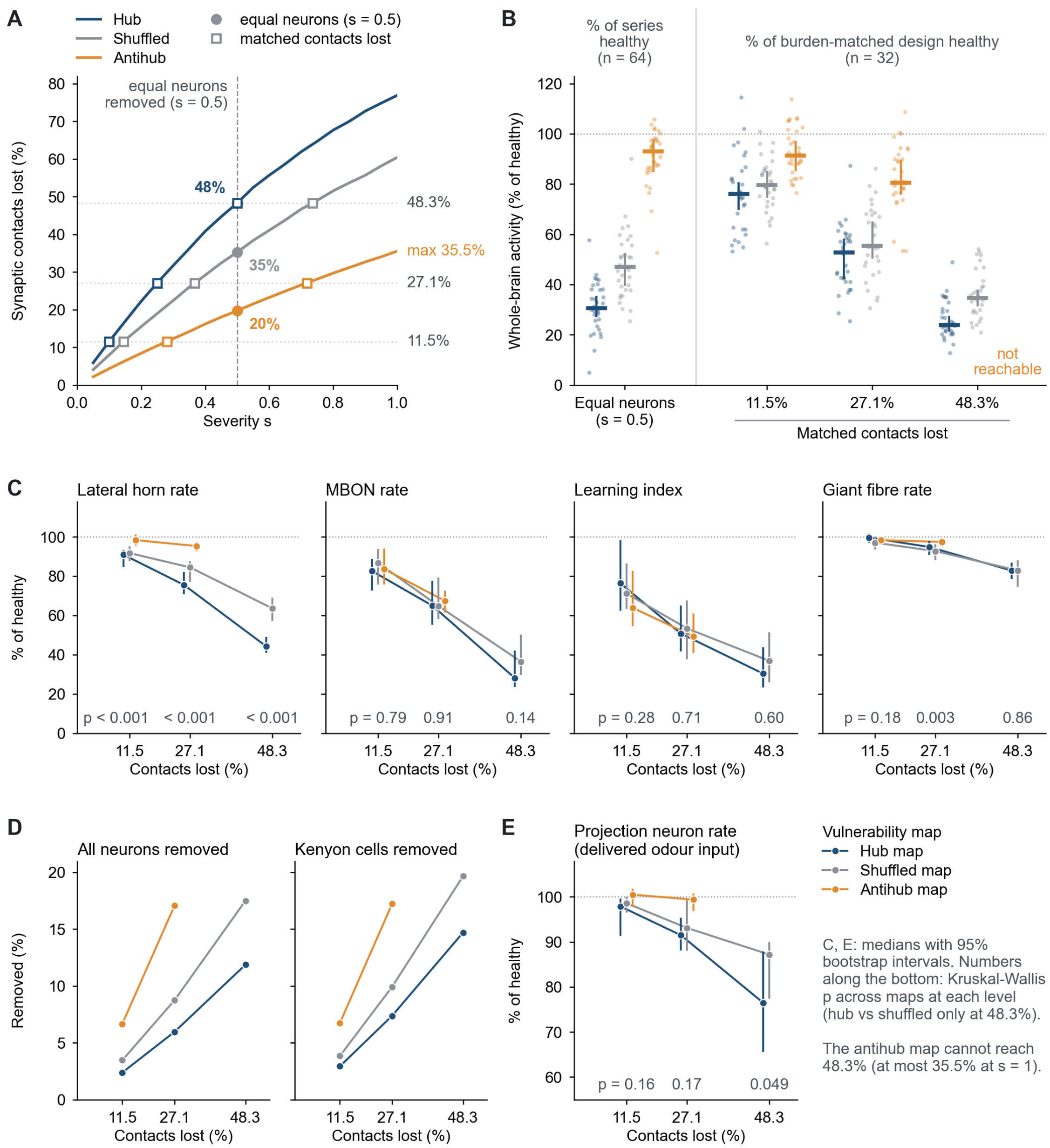

**Fig 4. The vulnerability maps compared at equal synaptic contacts lost.** Intervals are 95% percentile bootstrap intervals of the median as a percentage of the healthy median. (A) Synaptic contacts lost against severity under each map (mean of three random draws of the damage). Dashed line, s = 0.5, where the maps are compared at equal numbers of neurons removed in Fig 3C; dotted lines, the three matched levels (contacts lost under the hub map at s = 0.10, 0.25 and 0.5); squares, the severities at which each map reaches them. (B) Whole-brain activity during odour A. Points, flies (jittered); thick bars, medians; thin lines, intervals; horizontal dotted line (also in C and E), healthy median. Left, the map comparison, as a percentage of the 64 healthy flies of the severity series; right, the contact-matched comparison ("burden-matched design" on the axis), as a percentage of its own 32 healthy flies. Percentages are compared within a design; 32 flies per group. (C) Four other readouts at the matched levels, as percentages of the healthy median of the contact-matched comparison. The p values printed are uncorrected. Only whole-brain activity had a pre-specified criterion; the readouts in C and E are descriptive. (D) Median percentage of all neurons and of Kenyon cells removed per fly. (E) PN rate, plotted as in C.

## The maps differ less when matched for synaptic contacts lost

To find out, the maps were compared at equal fractions of contacts lost (the contact-matched comparison). The hub map at s = 0.10, 0.25 and 0.5 removes 11.5%, 27.1% and 48.3% of contacts (Fig 4A); the severity of the other two maps was raised until they removed the same fractions, the three matched levels. The antihub map removes only 35.5% at s = 1.0 and cannot reach 48.3%. The comparison comprised 32 new flies per map and matched level and 32 healthy flies, to which their percentages refer (S1 Text).

At 11.5% of contacts lost, whole-brain activity was 76% (70 to 81%), 80% (75 to 85%) and 91% (86 to 97%) of the healthy median under the hub, shuffled and antihub maps; at 27.1%, 53% (43 to 58%), 55% (51 to 65%) and 81% (76 to 89%); and at 48.3%, 24% (22 to 27%) under the hub and 35% (32 to 38%) under the shuffled map (Fig 4B). The hub-map median was the lowest at every level: the antihub map stayed above it ($p = 4 \times 10^{-6}$ at 11.5%), the shuffled map detectably so only at the highest level ($p = 1 \times 10^{-5}$; p = 0.12 and 0.09 at the lower two).

A criterion set before these flies were simulated, for whole-brain activity only, judges whether the maps differ by more than replicate groups do (Materials and methods). R is the median under the most active map divided by that under the least active. If R is at most 1.27, the replicate variation at s = 0.5 (S2 Fig), the maps differ within replicate variation, which is taken to mean that contacts lost account for their difference; if the whole interval of R lies above 1.27 and the hub map is the least active, they differ beyond it, by more than contacts lost account for; otherwise the outcome is unresolved. The outcome was within replicate variation at 11.5% (R = 1.20, 1.13 to 1.31), where even antihub-map flies as active as healthy flies would not have taken it beyond (R = 1.31, 1.25 to 1.43; S1 Text); beyond it at 27.1% (R = 1.52, 1.41 to 1.93); and unresolved at 48.3% (R = 1.45, 1.26 to 1.62). The difference between maps was thus clearly beyond replicate variation only at the middle level.

Did matching for contacts lost bring the other maps closer to the hub map? For the antihub map it did: measured as the fall below the healthy median, its deficit was a tenth of that under the hub map with equal numbers of neurons removed (10%, 3 to 21%) and over a third at its two matched levels (36% and 41%), where its flies had 2.8 to 2.9 times as many neurons removed as hub-map flies (Fig 4D), so their higher activity does not come from sparing neurons. For the shuffled map it depends on the hub-map group used as reference: shuffled-map flies were less active at 48.3% of contacts lost than with equal numbers of neurons removed (35% against 47% of healthy), but so was the hub-map group simulated with them (24% against 31% in Fig 4B, a difference between replicate groups), so within each comparison the shuffled/hub ratio hardly changed (1.53 against 1.45); against any one hub-map group it fell by about a quarter (S1 Text).

No criterion had been set for the other readouts (Fig 4C; S1 Text). Lateral horn rate, Kenyon cell rate and MN9 drive still differed between maps at every matched level, lowest under the hub map, but none beyond its own replicate variation when the criterion was applied afterwards; giant fibre rate no longer differed between the hub and shuffled maps; and MBON rate and the learning index did not differ detectably between maps ($p = 0.14$ to $0.91$), although at 48.3% a difference as large as that in whole-brain activity could have gone undetected.

The remaining difference between maps cannot be attributed to where the damage falls, for two reasons. First, matching the maps raised the severity, and so the loads and threshold shifts, of shuffled-map and antihub-map flies (mean shift 0.24, 0.35 and 0.68 mV under the hub, shuffled and antihub maps at 27.1%). The threshold shift partly offsets the losses, most strongly in the mushroom body (Fig 3A). That mushroom body output did not differ detectably between maps may therefore reflect equal contacts lost, or a larger deficit under the shuffled and antihub maps offset by their larger shifts, which may also raise whole-brain activity under these maps; a contact-matched comparison without the shift, not run here, would distinguish the two. Second, hub-map flies, the least active, also received the least odour input: the median PN rate was lowest under the hub map at every matched level (Fig 4E; 76.5% against 87% of the healthy median at 48.3%, Kruskal-Wallis $p = 0.049$), and it correlated with whole-brain activity across flies (Spearman $\rho = 0.42$ to 0.75; S1 Text). Contacts lost thus account for part of the difference between maps; how large the rest is, and whether it is caused by where the damage falls, is not established.

### Within single flies, lost synapses account for more of the whole-brain deficit than lost neurons

The group comparisons left open what, beyond contacts lost, makes a deficit large; within a single fly one can ask which part of its damage accounts for its deficit. Simulating a fly again with identical input spikes but part of its damage undone shows what that part causes for that input (the within-fly analysis; Materials and methods). In 48 flies of the severity series, 24 at each of $s = 0.3$ and 0.5, either all removed neurons were restored, each with its threshold shift and its own lost synapses, or all the synapses that surviving neurons had lost (thinned synapses) were restored, or the threshold shift was removed. These runs used a shortened protocol without training, and so give no learning index. Results are given as the fraction of the fly's deficit restored, where 0 is the damaged fly and 1 the same fly without pathology (its healthy copy).

Synapse loss accounted for more of the whole-brain deficit than neuron loss (Fig 5A): over both severities, restoring all thinned synapses recovered a median of 62% of it (45 to 71%) and restoring all removed neurons 39% (30 to 42%), fractions referred to here as the deficit due to synapse loss and to neuron loss. Per fly they summed to about the whole deficit (S1 Text). The deficit due to synapse loss was the larger at both severities (64% against 49% at $s = 0.3$, 60% against 30% at $s = 0.5$); only that due to neuron loss fell with severity (Mann-Whitney $p = 0.001$). Yet of the contacts lost at $s = 0.5$, removed neurons had taken more than the synapse-loss rule had (26.3% against 22.2% of all contacts; S1 Text). Neuron loss accounted for more only for giant fibre rate (84% against 16%) and, except at $s = 0.3$, for lateral horn rate (57% against 44%).

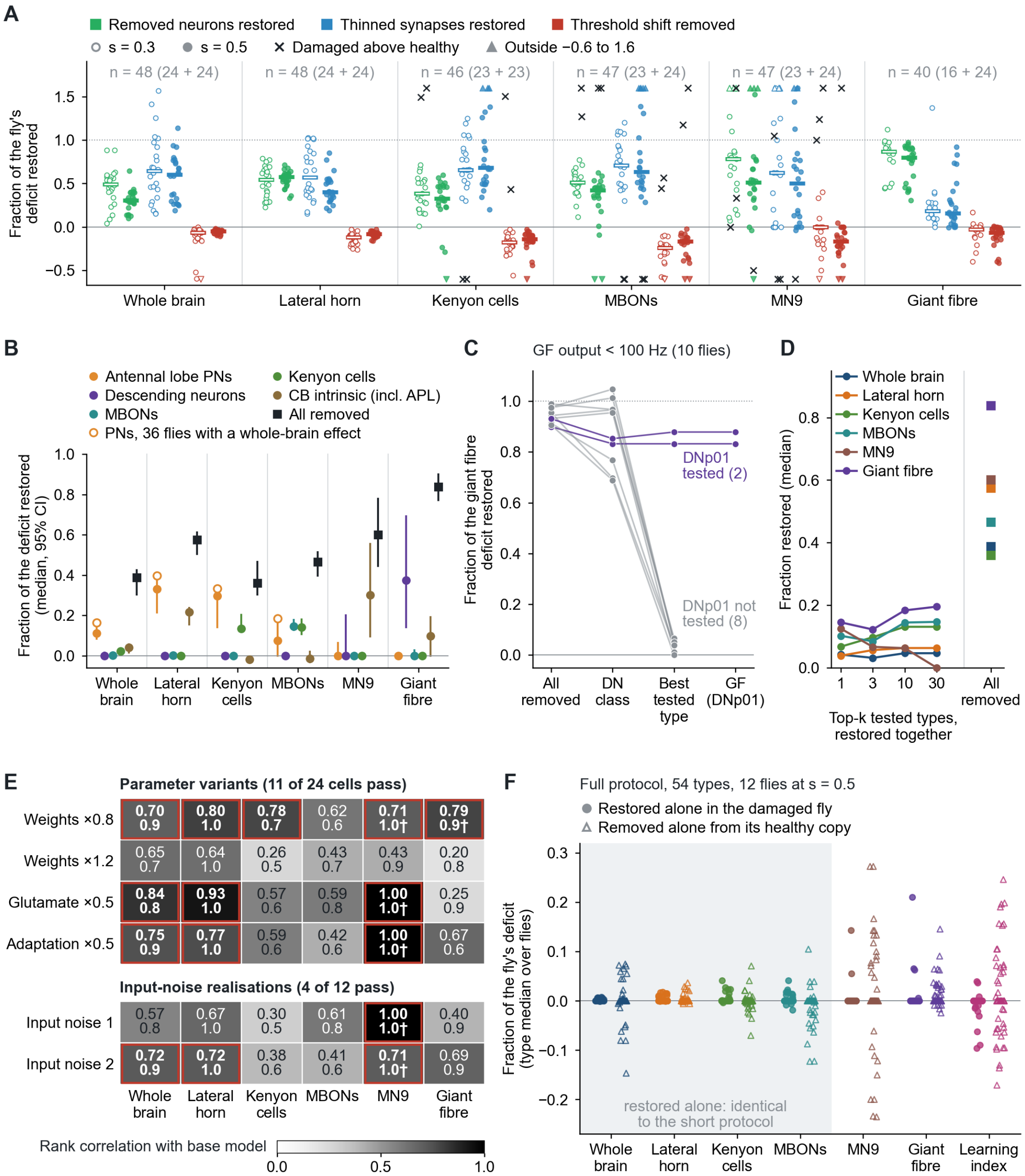

| | Whole brain | Lateral horn | Kenyon cells | MBONs | MN9 | Giant fibre |
|---|---|---|---|---|---|---|
| Weights ×0.8 | 0.70 0.9 | 0.80 1.0 | 0.78 0.7 | 0.62 0.6 | 0.71 1.0† | 0.79 0.9† |
| Weights ×1.2 | 0.65 0.7 | 0.64 1.0 | 0.26 0.5 | 0.43 0.7 | 0.43 0.9 | 0.20 0.8 |
| Glutamate ×0.5 | 0.84 0.8 | 0.93 1.0 | 0.57 0.6 | 0.59 0.8 | 1.00 1.0† | 0.25 0.9 |
| Adaptation ×0.5 | 0.75 0.9 | 0.77 1.0 | 0.59 0.6 | 0.42 0.6 | 1.00 1.0† | 0.67 0.6 |
| Input noise 1 | 0.57 0.8 | 0.67 1.0 | 0.30 0.5 | 0.61 0.8 | 1.00 1.0† | 0.40 0.9 |
| Input noise 2 | 0.72 0.9 | 0.72 1.0 | 0.38 0.6 | 0.41 0.6 | 0.71 1.0† | 0.69 0.9 |

**Fig 5. Within single flies, the deficit due to neuron loss comes mostly from input and command neurons.** Within-fly analysis in 48 flies of the severity series (24 each at s = 0.3 and 0.5). Values are the fraction of the deficit restored (0, the damaged fly; 1, its healthy copy; Materials and methods). A readout is omitted for a fly whose deficit is under 5% of the healthy value (n in A, by severity in brackets). A to E used the shortened protocol (printed as short protocol in F). (A) Restoring all removed neurons, restoring all thinned synapses or removing the threshold shift. Bars, medians (hollow, s = 0.3; solid, s = 0.5); crosses, the 10 fly-readout pairs with the damaged value above the healthy value; values outside −0.6 to 1.6, at the axis edge. (B) Restoring all removed neurons of one class, next to all removed neurons; medians over flies with 95% bootstrap intervals (CI); open circles, as keyed (S1 Text). Antennal lobe PNs, the projection neuron (PN) class; CB intrinsic, central-brain intrinsic neurons without a finer class label; APL, anterior paired lateral neurons. (C) The 10 flies with a giant fibre (GF) rate below 100 Hz (printed as GF output): deficit restored by all removed neurons, the descending-neuron (DN) class, the fly's most effective tested cell type (best tested type) and, in 2 flies (purple), DNp01 (both giant fibres). (D) Median fraction restored by the k = 1, 3, 10 or 30 tested types most effective in that fly, restored together, against all removed neurons (squares). (E) Robustness test: the ranking of 54 cell types in 12 flies at s = 0.5 under each variant (rows). "Base model" in this panel means the extended model with unchanged parameters, not the published model. Shading and upper number, Spearman correlation with it of the types' medians over flies; lower number, fraction of the ten highest-ranked types shared. Red outlines, cells meeting both requirements of the pre-specified robustness condition (correlation at least 0.7, overlap at least 0.6); daggers, cells that pass on the strength of 5 or fewer types with a non-zero median. (F) The same types and flies under the full circuit protocol; points, the types' medians over flies. Circles, the type's removed neurons restored alone in the damaged fly; triangles, only those neurons removed from the fly's healthy copy (fraction of the deficit reproduced; S1 Text).

Without its threshold shift, every fly with a deficit had lower whole-brain activity and lateral horn, Kenyon cell and MBON rates, the deficit growing by medians of 5%, 10%, 16% and 20% of its size, and MN9 drive and giant fibre rate were lower in most flies (Fig 5A; S1 Text). The shift therefore partly offsets the losses throughout the brain, most strongly in the mushroom body.

### The deficit due to neuron loss comes mostly from input and command neurons

Thinned synapses were restored only all together, so which of them mattered was not tested. Removed neurons were also restored in parts, one class and then one cell type at a time (the screen; a cell type is an annotated category of neurons, a class a broader one). Of a median of 4,982 cell types per fly that had lost neurons, 450 were tested, holding a median of 18% of the removed neurons: the 400 that had lost the most neurons within the most effective classes, and 50 random controls. Types with few neurons, among them the giant fibres and the projection neuron types that deliver odour A, were therefore rarely tested.

Over both severities, the antennal lobe projection neurons (the PN class) had the largest effect of any class on whole-brain activity, lateral horn rate and Kenyon cell rate (Fig 5B; S1 Text): a median of 99.5 removed neurons per fly, 0.65% of all removed, whose restoration recovered 11%, 33% and 30% of these deficits (8 to 17%, 21 to 40% and 14 to 33%). The effect came from the eight neurons that deliver odour A: it grew as the PN rate fell (for whole-brain activity, $\rho = -0.85$, $p = 2 \times 10^{-14}$) and was absent in the 12 flies with undiminished odour input (median PN rate 102% of the healthy copy's). In the other 36 flies (median PN rate 76%), the PN class recovered about half as much as all removed neurons did for whole-brain activity and two-thirds as much for lateral horn and Kenyon cell rates. A large part of the deficit due to neuron loss in the odour readouts is therefore loss of the odour input itself.

The descending neurons, which link the brain to the ventral nerve cord and include the giant fibres, recovered 37% of the giant fibre deficit (14 to 69%; Fig 5B). In the ten flies with a giant fibre rate below 100 Hz, about half the healthy rate, the deficit came from the loss of descending neurons and, where this could be tested, of the giant fibres themselves (Fig 5C): restoring all removed neurons recovered 90 to 99% of it and restoring the descending-neuron class 69 to 105%. In the two flies that had lost both giant fibres, restoring them recovered 83% and 88%. In the other eight, which still fired at 10 to 55 Hz, as after loss of the left giant fibre alone, the giant fibre type (DNp01) was not among the types tested, and the most effective tested type recovered at most 6.6% (S1 Text).

Beyond these input and command neurons, single tested cell types recovered little (Fig 5D). Most single-type restorations (91 to 98% across readouts) left the readout exactly as in the damaged fly. The most effective tested type in a fly recovered a median of 4.2% of the whole-brain deficit (2.2 to 6.4%), and the ten most effective together 4.6%, hardly more and 10% of what all removed neurons recovered (medians over flies; S1 Text).

Nor could the tested types be ranked reliably by their effect. Of two conditions set before the screen for reading such a ranking as more than wiring or chance (Materials and methods), the first, that wiring alone should not predict which types matter, was met (median leave-one-type-out $R^2$ of five wiring measures between −0.02 and 0.05), but means little because most types changed nothing. The second, that the ranking should hold under changed parameters and input noise, was not (Fig 5E): for 54 types restored again in 12 of the $s = 0.5$ flies under six such variants, it agreed as required with the ranking of the unchanged model in 15 of the 36 combinations of readout and variant (S1 Text). Under the full circuit protocol the same types each recovered at most 4.2% of any odour readout deficit and 3.9% of the learning-index deficit (medians over flies; Fig 5F; S1 Text).

## Loss of single neurons produces discrete circuit phenotypes

Two changes in the severity series were discrete rather than graded (Fig 6A to 6C; S1 Text): from $s = 0.05$ upward some flies had a giant fibre rate below 100 Hz, and from $s = 0.10$ upward some had a silent MN9. Both giant fibres have a connectivity rank above the 99.6th percentile, so each is removed with nearly the highest probability of any neuron, and chance decides which flies lose one. Of the 320 damaged flies, all 66 that had lost the left giant fibre were below 100 Hz, against 1 of the 254 that kept it (Fig 6A). The drop itself is expected, because the rate averages the two giant fibres and looming on the left drives mainly the left one; what the counts show is that a single neuron splits a severity group in two.

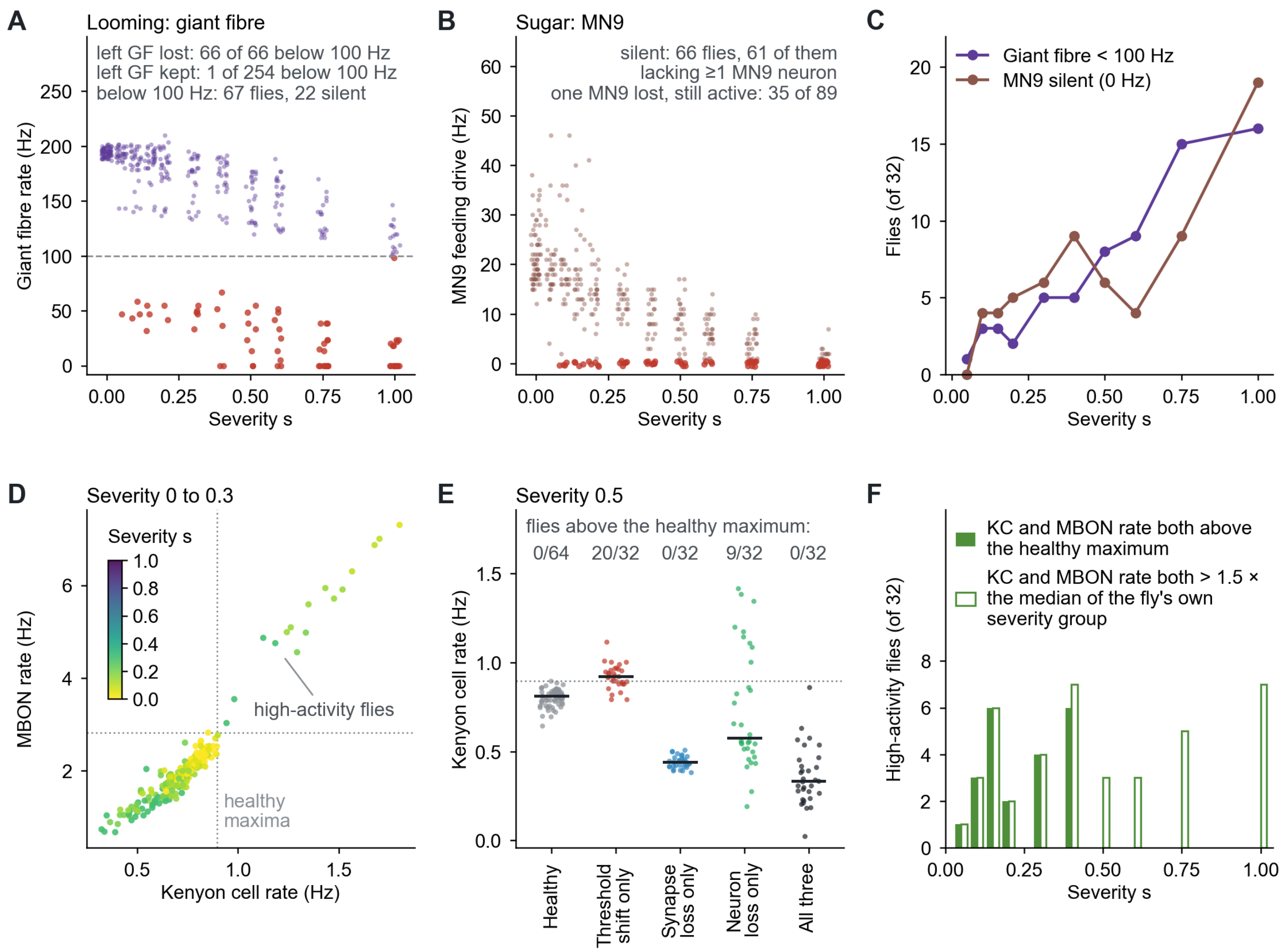

**Fig 6. Loss of single neurons produces discrete circuit phenotypes.** A to D and F, severity series. (A) Giant fibre (GF) rate; red, below 100 Hz (dashed line). (B) MN9 feeding drive; red, silent flies (0 Hz), with a small vertical jitter. The counts printed in A and B give the flies by the neurons they lost. (C) Flies, of 32, with each phenotype at each severity. (D) MBON rate against Kenyon cell rate during odour A for the 224 flies from s = 0 to 0.3; dotted lines, highest healthy values. (E) Rule ablation at s = 0.5: Kenyon cell rate during odour A in the 64 healthy flies and under each rule alone and all three rules (32 flies each). Bars, medians; dotted line, highest healthy value. (F) Flies (of 32) at each severity whose Kenyon cell (KC) and MBON rates both exceeded the highest healthy values (filled bars, the high-activity flies) or 1.5 times the median of their own severity group (open bars; S1 Text).

Below 100 Hz the response was not always lost: of the 67 such flies, 22 were silent and 45 still fired (median 38 Hz), consistent with a response of the right giant fibre alone. Such flies could still take off: of the 15 flies of the arena cohort below 100 Hz that met a looming threat, 13 took off. A take-off, a short burst of giant fibre spikes, is scored with or without a threat, so this shows that the burst can still be fired, not that escape succeeds (S1 Text). A rate below 100 Hz is thus a circuit phenotype, the mark of a lost left giant fibre, not in itself a failure to escape. A silent MN9 followed the loss of an MN9 neuron less closely (counts in Fig 6B; S1 Text).

The mushroom body showed the opposite discrete phenotype: in a few flies (the high-activity flies), Kenyon cell and MBON rates both exceeded the highest healthy values, 1 to 6 of 32 at each severity from 0.05 to 0.4 and none from s = 0.5 (Fig 6D and 6F; S1 Text). Of the rules applied alone (rule ablation, s = 0.5), only neuron loss moved flies far above the healthy maximum Kenyon cell rate (Fig 6E). A candidate cause is loss of the anterior paired lateral (APL) neurons, a pair that inhibits all Kenyon cells [38, 39]: it was the cause in the one fly in which the within-fly analysis tested it. That fly was high for its group but below the healthy maxima, so for the high-activity flies APL remains a candidate (S1 Text).

### Separation from healthy flies follows the change relative to trial-to-trial spread

To find which readouts reveal damage milder than the s = 0.5 of most comparisons above, the analysis measured how well each separates a severity group from the healthy flies as the area under the receiver operating characteristic curve (AUC), the probability that the readout orders a damaged and a healthy fly correctly, and ranked readouts by the first severity at which the AUC reached 0.80 (Fig 7A and 7B; S1 Text). In the severity series, whole-brain activity and lateral horn rate reached it at s = 0.10 (bootstrap range 0.05 to 0.15) and the other readouts between s = 0.2 and 0.4, the learning index and PN rate last; the learning index divided by MBON rate never did. Whole-brain activity had a higher AUC than the learning index at every severity from 0.05 to 0.3 (at s = 0.05, 0.75 against 0.56, close to the chance level of 0.55; difference 0.19, 95% paired-bootstrap interval 0.03 to 0.31), and so did lateral horn rate (S1 Text).

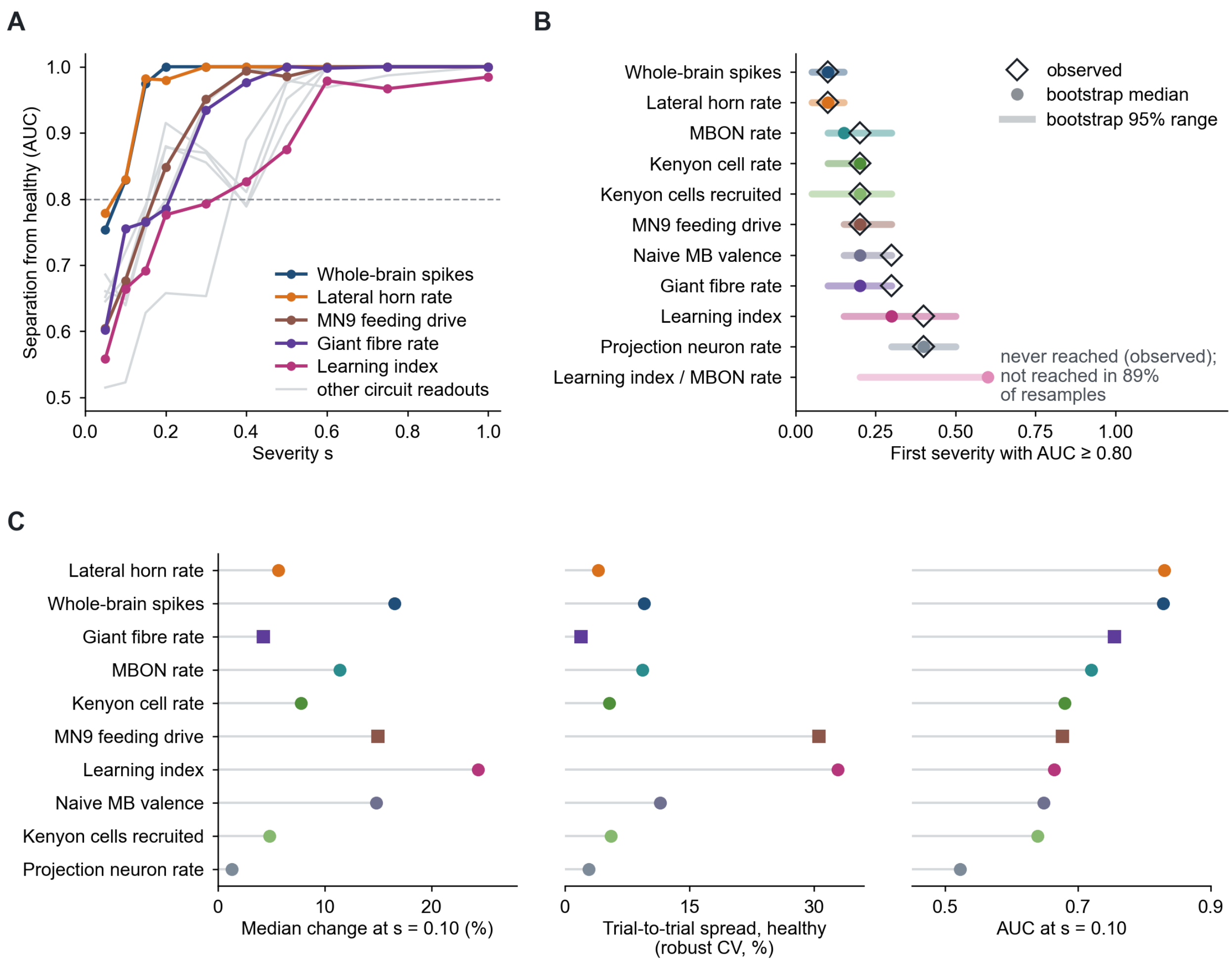

**Fig 7. Separation from healthy flies follows the change relative to trial-to-trial spread.** Separation is the area under the receiver operating characteristic curve (AUC) between a group and its healthy flies, taken in whichever direction exceeds 0.5. (A) AUC against severity for the ten readouts of Fig 2 (five in colour, as keyed; the others grey); dashed line, AUC 0.80. (B) First severity at which each readout reaches AUC 0.80. Diamonds, the full sample; dots and bars, median and 2.5 to 97.5 percentile range over the bootstrap resamples that reached it. The learning index divided by MBON rate (printed with a slash) never reached it in the full sample. (C) At s = 0.10: absolute change of the group median as a percentage of the healthy median; trial-to-trial spread in healthy flies as a robust coefficient of variation (CV, %); and AUC. Squares, the two motor readouts with discrete phenotypes (Fig 6).

An AUC generally rises with a readout's change, the percentage shift of the group median from the healthy median, and falls with its spread, the variation among healthy flies relative to their median (robust coefficient of variation), which ranged from 1.9% for giant fibre rate to 32.8% for the learning index (Fig 7C). The healthy flies are one network and differ only in their input noise, so their spread is trial-to-trial spread, not variation between individuals. Whole-brain activity separated at low severity because it combines a large change (16.6% at s = 0.10) with a moderate spread (9.6%), and lateral horn rate because its spread is small (4.0%, against a fall of only 5.6%). At s = 0.10 the learning index changed most, by about a quarter, but less than its spread (AUC 0.66, against 0.83 for the other two). Across the eight graded readouts, those of Fig 2 other than MN9 drive and giant fibre rate with their discrete changes, AUC at s = 0.10 followed the absolute change divided by the spread ($\rho = 0.74$, exact permutation $p = 0.046$, $n = 8$), not the change alone ($\rho = 0.38$) or the inverse spread alone ($\rho = -0.07$), as it did at s = 0.05 and 0.2 (S1 Text).

The ranking was repeated in the 176 flies of the arena cohort (S4 Fig). Whole-brain activity and lateral horn rate again reached an AUC of 0.80 at s = 0.10, but the order of the other readouts was not reproduced (S1 Text). The arena readouts time feeding and jumps per threat declined with severity but first reached that AUC at s = 0.5, the highest severity in this cohort and well above that of the two population readouts.

## Discussion

Three rules of amyloid-like pathology, all scaled by a fly's severity (s), were applied to a simulation of the whole adult fly central nervous system: synapse loss and neuron loss, which reduced circuit activity, and a lowered spike threshold (the threshold shift), which raises excitability and partly offset them. In this model the number of neurons lost was a poor measure of damage: what the loss of neurons cost depended partly on how many synaptic contacts went with them and on whether they delivered a stimulus or commanded a response. Readouts (the measured quantities) detected milder damage the more they changed relative to their variation between trials.

### Neuron counts, synapses and hubs

With equal numbers of neurons removed, whole-brain activity (all spikes during an odour) was three times as high when the damage fell most on poorly connected neurons (the antihub map) as when it fell most on highly connected ones (the hub map) at s = 0.5; damage placed at random (the shuffled map) lay between. A structural analysis would not predict so large an effect: in FlyWire, the connectome of a female fly brain, the network stays connected even when neurons are removed in order of degree (number of partners), from the highest down to a degree of about 50 [40], and by that measure is robust to the loss of its hubs.

The obvious explanation, that contacts rather than neurons are the unit of damage, proved partly right when the maps were compared at three matched levels of contacts lost. In whole-brain activity the antihub map came much closer to the hub map, the shuffled map only when both comparisons were set against the same hub-map group (Results). By a pre-specified criterion, the remaining difference clearly exceeded the variation between replicate groups of one condition at the middle level (27.1% of contacts lost); it lay within that variation at the lowest level, where it could hardly have exceeded it, and was unresolved at the highest.

What remains cannot be attributed to where the damage falls: the maps matched for contacts lost still differed in their threshold shifts, which were larger under the shuffled and antihub maps and raise activity, and in the odour input delivered, which was lowest under the hub map (Results). Contacts lost thus explain part of the difference between maps; how much remains, and whether a neuron's place in the network causes it, is not established.

This partial role of contacts lost resembles the human observation that synapse density tracked cognitive impairment better than plaque or tangle counts [1], but only loosely: there the poorer predictor was a count of lesions, here a count of lost neurons, and the model's result holds for assumed rule sizes. Contacts lost are not a complete measure either: removed neurons took with them more contacts than the synapse-loss rule did, yet within single flies they accounted for less of the whole-brain deficit than the synapses lost by surviving neurons (Results).

Weighting damage by connectivity is borrowed from human imaging [5–7] and from models of activity-dependent degeneration [9]. It is untested in flies, where single-cell data on tau and α-synuclein toxicity point to vulnerability set by cell identity [41], and in human Alzheimer's cortex early losses concentrate in particular inhibitory neuron types [42]. A better starting assumption would set vulnerability by cell type from single-cell expression data, with a term for activity.

### Where the deficit lies within single flies

Simulating a fly again under identical input with part of its damage undone (the within-fly analysis) has no exact counterpart in animals. The resulting split of a deficit between removed neurons and the synapses lost by surviving neurons holds for the assumed sizes of the rules (Table 1) and would change with them.

Where removed neurons accounted for much of a deficit, it came mostly from a few groups (Fig 5): the projection neurons that deliver odour A and the descending neurons that include the giant fibres, which command escape. The giant fibres and most of these projection neurons are among the most connected neurons, but connectivity alone did not make a cell type important (S1 Text).

Beyond these input and command neurons, cell types restored one at a time (the screen), or the most effective of them together, recovered little, and their ranking did not hold when a model parameter was changed or the random timing of the input was redrawn. Only the few effects that held under redrawn input, such as those of the cell types PVLP151 on giant fibre rate and GNG120 on MN9 drive (S1 Text), are worth testing in flies. Nor does the screen show that the rest of the deficit due to neuron loss is spread over many cell types: it chose them by the number of neurons lost and rarely tested those with few.

Only sensory neurons are exempt from pathology, so the projection, visual and dopaminergic neurons that the model drives to deliver the odour, the looming stimulus and the teaching signal can be removed, and part of every odour- and looming-evoked deficit is this loss of input (Results), most of all in odour-evoked activity and less in learning and in the rate of the mushroom body output neurons (MBON rate; S1 Text).

In amyloid-β42 flies with pan-neuronal expression these input neurons, and the receptor neurons the model spares, would also be exposed, so a reduced odour response or learning score could reflect in part a weaker input or teaching signal. Imaging projection neuron or dopaminergic neuron responses in the same animals would distinguish the two, and confining expression to the mushroom body largely avoids the problem.

Under the hub map a few highly connected neurons, such as the giant fibres, are among the likeliest to be removed at every severity, and their loss adds discrete phenotypes to a graded decline. A giant fibre rate below 100 Hz, about half the healthy rate, is a change in circuit activity, not in itself a failure to escape: in the model, recording giant fibre responses to looming would split flies of one severity into two groups, whereas counting take-offs mostly would not.

In the mushroom body, a few flies at low severity had Kenyon cell and MBON rates above the healthy range. Loss of the two inhibitory anterior paired lateral (APL) neurons is a candidate cause, shown in the one fly in which it was tested, whose rates were high for its severity group; restricting amyloid-β42 expression to APL would test it. More generally, a few animals with an unusual phenotype may reflect the loss of a particular neuron rather than noise.

### Early markers

Because readouts detected milder damage the more they changed relative to their trial-to-trial spread, a population activity measure with a small spread, such as a lateral horn calcium signal, is a better candidate for early detection in flies than a learning score or behavioural measures: in the model, the learning index and behaviour in a virtual arena varied far more between trials (S1 Text). Loss of input does not explain the detection of damage at low severity: at $s = 0.10$ the median PN rate had changed by only 1.3% (S1 Text). The principle, change relative to spread, should transfer to animals; the model's ranking may not, because its healthy flies are one network, whose spread is variation under random input, whereas real animals also differ in wiring, excitability and internal state.

The learning index is computed from MBON firing, not from behaviour, and most of its deficit is a loss of that firing (Results), so a reduced learning score in amyloid-β42 flies can arise without a specific defect in plasticity.

### The threshold shift partly offsets the losses

The offset was partial, and the model's shift is smaller than the change in excitability measured in cultured neurons of amyloid-β42 flies [11], so a shift of the measured size could offset more; raised excitability was also protective in a neural mass model of activity-dependent degeneration [43]. On one point the model disagrees with flies: it predicts that a lowered spike threshold alone raises MBON rate and the learning index (Fig 3A; S1 Text), whereas loss of Kv4 function alone impaired larval learning [11]. In amyloid-β42 flies, restoring the potassium channel Kv4, whose loss raises excitability, also slowed the loss of Kenyon cells [11], so hyperexcitability may contribute to neuron loss, which the model's independent rules cannot represent. The model further lacks homeostatic compensation [44], elimination of hyperactive neurons [45], sleep [46] and the failing inhibition to which amyloid-associated hyperactivity is attributed in mammals [47, 48]. That the shift only offset the losses therefore reflects how the model was built, not the biology of hyperexcitability.

## Limitations

The main limitation is that the damage goes beyond the range in which the base model, the published model that this model extends, was shown to be robust: a uniform change of plus or minus 30% in synaptic weight, with its sensory input readjusted [19]. Under the hub map, 48% of all synaptic contacts were lost at s = 0.5, the severity of many of the comparisons, and 77% at s = 1.0 (S1 Text). The fraction of contacts lost passes 30% between s = 0.25 and s = 0.3, so results from about s = 0.3 upward, the within-fly analysis included, describe this model under large perturbation and do not inherit the base model's validation. The lowest severity at which readouts detected damage, s = 0.10 (11.5% of contacts lost), lies below that point. The spike-frequency adaptation and feedforward antennal lobe added in this study were not part of the check of the simulator against the published code, and the latter removes recurrent circuitry and places antennal lobe computation outside the model's scope.

Severity is tied to data at one point: the Kenyon cell loss reported at 40 days in flies expressing amyloid-β42 in the mushroom body alone [11] is reached between s = 0.5 and 0.6. This measurement shows which severity matches that loss, not how strong any rule should be: the sizes of all three rules are assumed, and only their direction follows observations in amyloid-β flies. The synapse-loss rule is uniform across cell types, for want of per-type synapse counts in amyloid-β42 flies, which electron microscopy could now supply [49]; vulnerability follows synapse count rather than measured susceptibility; the rules act independently; and pathology does not propagate between neurons (Table 1).

Loss of input could be distinguished from damage to the circuit only approximately: the PN rate averages all 686 projection neurons, only a few of which deliver the odour, and no simulation damaged the whole network while sparing the input neurons. Replicate groups of 32 flies, simulated under one condition, differed by about a quarter in whole-brain activity and by more in MBON rate and the learning index (S2 Fig), so smaller differences between two groups cannot be distinguished from replicate variation.

The within-fly analysis tested 450 of a median of 4,982 cell types that had lost neurons in a fly, measured the learning index in only 12 flies, and restored neurons with their threshold shift and their own lost synapses still in place. Cell types other than the input and command neurons are therefore only candidates, and which lost synapses matter was not tested (Results). Each step of the stimulus sequence starts from the state left by the step before, so the learning index also picks up changes in earlier steps, and odour A was always the trained odour, with no reverse pairing as a control.

Glutamate is treated as inhibitory throughout, following the base model [19], although glutamatergic signalling in the fly is not uniformly inhibitory. The published weight per synaptic contact was retained although MaleCNS, the connectome used here, has more input synapses per neuron than FlyWire, which at face value calls for a lower weight; the effect on the results was not determined (Materials and methods). Glia, neuromodulation apart from dopamine-gated plasticity, homeostatic plasticity, electrical synapses, spontaneous activity and developmental compensation are absent, and the connectome comes from a single male. In the virtual arena, each simulated brain is coupled to a simple model body, the odour readouts failed as a measure of memory, and a take-off requires giant fibre spikes although real flies also take off without them [33], so conclusions about behaviour are limited.

## Predictions

A connectome-constrained disease model makes the path from rule to circuit readout explicit, and each rule, like the choice of which neurons are damaged most, can be replaced as measurements become available. The model predicts that deficits in amyloid-β42 flies follow the number of synaptic contacts lost, those of lost neurons included, more closely than the number of neurons lost, and that Kenyon-cell-restricted expression produces a much smaller mushroom body deficit than pan-neuronal expression at the same Kenyon cell loss. If pan-neuronal expression removes projection neurons, as the model assumes, their summed response to odour should fall, and while damage is still mild a minority of flies should show Kenyon cell and MBON rates above the healthy range, with loss of APL a candidate cause (S1 Text). A reduced learning score should come with reduced MBON responses to the trained odour before training. Population-averaged activity should distinguish affected flies earlier than feeding and take-off measures do, provided its variation across animals is small relative to the change. The mismatch between damage and deficit is a human observation, and the model's answers do not transfer to people directly, but a fly model shows how such a mismatch can arise: the same number of lost neurons produces different deficits depending on how many synapses go with them and on whether they include the neurons that deliver the input or single cells such as a command or inhibitory neuron.

## Materials and methods

Every result reported here comes from simulation or from analysis of published connectome data; no animal was used and no new measurement was made. Table 1 shows what in each pathology rule follows fly data and what is assumed; no rule parameter was fitted. Further methods and secondary results are given in S1 Text.

### Connectome

The network was built from MaleCNS v1.0, the connectome of the central nervous system of one adult male *Drosophila melanogaster* [18], and keeps synapses with a confidence score of at least 0.5 and every reconstructed cell with a superclass annotation, the broadest category of neuron (166,700 neurons). Each connection was made excitatory (+1) or inhibitory (−1) by the consensus transmitter prediction for its presynaptic neuron: acetylcholine excitatory, γ-aminobutyric acid (GABA) and glutamate inhibitory, glutamate as in the published whole-brain model [19], referred to here as the base model. The 11,609 neurons with another prediction or none, dopaminergic neurons among them, receive input but have no output connections. The network has 24,469,412 connections made of 120,260,398 synaptic contacts (single synapses). Two releases of FlyWire, the connectome of one adult female brain [15, 31] (electron microscopy volume [50]; synapses [51, 52]; transmitters [53]), were also used: v630, on which the base model was published [19], to check the simulator written for this study, and v783 to compare synapse counts (S1 Text).

## Neuron model and simulation engine

Each neuron is a leaky integrate-and-fire neuron with the parameters of the base model [19]: resting and reset potential −52 mV, threshold −45 mV, membrane time constant 20 ms, synaptic time constant 5 ms, refractory period 2.2 ms, synaptic delay 1.8 ms and unitary synaptic weight 0.275 mV per contact, integrated at 0.1 ms. Its membrane potential v follows $dv/dt = (v_0 - v + g)/\tau_m$ ($v_0$, resting potential; $\tau_m$, membrane time constant); the synaptic input g, in mV, decays with the synaptic time constant. After the synaptic delay, a presynaptic spike changes g of the postsynaptic neuron by the unitary weight times the number of contacts, with the sign of the connection. Stimulated neurons, which deliver a stimulus, received independent Poisson input, random spike trains whose timing is the input noise; as in the base model, each input spike raised the membrane potential by 68.75 mV (250 unitary weights) and stimulated neurons had no refractory period, so an intact stimulated neuron fires at nearly the input rate. Non-stimulated neurons receive no background input, so the network is silent at rest. The simulator, called the engine here, advances many copies of the network in parallel; each copy, with its own weights and thresholds, is one simulated fly. Like the base model, it discards synaptic input that arrives while a neuron is refractory (S1 Text).

## Engine validation

The engine was checked by running the sugar-stimulation example distributed with the published code [19, 29] (base model on FlyWire v630; 21 sugar-sensing gustatory receptor neurons of one side at 150 Hz; 30 trials of 1 s) in the engine and, with the unmodified code, in Brian2 2.10.1, the simulator of the base model. Firing rates agreed neuron by neuron (Pearson correlation 0.9997 across the 438 neurons active in either run; S1A Fig; further checks in S1B to S1D Fig and S1 Text).

On MaleCNS, right-side putative sugar-sensing neurons drove mainly one of the two MN9 feeding motor neurons (S1 Text), so the circuit protocol, the stimulus sequence of every fly, gives sugar on both sides and MN9 feeding drive averages the two neurons.

### Model extensions: adaptation and a feedforward antennal lobe

On MaleCNS the base model kept firing after a stimulus had ended (persistent activity; S1 Text). Before the model was changed, three requirements were set: activity ends after each stimulus; sugar still drives MN9, and looming (an expanding dark shape) the giant fibres, two descending neurons that command the escape take-off; and one odour recruits roughly 5 to 20% of Kenyon cells, the range reported in experiments [39, 54].

First, spike-frequency adaptation was added to non-sensory neurons: a current a, raised by $\Delta a$ at each spike and decaying with time constant $\tau_a$, is subtracted in the membrane equation, $dv/dt = (v_0 - v + g - a)/\tau_m$, so a neuron fires less under sustained input. Of the eight settings tested, $\tau_a = 1$ s with $\Delta a = 1$ mV kept the most MN9 drive among those that ended persistent activity after sugar and looming, and it was adopted, although it weakened both motor responses and left persistent activity after the odour; no setting met the third requirement (Results; S1 Text).

Second, the antennal lobe, the first olfactory relay, was made feedforward: uniglomerular projection neurons (PNs) are driven directly, and every synapse between two neurons that both belong to the antennal lobe classes of the connectome annotation (labelled ALPN, ALLN, ALIN and ALON) or to the olfactory, hygrosensory and thermosensory receptor neuron types is set to zero; everything downstream of the PNs is unchanged. With PNs driven at 100 Hz for 300 ms, a mean of 1 neuron per run remained active 500 to 700 ms after odour A (Circuit protocol). The protocol drives PNs harder and longer, at 150 Hz for 1 s, where the test was not repeated, so the end of activity after its stimuli is inferred, not measured;

persisting activity would carry into the next step, which starts without a reset (S1 Text). Under the protocol, flies without pathology met the third requirement (Results).

## Synaptic weight

The unitary weight is that of the base model on FlyWire, although MaleCNS reports more input synapses per neuron: over 6,808 cell types matched with FlyWire v783, the median ratio was 1.755, and dividing the weight by it would give 0.157 instead of 0.275 mV (S1 Text). The lower weight weakened the sugar response about fourfold without ending persistent activity after the odour, so 0.275 mV was kept. Whether the conclusions hold at the lower weight is an open question.

## Pathology

Each neuron i has a load between 0 and 1, $l_i = s \times u_i$, where s is the severity, one number per fly, and $u_i$ is set by the vulnerability map (below). Three rules damage a neuron in proportion to its load, each modelled on an observation in amyloid-β flies, which make human amyloid-β (Table 1). **Threshold shift** [11]: the spike threshold is lowered to $-45$ mV $-$ 2 mV $\times l_i$. **Synapse loss** [12, 13]: each outgoing synaptic contact of neuron i is kept independently with probability $1 - 0.6 \times l_i$. **Neuron loss** [10, 11]: neuron i is removed with probability $0.5 \times l_i$ by setting its threshold to infinity, so it still receives input but never fires.

**Which neurons can be damaged.** Only neurons of the sensory superclasses (17,937, 10.8% of the network) are exempt, a modelling choice, although pan-neuronal expression in the fly also reaches sensory neurons; pathology in every other neuron is called pan-neuronal pathology here. The PNs of both odours, the 165 left-side LPLC2 and LC4 visual neurons that deliver the looming stimulus and the 316 protocerebral anterior medial (PAM) dopaminergic neurons driven during training can be damaged like any other neuron (S1 Text), so part of every odour- and looming-evoked deficit is loss of input: less of the stimulus reaches the network.

**Vulnerability maps.** In the default vulnerability map, the hub map, $u_i$ is the neuron's connectivity rank, its percentile among all 166,700 neurons by total number of synaptic contacts, inputs plus outputs (S1

Text), so the most connected neurons are damaged most. Two comparison maps reassign the same set of load values among the non-sensory neurons, at random in each fly (the shuffled map) or in reverse order of connectivity (the antihub map), so the three maps remove the same expected number of neurons. Kenyon-cell-only pathology damages only the 4,064 Kenyon cells, each with its hub-map load. At $s = 0$ no neuron is damaged: all healthy flies are the same intact network and differ only in input noise.

### Anchoring to data and assumptions

Only the direction of each rule follows measurements in flies; its size is assumed. Severity does not correspond to an age, and the removal probability at full load, 0.5, was chosen beforehand. One measurement relates severity to real flies (Results): with amyloid-β42 expressed in the mushroom body alone, about 17% of Kenyon cells are lost at 40 days [11]. By linear interpolation between the simulated severities 0.5 and 0.6, the model reaches this loss at $s \approx 0.56$ to 0.58, so figures shade severities of 0.55 and above as extrapolation. The threshold shift is weaker than the measured change in excitability (Table 1; S1 Text).

### Circuit protocol

Every fly ran the circuit protocol, one continuous sequence lasting 15.8 s (Fig 1B):

1. odour A for 1 s, 1-s gap, odour B for 1 s, 1-s gap;
2. sugar: the putative sugar-sensing gustatory neurons (types LB3c and LB3b) on both sides driven for 0.5 s, 1-s gap;
3. looming: the LPLC2 and LC4 neurons on the left driven for 0.3 s, 1-s gap;
4. training: odour A paired with activation of the PAM neurons, the teaching signal, for 1 s, 2-s gap, odour B alone for 1 s, 2-s gap;
5. memory test: odour A for 1 s, 1-s gap, odour B for 1 s.

Odour A drove the uniglomerular PNs of the glomeruli DM1, DM4 and VA2 (8 PNs) and odour B those of DL5, DM2 and VM2 (9 PNs), on both sides at 150 Hz; gustatory and looming neurons were driven at 150 Hz and PAM neurons at 100 Hz. The state of the network is not reset between steps, so each step starts with the adaptation left by the one before (S1 Text).

**Learning.** In the mushroom body, dopaminergic neurons and mushroom body output neurons (MBONs) are grouped into compartments [55, 56]. Dopamine-gated depression (weakening) of Kenyon cell to MBON synapses, which was shown in compartments of protocerebral posterior lateral 1 (PPL1) dopaminergic neurons [57], was implemented and applied to the compartments of the 15 PAM types (316 neurons), the only compartments whose synapses change. Dopaminergic neurons have no output connections in the model (Connectome), so the teaching signal acts on the network only through this mechanism. A PAM type counted as active if its mean rate during pairing, removed neurons counted as silent, reached 5 Hz; a Kenyon cell to MBON synapse was then multiplied by 0.4 if the MBON lay in the compartment of an active type and the Kenyon cell fired during pairing (S1 Text).

**Readouts.** Readouts, the measured quantities, are defined in Table 2. Valence, V, is the sum of MBON rates during an odour, glutamatergic MBONs signed −1 and the others +1 [37]; $V_{pre}$ and $V_{post}$ are its values in step 1 (naive valence) and in step 5. The learning index is LI = $(V_{post,A} - V_{pre,A}) - (V_{post,B} - V_{pre,B})$.

## Virtual arena

In the arena cohort (Simulated cohorts), each simulated brain was coupled for 120 s, after the circuit protocol, to a simple model body in its own copy of a circular arena with two food patches, two odour plumes and looming threats (S1 Text). A fly on a food patch fed while its MN9 rate exceeded 10 Hz, took off when its giant fibres fired at least two spikes within 30 ms, with or without a threat, and steered up or down the odour gradient according to its valence relative to its own naive valence.

## Simulated cohorts

The severity series comprised 64 healthy flies and 32 at each of ten severities (0.05, 0.10, 0.15, 0.2, 0.3, 0.4, 0.5, 0.6, 0.75 and 1.0; 384 flies; Fig 2) under the default model (pan-neuronal pathology, hub map, all three rules). Three further designs, each a set of flies built for one question, had 32 flies per condition at

s = 0.5 and 1.0: the rule ablation (all three rules, both loss rules without the threshold shift, and each rule alone; 320 flies; Fig 3A), the map comparison (the three maps; 192 flies; Fig 3C) and Kenyon-cell-only pathology (64 flies; Fig 3B). The severity series, the rule ablation and the map comparison thus share the default model at these severities: three replicate groups with independent random draws of the damage, whose ratio of largest to smallest median (replicate variation) shows how far group medians differ by chance (S2 Fig; S1 Text).

The arena cohort comprised 32 healthy flies and 24 at each of six severities from 0.05 to 0.5 (176 flies; S4 Fig), all under the hub map. Each fly ran the circuit protocol with two steps added before training: a state reset, which returns the dynamic state of the network to rest and leaves the weights unchanged, and a naive-valence probe, odour presentations that give the reference valence for steering in the arena, which it entered after a second reset (S1 Text). The severity series has neither step, which limits comparison of the learning index between the two designs.

**Healthy reference of each design.** Percentages of the healthy value refer to a design's own healthy flies where it has them, otherwise to the 64 healthy flies of the severity series, and in the within-fly analysis, which simulated 48 flies of the severity series again, to the same fly without pathology (its healthy copy). Healthy groups differ by chance, so percentages are compared within a design (S1 Text).

## Maps matched for synaptic contacts lost

The contact-matched comparison compares the maps at equal fractions of synaptic contacts lost instead of equal numbers of neurons removed. Contacts lost are those removed by the synapse-loss rule plus the remaining contacts of every connection to or from a removed neuron; connections within the antennal lobe, set to zero in the model, are not counted. For three flies per map the damage was drawn, without simulation, at severities from 0.05 to 1.0 in steps of 0.05, and the contacts lost were interpolated. The matched levels are the fractions lost under the hub map at s = 0.10, 0.25 and 0.5: 11.546%, 27.081% and 48.282% (Fig 4A). The shuffled map reached them at higher severities (s = 0.1452, 0.3676 and 0.7348), as did the antihub map for the lower two (s = 0.2809 and 0.7175), so flies under these two maps had larger loads and threshold shifts. The antihub map lost only 35.5% at s = 1.0 and has no group at the highest level. Each map at each level it reached, and a healthy group, had 32 flies (288 flies; 1,424 with all other designs); group medians of contacts lost were within 0.14 percentage points of their targets.

**Pre-specified criterion.** Before the simulations were run, one question and its analysis were fixed: at equal contacts lost, do the maps still differ by more than replicate groups do? It concerns one readout, whole-brain activity (all spikes during odour A), at each matched level: three maps at 11.5% and 27.1%, and the hub and shuffled maps at 48.3%. The statistic is the ratio of the highest to the lowest map median, with a 95% bootstrap interval (flies of each map resampled independently). It is compared with 1.27, the same ratio among the three replicate groups at s = 0.5 (S2 Fig). The outcome is "within replicate variation" when the ratio is at most 1.27, whatever its interval, read as contacts lost accounting for the difference between maps; "beyond replicate variation" only when the ratio and also its whole interval exceed 1.27 and the hub map has the lowest median, read as a difference that contacts lost do not account for; and "unresolved" otherwise (wording before the run: S1 Text). The ratio is at least 1 by construction, so its interval is compared with 1.27 and is no test of whether the maps differ at all. If antihub-map flies are no more active than healthy flies, the ratio has a ceiling, low where hub-map flies have lost little activity; after the run it was computed at 11.5% and 27.1%, to see whether the second outcome could be reached. All other readouts and comparisons had no pre-specified criterion and are reported descriptively (S1 Text).

### Within-fly analysis: undoing parts of a fly's own damage

To find which part of a fly's damage produces its deficit, the within-fly analysis simulated damaged flies again with one part undone and the rest kept: the first 24 flies by random-number seed at each of s = 0.3 and 0.5 of the severity series. Either removed neurons were restored (keeping their threshold shift and their own synapse loss), or the synapses that surviving neurons had lost (thinned synapses) were restored by using the healthy weights, or the threshold shift was removed. All versions of a fly started from a reset state with the same Poisson input, so they differ only through the undone damage. The shortened protocol used here, odour A, sugar and looming (3.8 s), has no training and so yields no learning index (S1 Text).

**Fraction of the deficit restored.** For a version of a fly with readout $y_{restored}$, $f = (y_{restored} - y_{damaged})/(y_{healthy} - y_{damaged})$, where $y_{healthy}$ is the readout in its healthy copy (simulated with the same input): f is 0 for the damaged fly and 1 for the healthy copy. A readout was left out for a fly when $|y_{healthy} - y_{damaged}|$ was at most 5% of $y_{healthy}$ (S1 Text).

**Screen.** The screen restored removed neurons one group at a time: one class (the connectome's class label, or superclass where none is given) or one cell type, a narrower category. Every class that had lost a neuron was restored singly, as were 450 cell types per fly: the 400 with the most neurons lost in the classes with the largest effects on six readouts (whole-brain activity, lateral horn, Kenyon cell and MBON rates, MN9 drive and giant fibre rate) and 50 random types from other classes as controls (S1 Text). Types with few neurons, such as the giant fibres (DNp01), the anterior paired lateral (APL) neurons and the projection neuron types of odour A, were therefore rarely tested. The most effective types were also restored together (S1 Text).

**Effect relative to all removed neurons.** The fraction restored by a class or a set of types was divided, fly by fly, by that of all removed neurons; medians are over flies (S1 Text).

**Pre-specified conditions.** Two conditions were fixed before the screen for treating its ranking of cell types (by fraction of the deficit restored) as a result about network activity. First, the ranking should not be predictable from the wiring: the fractions were regressed on five measures of the wiring of each type's lost neurons (S1 Text), and the median leave-one-type-out $R^2$ across flies had to be below 0.5. Second, the ranking should be robust (the robustness test): the ten most effective types per readout over the 48 flies (44 distinct types) and 10 control types (54 types) were restored singly in the first 12 flies at s = 0.5 under the extended model, the base model with both extensions, with unchanged parameters (the unchanged model) and under six variants, four with a parameter changed (all weights ×0.8 and ×1.2, glutamatergic weights ×0.5, adaptation increment ×0.5) and two with new input noise. In each variant and for each of the six readouts (36 cases), the Spearman correlation of the types' medians over flies with those of the unchanged model had to reach 0.7, and the fraction of the ten highest-ranked types shared with it 0.6. No giant fibre, APL or odour A projection neuron type was in this set, so the test does not cover them.

**Full-protocol test.** On the full circuit protocol, the same 54 types were restored alone in the same damaged flies and, separately, their lost neurons were removed alone from the healthy copies (S1 Text).

## Statistics

**General.** Group medians, given as percentages of the healthy median (Simulated cohorts), have 95% bootstrap intervals in which the group and its healthy flies were resampled independently; ratios of two group medians were bootstrapped in the same way, and median fractions of the deficit restored (within-fly analysis) by resampling flies. Groups were compared in pairs with the two-sided Mann-Whitney U test, and across maps, replicate groups or severity groups with the Kruskal-Wallis test; correlations are Spearman rank correlations unless stated. Because the number of simulated flies is a free choice, p values describe how clearly a difference can be detected at that sample size, and effect sizes are reported with intervals throughout. Unless a correction is stated, p values are not adjusted for multiple comparisons.

**Trends and replicate groups.** Trends with severity were tested as Spearman correlations between severity and readout across the 384 flies of the severity series, healthy flies included, with Holm correction across the twelve readouts tested (Table 2). The three replicate groups of the default model were compared with the Kruskal-Wallis test (S2 Fig).

**Rule ablation and Kenyon-cell-only pathology.** Each condition of the rule ablation was compared with healthy flies and, where stated, with another condition. Kenyon-cell-only pathology was compared with pan-neuronal pathology at the same severity, which removed the same fraction of Kenyon cells (14.72% against 14.67% at $s = 0.5$).

**Motor and high-activity phenotypes.** Two motor phenotypes were counted: a giant fibre rate below 100 Hz, about half the healthy rate, and a silent MN9, exactly 0 Hz. High-activity flies have Kenyon cell and MBON rates both above the maxima of the 64 healthy flies (S1 Text).

**Separation from healthy flies.** Separation of a severity group from healthy flies was measured as the area under the receiver operating characteristic curve (AUC; the Mann-Whitney statistic, ties counted as one half), the probability that a readout ranks a damaged and a healthy fly drawn at random in the right order, reported direction-free as max(AUC, 1 − AUC). Its chance level, the mean without any group difference, is 0.55 for 32 against 64 flies (S1 Text). The lowest severity at which the observed AUC reached 0.80 is reported, with the median and percentile range of that severity over the bootstrap resamples that reached it. Readouts measured in the same flies were compared by bootstrapping their AUC difference over flies; one was taken to separate better when the interval of the difference excluded zero. Because healthy flies are one network, the variation of a readout among them is its trial-to-trial spread under input noise, given as the robust coefficient of variation (interquartile range/1.349/|median|). Across the eight graded readouts (those of Fig 2 other than MN9 drive and giant fibre rate), AUC was correlated with the absolute percentage change of the group median divided by the spread, with an exact permutation p value (S1 Text).

**Arena cohort.** Its readouts were analysed in the same way, against its own healthy flies (S1 Text).

## Software, code and data

Simulations ran on a Linux server (two Intel Xeon Gold 6126 processors, 48 threads) in Python 3.13.5 with numba 0.68.0 [58], NumPy 2.5.3, SciPy 1.18.1 and pandas 3.0.6; engine validation used Brian2 2.10.1 [30]. Statistics and figures were computed in Python 3.13.3 with NumPy 2.2.5, SciPy 1.15.3, pandas 2.2.3 and Matplotlib 3.10.3. The connectomes are not redistributed: MaleCNS v1.0 (CC BY 4.0; S1 Text) is available from its publishers (https://male-cns.janelia.org/download/), and the FlyWire v630 and v783 tables (FlyWire data, CC BY-NC 4.0) are those distributed with the base model [29]. The simulation, analysis and figure code (MIT licence), the scripts that build the MaleCNS network, and the per-fly result table of every design are at https://github.com/taehojo/adfly-connectome (release v1.0.0; result tables under CC BY 4.0, the six files derived from FlyWire data under CC BY-NC 4.0). The repository's README documents how the reported numbers, figures and runs are reproduced.

### Use of artificial intelligence tools

This study was carried out with Claude API credits provided through the AI for Science program of Anthropic. Claude (Anthropic; used through Claude Code) assisted with coding, running the simulations and analyses, literature search and drafting of the text. The author verified its output: the simulation engine against an independent simulator running the base model (S1 Fig), the reported numbers against the result files, and the references against publisher records and the claims they support. The author is responsible for the content of the article.

## Acknowledgments

The author thanks the groups that released the MaleCNS and FlyWire connectomes and the authors of the base model for making their data and code openly available, and acknowledges the Indiana University Pervasive Technology Institute for providing storage resources that have contributed to the research results reported within this paper. The author also thanks Antonio Lee for an earlier pilot synapse-loss simulation on the MaleCNS connectome and for comments on a draft of this article.

## Funding

This study was funded by the Department of Radiology and Imaging Sciences, Indiana University School of Medicine, through start-up funds to T.J., and by Anthropic through Claude API credits from its AI for Science program. The funders played no role in study design, data collection, analysis and interpretation of data, or the writing of this manuscript.

## Competing interests

T.J. has received Claude API credits from Anthropic through its AI for Science program, which were used in this study; Anthropic develops Claude and Claude Code, the AI coding assistant used in this study. T.J. is the author of *Build with Claude Code* (Packt, 2026).

## Supporting information

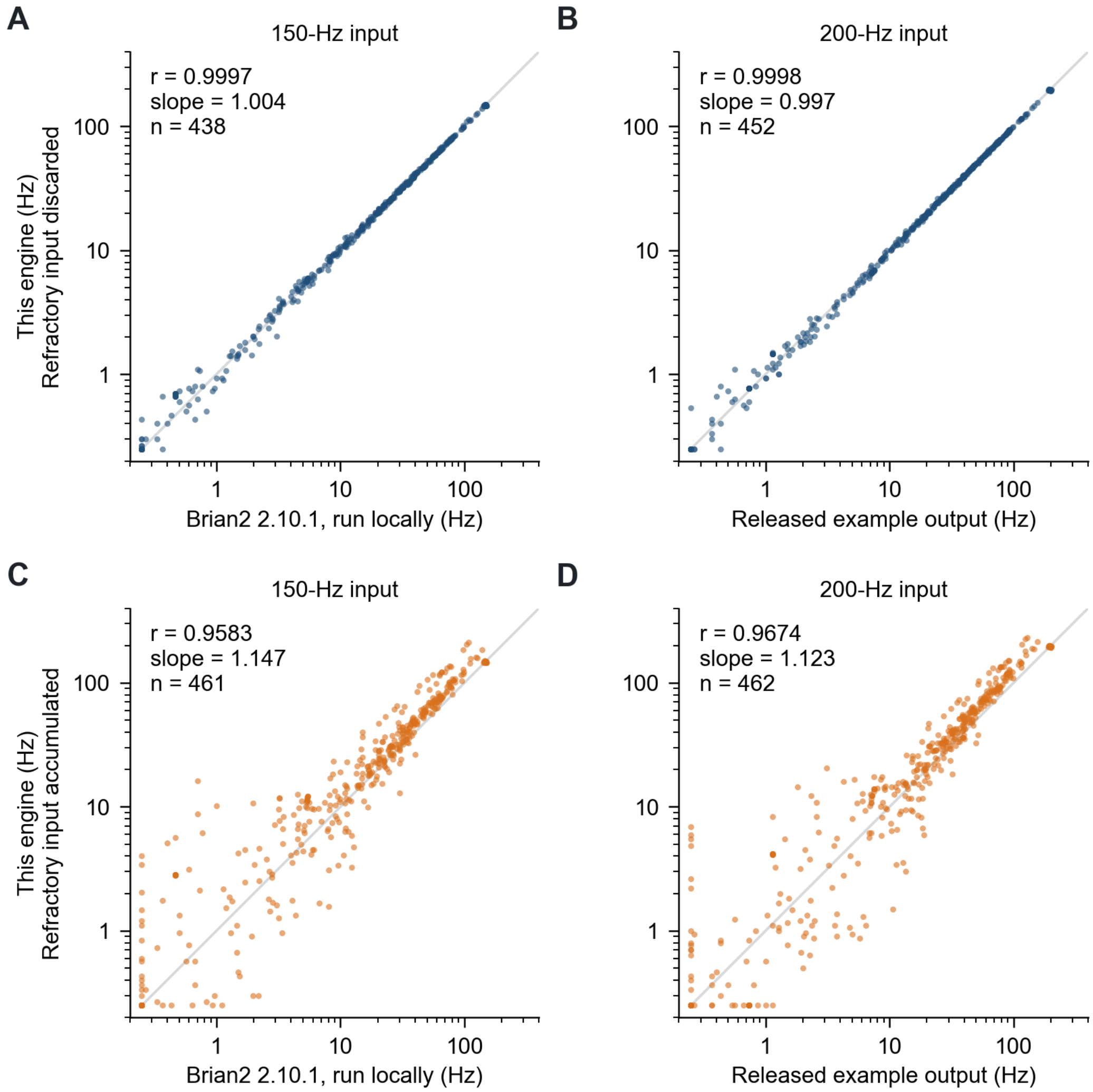

**S1 Fig. The simulation engine against the reference simulator.** The base model, without the two extensions, on the FlyWire v630 network (127,400 neurons), driven through 21 sugar gustatory receptor neurons of one side for 30 trials of 1 s at 150 Hz (A, C) or at 200 Hz, the input rate of the released example output (B, D). Per-neuron firing rates from the engine (y axis) against Brian2 2.10.1 running the unmodified published code locally (A, C) [29, 30] and against the example output released with that code (B, D), on logarithmic axes. (A, B) Synaptic input arriving during the refractory period is discarded, the setting used throughout; (C, D) it is accumulated. Each point is a neuron active in either run; rates below 0.25 Hz, including silence, are drawn at 0.25 Hz; grey line, identity. Panels give the Pearson correlation r and the slope of the engine rate regressed on the reference rate, both on linear rates, and the number of neurons n.

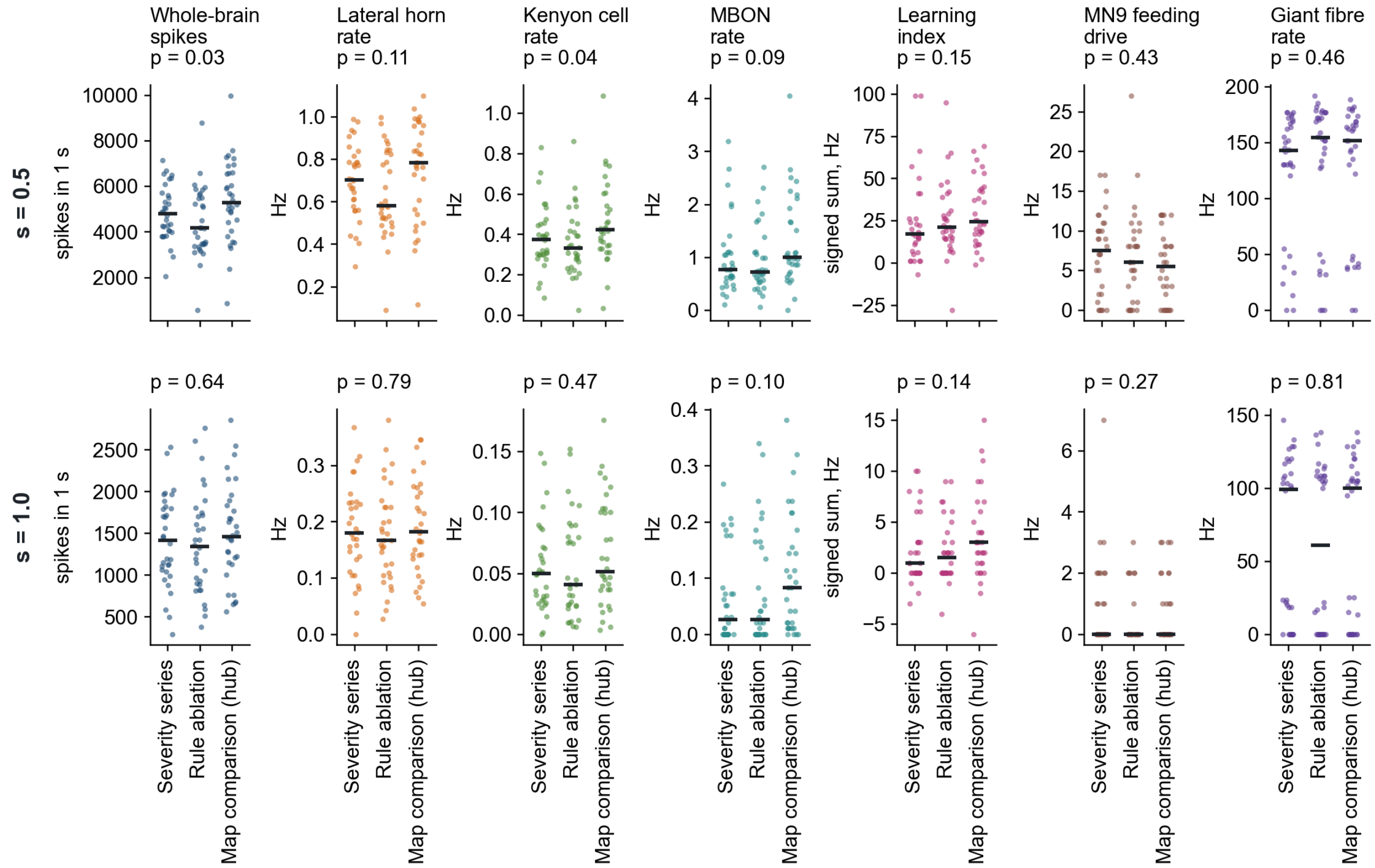

**S2 Fig. Three replicate groups of the default model.** The default model (pan-neuronal pathology, hub map, all three rules) simulated in three groups of 32 flies with independent random draws of the damage (severity series, rule ablation with all three rules, map comparison with the hub map), at s = 0.5 (top) and s = 1.0 (bottom). Columns are whole-brain activity (plotted as whole-brain spikes), lateral horn, Kenyon cell and MBON rates during odour A, learning index, MN9 feeding drive and giant fibre rate. Points are flies (jittered), bars medians; p, Kruskal-Wallis test across the three groups, uncorrected.

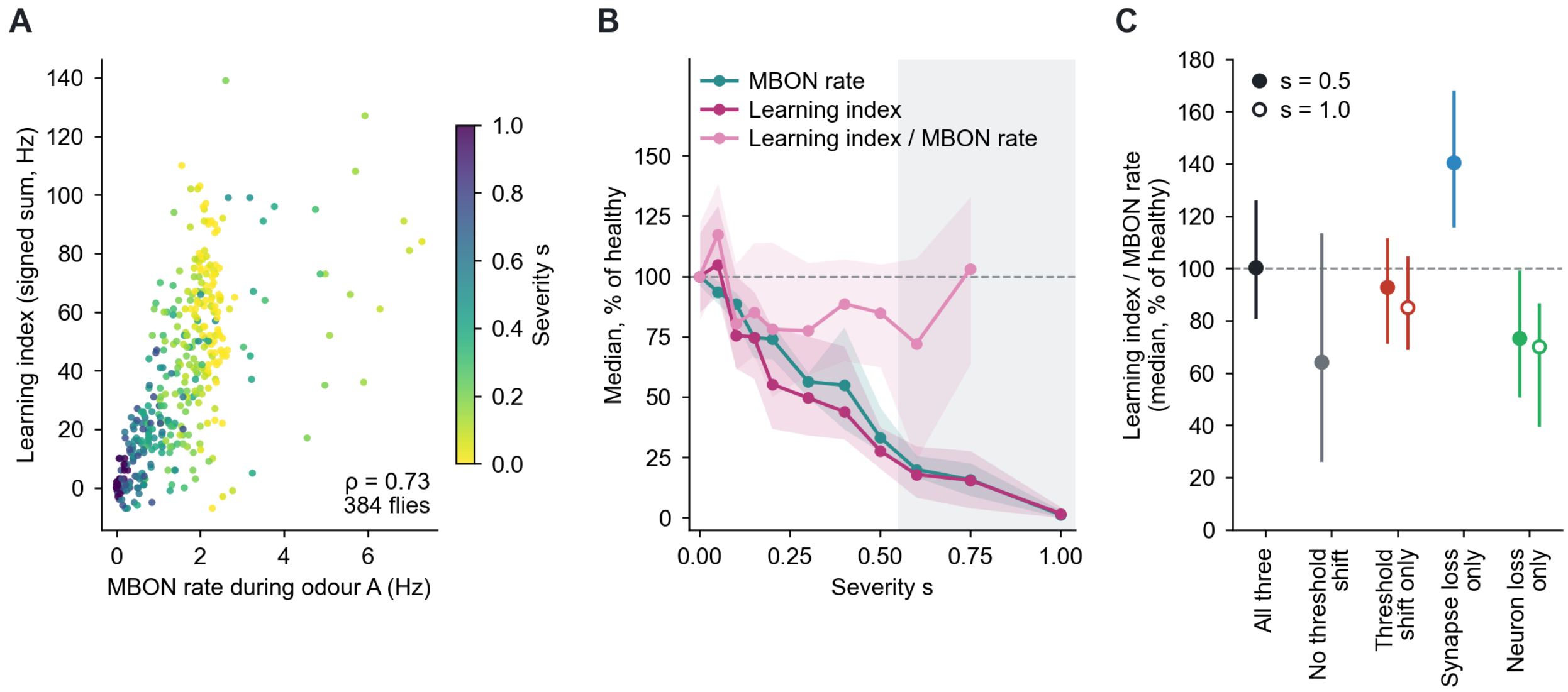

**S3 Fig. The learning deficit mostly follows mushroom body output.** Percentages are relative to the median of the 64 healthy flies of the severity series, with 95% bootstrap intervals (shaded bands in B, vertical lines in C); dashed lines, healthy median. (A) Learning index against MBON rate during odour A for the 384 flies of the severity series, coloured by severity; ρ, Spearman correlation pooled over all severities. (B) Medians of MBON rate, learning index and learning index divided by MBON rate in the severity series; grey, s ≥ 0.55 (extrapolation). The ratio is defined in flies with an MBON rate of at least 0.5 Hz and plotted where defined in at least 8 of 32 flies; from s = 0.4 it describes only the flies that kept such a rate, which are not typical of their group (S1 Text). (C) Learning index divided by MBON rate under the rule ablation (filled, s = 0.5; open, s = 1.0), shown where defined in at least 8 of 32 flies.

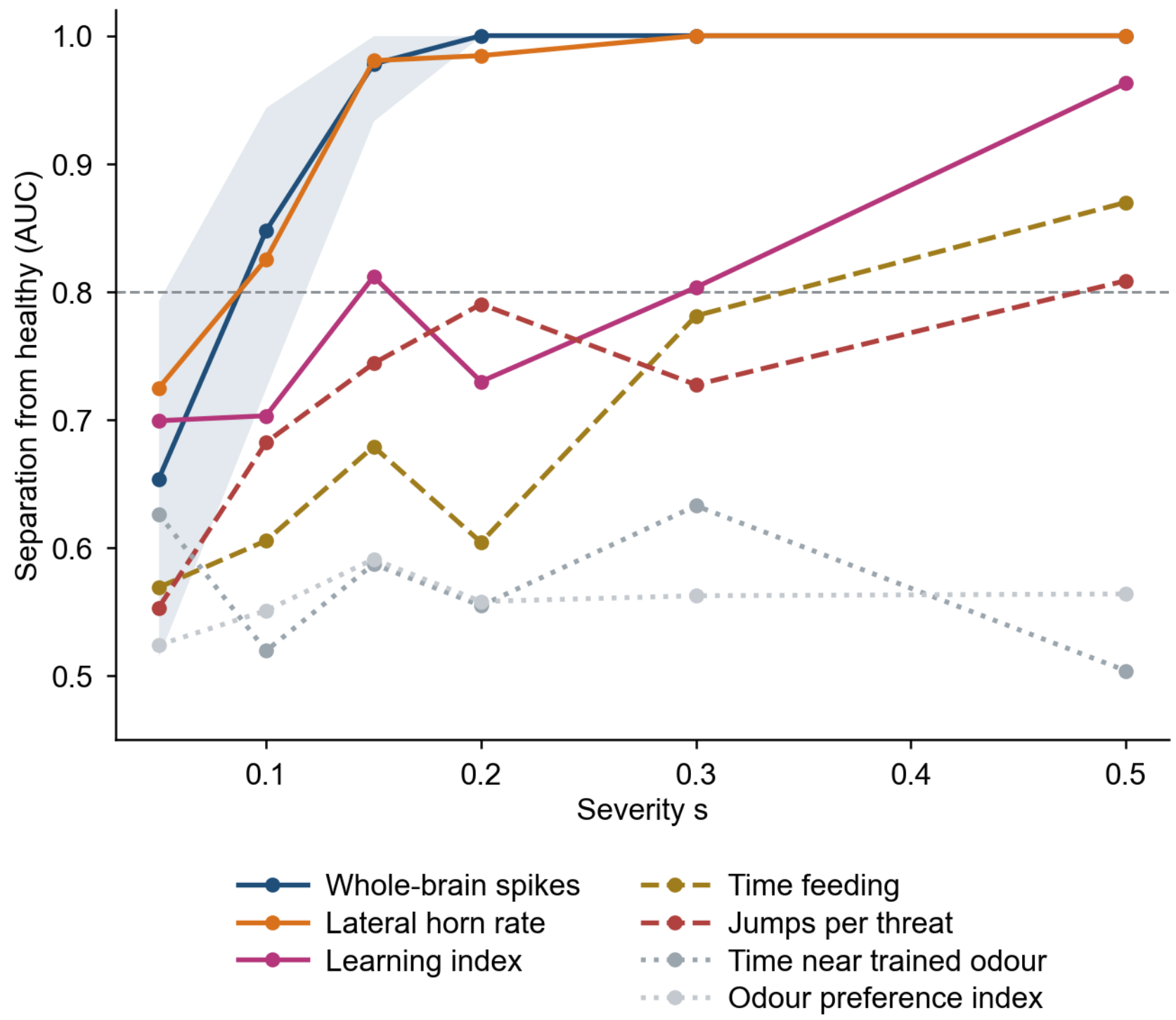

**S4 Fig. Circuit readouts against virtual-arena behaviour in the same flies.** Arena cohort of 176 flies (32 healthy, 24 at each of six severities), measured on the circuit protocol and then for 120 s in the virtual arena. Separation is the area under the receiver operating characteristic curve (AUC) between a severity group and the cohort's own healthy flies, taken in whichever direction exceeds 0.5; with no difference it averages 0.56 for 24 against 32 flies. Solid lines, three circuit readouts; dashed and dotted lines, four arena readouts; horizontal dashed line, AUC 0.80. Jumps per threat is defined in flies that met a threat (31 healthy, 22 to 24 per group) and the odour preference index in flies that entered a plume (32 healthy, 21 to 24 per group). Shaded band, 95% bootstrap interval for whole-brain activity (plotted as whole-brain spikes).

**S1 Text. Supplementary methods and results.** Selection rules, parameter values, resample counts, sensitivity checks and secondary results, organised in the order of the main text, with two further sections that give in full the learning-index and virtual-arena results summarised there (S3 Fig and S4 Fig).

# S1 Text. Supplementary methods and results

S1 Text gives the secondary results, full numeric series, sensitivity checks, selection rules, parameter values and resample counts behind the main text. The Supplementary results follow the order of the Results, with two further sections, "The learning deficit mostly follows mushroom body output" (S3 Fig) and "Circuit readouts against virtual-arena behaviour in the same flies" (S4 Fig), which give in full the learning-index and virtual-arena results that the Results summarise; the section "Pathology confined to Kenyon cells impairs only mushroom body readouts" belongs to the opening paragraph of "The number of neurons lost is a poor measure of damage" in the Results, and the section "Within single flies, the deficit due to neuron loss comes mostly from input and command neurons", which has the title of Fig 5, covers the two subsections of the Results on the within-fly analysis. The Supplementary methods follow the headings of the Materials and methods, with a final section of figure details. Tables in this text are labelled with letters. The names of the rules, the vulnerability maps, the sets of flies and the readouts are those of the main text, where each is defined at first use (Results, Table 2 and Materials and methods); a name that this text uses often is explained again where it first occurs. Abbreviations are those of the main text: PN, projection neuron; MBON, mushroom body output neuron; PAM, protocerebral anterior medial (dopaminergic neurons); AUC, area under the receiver operating characteristic curve. Two more are used in this text only: ORN, olfactory receptor neuron; SD, standard deviation. Unless stated otherwise, percentages are group medians relative to the healthy reference of the design (Materials and methods), with 95% bootstrap intervals in parentheses.

## Supplementary results

### A whole-CNS model carrying graded amyloid-like pathology

The engine validation (S1 Fig), the gustatory check on MaleCNS, the synaptic weight, the two extensions and the stimulated neurons that can be damaged are reported under Supplementary methods. In the severity series, the main set of flies, the fraction of neurons removed rose from 1.2% at s = 0.05 to 23.8% at s = 1.0, and the fraction of Kenyon cells removed from 1.5% to 29.4% (Fig 1D).

## Circuit readouts decline with severity

**Trend tests.** Across the 384 flies of the severity series, the Spearman correlation with severity was −0.95 for whole-brain activity, −0.94 for lateral horn rate, −0.87 for naive valence, −0.85 for Kenyon cell rate, −0.82 for MBON rate and for Kenyon cell recruitment, −0.78 for giant fibre rate, −0.77 for MN9 drive and −0.71 for the learning index and for the PN rate (Fig 2). For giant fibre latency it was +0.50 (362 flies with a defined latency), with the largest Holm-corrected p of these eleven tests ($1 \times 10^{-23}$). The per-severity medians reversed slightly at low severity for the learning index (105% of healthy at s = 0.05), the PN rate (101% at s = 0.10) and naive valence (71% at s = 0.15, 73% at s = 0.2).

**Medians at s = 0.5.** Whole-brain activity was 28% of healthy (24 to 31%), lateral horn rate 48% (42 to 53%), PN rate 78% (74 to 85%), Kenyon cell rate 46% (38 to 50%), Kenyon cells recruited 59% (51 to 65%), MBON rate 33% (27 to 46%), naive valence 22% (19 to 27%), learning index 28% (21 to 38%), MN9 feeding drive 38% (24 to 50%) and giant fibre rate 73% (67 to 82%). Removed Kenyon cells count as unrecruited, so part of the fall in recruitment is the loss itself; counted among surviving Kenyon cells, recruitment was 70% of healthy (59 to 77%).

**The projection neuron and PAM rates as delivered input.** In healthy flies the total spike count of all 686 antennal lobe projection neurons during odour A (median 1,124) was below the 1,200 input events expected for the 8 stimulated projection neurons in every fly. A line fitted to that total against the expected surviving fraction of the stimulated neurons, in healthy flies and under neuron loss alone, left a part not proportional to the surviving stimulated neurons close to zero (−4% of the healthy total, 95% bootstrap interval −19 to 11%), so the readout is mostly the output of those eight cells. The PN rate fell to 54% of healthy (51 to 68%) at s = 1.0, in line with the expected survival of the stimulated neurons (mean rate 78% of healthy against 78% expected at s = 0.5, and 58% against 55% at s = 1.0). Within each severity group, flies with a lower PN rate had a lower Kenyon cell rate (median Spearman ρ 0.64 over the ten groups). At s = 0.5, flies with a lower PN rate had, besides lower whole-brain activity (main text), a lower lateral horn rate (ρ = 0.79), Kenyon cell rate (0.69) and, more weakly, MBON rate (0.52); the PN rate did not correlate with MN9 drive (ρ = −0.02) or giant fibre rate (0.10). Whole-brain activity includes the spikes of the projection neurons themselves, a median of 6% of the count in healthy flies, 19% at s = 0.5 and 48% at s = 1.0, so the main text also gives its correlation with the PN rate with PN spikes excluded.

The PAM pairing rate had the most consistent trend of any quantity measured (ρ = −0.97). It fell to 88% of healthy (87 to 89%) at s = 0.5 and 75% (74 to 76%) at s = 1.0, close to the expected fraction of stimulated PAM neurons removed (12% and 24%), and within a severity it did not track the learning index (median ρ = 0.01 over the ten groups, range −0.27 to 0.39). A PAM type allows the synaptic change of training when its mean rate during pairing reaches 5 Hz, and surviving PAM neurons fired at about 99 Hz (healthy median 98.9 Hz), so a type falls below that rate only when about 95% of its neurons are removed.

**Motor phenotypes and giant fibre latency.** Counts of flies with a giant fibre rate below 100 Hz or a silent MN9 at every severity, and the graded decline of the other flies, are tabulated below with the discrete phenotypes. At s = 0.5, 8 of 32 flies had a giant fibre rate below 100 Hz and the other 24 a median of 153 Hz (79% of the healthy median); MN9 was silent in 4 to 9 of 32 flies at each severity from 0.10 to 0.75. Giant fibre first-spike latency had a median of 3 ms at every severity up to 0.75 and 4 ms at s = 1.0. Its rising trend reflects the loss of 2-ms responses (28 of the 64 healthy flies, none from s = 0.4 upward except one at s = 0.6) and the appearance of latencies above 3 ms, seen in no healthy fly but, among flies with a defined latency, in up to 16% at each severity up to 0.4, 20 to 32% from s = 0.5 to 0.75 and 61% at s = 1.0. Latency is undefined in the 22 flies whose giant fibres never fired.

**The highest severity, s = 1.0 (extrapolation).** Whole-brain activity was 8.2% of healthy (6.4 to 10.0%), and 4.5% without PN spikes. Lateral horn rate was 12% (10 to 15%), Kenyon cell rate 6.1% (3.9 to 7.5%) and Kenyon cell recruitment 13% (10 to 15%). MBON rate was 1.1% of healthy (0 to 3.0%) and exactly 0 in 13 of 32 flies, the learning index 1.6% (0 to 4.1%) and at or below 0 in 15, and naive valence 0%. MN9 drive had a median of 0, with 19 of 32 flies silent. The giant fibre median, 51% of healthy (10 to 56%), falls between two modes: 16 of the 32 flies were below 100 Hz, and the others fired at a median of 118 Hz (60% of healthy).

**Replicate groups of the default model.** Across the three replicate groups of the default model, which share one condition and differ only in their random draws of the damage (S2 Fig), whole-brain medians at s = 0.5 were 4,784, 4,164 and 5,288 spikes, a ratio of 1.27 between the highest and the lowest, their replicate variation (Kruskal-Wallis p = 0.03). Kenyon cell rate had the same ratio (1.268, against 1.270 for whole-brain activity; p = 0.04). The ratio was larger for lateral horn rate (1.35), MN9 drive (1.36), MBON rate (1.39), the learning index (1.44) and naive valence (1.48), smaller for Kenyon cell recruitment (1.16), PN rate (1.18) and giant fibre rate (1.08), and 1.01 for the PAM pairing rate. The whole-brain and Kenyon cell rate tests were the only two of the 14 shown in S2 Fig below 0.05, a number compatible with chance (0.7 expected); Kenyon cell recruitment, not shown in S2 Fig, also gave p = 0.04 at s = 0.5. At s = 1.0 the whole-brain ratio was 1.09. The ratios for MBON rate (3.2) and the learning index (3.0) at that severity come from medians near zero and do not measure replicate variation, the median MN9 drive was 0 in all three groups, and the giant fibre medians (51%, 31% and 51% of healthy) fell between the two modes, each group having 16 of 32 flies below 100 Hz.

### Synapse loss and neuron loss produce the deficit; the threshold shift partly offsets it

**All three rules together.** Unless stated otherwise, p values in this and the next section are two-sided Mann-Whitney tests. In the rule ablation, which applies each rule alone and in combination, all three rules together at s = 0.5 also reduced lateral horn rate to 40% of healthy (36 to 52%), Kenyon cell rate to 41% (35 to 49%), MN9 drive to 30% (20 to 42%) and giant fibre rate to 79% (69 to 89%).

**Loss rules alone and combined (s = 0.5 unless stated).**

1. Whole-brain activity: synapse loss alone 36%, neuron loss alone 47%, both loss rules 26% (24 to 29%).
2. Lateral horn rate: synapse loss alone 70% (69 to 72%), neuron loss alone 65% (57 to 72%); p = 0.06.

3. MBON rate: synapse loss alone 36%, with all 32 flies below the lowest healthy value; neuron loss alone 84% ($p = 0.27$ against healthy flies, $p = 3 \times 10^{-8}$ against synapse loss alone), with 14 of 32 flies below the lowest healthy value and 11 above the highest; both loss rules 19% (13 to 28%; $p = 4 \times 10^{-4}$ against synapse loss alone), so neuron loss, added to synapse loss, deepened the MBON deficit.
4. Kenyon cell rate: neuron loss alone 71% (67 to 104%; $p = 0.03$ against healthy flies), with 17 flies below and 9 above the healthy range; synapse loss alone 54% (52 to 56%). All 9 flies above the healthy Kenyon cell range were among the 11 above the healthy MBON range, 8 of them with Kenyon cell rates above 1.2 times the healthy maximum; with both rates above the highest healthy values, they are high-activity flies, examined below.
5. Learning index: neuron loss alone 52% (38 to 73%), synapse loss alone 47% (42 to 57%; $p = 0.55$), both 12% (7 to 18%).
6. MN9 drive: neuron loss alone 25% (15 to 59%), synapse loss alone 45% (41 to 51%); $p = 0.23$.
7. Giant fibre rate: neuron loss alone 83% (70 to 89%), synapse loss alone 91% (91 to 93%); $p = 3 \times 10^{-5}$.
8. PN rate: 78% under neuron loss alone, 103% under synapse loss alone and 101% under the threshold shift alone. PAM pairing rate: 88%, 100% and 101%.
9. At $s = 1.0$, synapse loss alone took mushroom body output close to zero (MBON rate 0.4%, learning index 1.6%), whereas neuron loss alone left them at 24% (19 to 41%) and 19.5% (14 to 26%) ($p = 7 \times 10^{-11}$ and $4 \times 10^{-6}$).

**Synaptic contacts removed by each loss rule.** In the 32 hub-map flies at $s = 0.5$ of the contact-matched comparison, in which the maps are compared at equal fractions of contacts lost, with all rules applied, the synapse-loss rule removed a median of 22.2% of all synaptic contacts and the connections of removed neurons a further 26.3% (48.4% in total), more than the synapse-loss rule in each of the 32 flies. Neuron loss alone therefore removes at least as many contacts as synapse loss alone.

**Threshold shift.** At s = 0.5 the threshold shift alone raised all five readouts in Fig 3A and the lateral horn and Kenyon cell rates, with MBON rate at 139% of healthy (132 to 145%), the learning index at 120% (102 to 159%) and MN9 drive at 128% (105 to 160%), whereas the PN and PAM rates did not change detectably ($p = 0.46$ and 0.10). At s = 1.0 it raised whole-brain activity to 243% (228 to 266%), against 148% at s = 0.5.

Removing the shift from all three rules at s = 0.5 gave $p = 0.51$ for whole-brain activity (26% against 24% of healthy), 0.18 for lateral horn rate, 0.96 for Kenyon cell rate, 0.18 for MN9 drive, 0.37 for giant fibre rate, 0.006 for MBON rate and $9 \times 10^{-4}$ for the learning index. The PN rate differed between these two groups of different flies (76.2% of healthy with the shift against 84.9% without; $p = 0.031$), in the direction opposite to an effect of the shift. At s = 1.0 the same comparison changed MBON rate (1.1% with the shift against 0.2% without; $p = 0.02$) and the learning index (2.4% against 0%; $p = 0.03$), all near zero, but not whole-brain activity ($p = 0.41$). With the per-fly variation of whole-brain activity at s = 0.5 (SD 7.6% of healthy), 32 against 32 flies detect a difference of about 5% of healthy with 80% power.

**Within-fly effect of the threshold shift.** Values are medians, with 95% bootstrap intervals, of the change in each fly's deficit when the fly was simulated again without its threshold shift (within-fly analysis), over the flies with a deficit in that readout (Materials and methods):

1. Whole-brain activity: deepened by 5.2% of the deficit (4.0 to 7.0%); lower in all 48 flies, by a median of 3.6% of the fly's healthy value (3.0 to 4.3%).
2. Lateral horn rate: 10.2% (8.6 to 11.4%); lower in 48 of 48 flies.
3. Kenyon cell rate: 15.5% (11.3 to 17.7%); lower in 46 of 46 flies.
4. MBON rate: 20.4% (16.7 to 24.1%); lower in 47 of 47 flies.
5. MN9 drive: 11.1% (0 to 20.0%); lower in 32 of 47 flies, unchanged in 11 and higher in 4.
6. Giant fibre rate: 5.5% (2.3 to 9.7%); lower in 29 of 40 flies, unchanged in 7 and higher in 4.

Without the flies whose damaged value was above the healthy value (2 for Kenyon cell rate, 4 for MBON rate and 4 for MN9 drive), the medians were 15.9% (44 flies), 21.1% (43) and 12.5% (43). These runs had no training, so the effect of the threshold shift on the learning index rests on the group comparison above.

### Pathology confined to Kenyon cells impairs only mushroom body readouts

**Neurons removed.** At s = 1.0 Kenyon-cell-only pathology removed 29.6% of Kenyon cells and 0.72% of all neurons, against 29.4% and 23.8% in the pan-neuronal severity series.

**Readouts against healthy flies.**

1. At s = 0.5: learning index 66% of healthy (54 to 90%; $p = 6 \times 10^{-5}$), MBON rate 76% (71 to 79%; $p = 5 \times 10^{-13}$), naive valence 85% (78 to 91%; $p = 2 \times 10^{-6}$), Kenyon cell recruitment 95% (92 to 97%; $p = 1 \times 10^{-5}$).
2. At s = 1.0: learning index 39% (30 to 51%), MBON rate 49% (46 to 52%), naive valence 63% (59 to 68%), Kenyon cell recruitment 88% (85 to 90%).
3. Mean Kenyon cell rate, which counts removed cells as silent, did not change detectably at s = 0.5 (99%; p = 0.89) and was slightly lower at s = 1.0 (96%; p = 0.005), so the surviving cells fired more; recruitment counted among surviving Kenyon cells rose to 111% (107 to 114%) and 124% (121 to 127%) of healthy.
4. The learning index divided by MBON rate was 89% of healthy at s = 0.5 (p = 0.51) and 79% at s = 1.0 (p = 0.02).
5. Outside the mushroom body, whole-brain activity and lateral horn, PN, MN9 and giant fibre rates did not differ detectably from healthy flies at either severity (p from 0.11 to 0.87).

**Comparison with pan-neuronal pathology at equal Kenyon cell loss.** At s = 0.5, Kenyon cell recruitment was 95% against 59% of healthy ($p = 3 \times 10^{-9}$) and Kenyon cell rate 99% against 46% ($p = 6 \times 10^{-11}$). In percentage points below the healthy median, damage confined to the Kenyon cells reproduced about half of the pan-neuronal learning-index deficit at s = 0.5 (34 of 72 points) and 62% at s = 1.0 (61 of 98), a third of the MBON deficit at s = 0.5 (24 of 67 points) and half at s = 1.0 (51 of 99), 13% of the fall in Kenyon cell recruitment (5.3 of 41 points) and almost none of the fall in Kenyon cell rate (0.7 of 54 points), both at s = 0.5. Damage outside the Kenyon cells accounts for the remainder of each deficit only if the two sources add, which was not tested; which neurons outside the Kenyon cells contribute is not known from these designs, because no condition damaged them alone.

### The number of neurons lost is a poor measure of damage

**Neurons removed and tests.** At s = 1.0, as at s = 0.5, the fraction of neurons removed did not differ between maps (23.8%; Kruskal-Wallis p = 0.62). The fraction of Kenyon cells removed at s = 0.5 did differ (hub 15.0%, shuffled 13.2%, antihub 12.1%; Kruskal-Wallis $p = 6 \times 10^{-18}$). Unless stated otherwise, p values in this section are Kruskal-Wallis tests across the three maps, and comparisons of two maps are two-sided Mann-Whitney tests.

**Readouts at s = 0.5.** Values are for the hub, shuffled and antihub maps in that order; whole-brain activity and MBON rate are given in the main text.

1. Lateral horn rate: 53%, 73% and 98%; $p = 1 \times 10^{-15}$.
2. Kenyon cell rate: 52%, 69% and 98%; $p = 1 \times 10^{-10}$.
3. MN9 drive: 28%, 60% and 88%; $p = 5 \times 10^{-11}$. Among flies with an active MN9: 40%, 60% and 90%; $p = 2 \times 10^{-9}$. MN9 was silent in 9, 1 and 1 of 32 flies.
4. Giant fibre rate: 78%, 90% and 96%; $p = 3 \times 10^{-9}$. Below 100 Hz in 9, 2 and 1 of 32 flies; among flies at or above 100 Hz: 84%, 91% and 96%; $p = 2 \times 10^{-7}$.
5. Kenyon cell recruitment: 64%, 75% and 94%. Hub against shuffled: p = 0.10.
6. Learning index: 40% (30 to 63%), 50% (36 to 69%) and 57% (46 to 74%); p = 0.18.
7. PN rate: 90%, 89% and 100%. Hub against shuffled: p = 0.55. Below 95% of healthy in 24, 20 and 9 of 32 flies.

8. PAM pairing rate: 88%, 87% and 86%.

Even the antihub map, which left whole-brain activity and lateral horn rate close to healthy values (93% and 98%), lowered MBON rate to 78% and the learning index to 57% (Mann-Whitney $p = 1 \times 10^{-11}$ and $9 \times 10^{-7}$ against healthy flies).

**The highest severity, s = 1.0 (extrapolation).** Whole-brain activity was 8.5%, 27% and 88% of healthy under the hub, shuffled and antihub maps, and MBON rate 3.6%, 29% and 52%. Giant fibre rate fell below 100 Hz in 16, 9 and 0 flies, and MN9 was silent in 20, 16 and 1. The learning index differed between maps at this severity (5%, 19% and 41%; $p = 5 \times 10^{-7}$).

**Map differences against replicate variation.** At s = 0.5 the ratio of the shuffled-map and hub-map medians, larger over smaller, exceeded the replicate variation of the default model (S2 Fig) for whole-brain activity (1.53 against 1.27), MN9 drive (2.18 against 1.36), naive valence (1.71 against 1.48), Kenyon cell rate (1.33 against 1.27) and giant fibre rate (1.15 against 1.08), and barely for lateral horn rate (1.37 against 1.35). It equalled replicate variation for Kenyon cell recruitment (1.16 against 1.16) and lay within it for MBON rate (1.19 against 1.39), the learning index (1.24 against 1.44) and the PN rate (1.01 against 1.18). For the PAM pairing rate both were close to 1 (1.02 against 1.01).

**Synaptic contacts lost under the three maps.** The hub, shuffled and antihub maps removed 48%, 35% and 20% of all synaptic contacts at s = 0.5, and 77%, 60% and 36% at s = 1.0. These fractions were estimated from three random draws of the damage per map and severity (Fig 4A); contacts lost were not computed for the flies of the map comparison themselves.

## The maps differ less when matched for synaptic contacts lost

**Healthy reference of this design.** In the contact-matched comparison the three maps are compared at equal fractions of synaptic contacts lost, the matched levels. Its 32 healthy flies had a median learning index 1.24 times that of the healthy flies of the severity series (Supplementary methods, Simulated cohorts, Healthy groups), so the hub map at 48.3% gave 30% of the healthy learning index of this design but 37% of that of the severity series.

**Whole-brain activity at each matched level.** R is the ratio of the highest to the lowest map median. At 11.5% of contacts lost, with the hub map at 76% of healthy, R could exceed 1.31 only if antihub-map flies were more active than healthy ones. Applied with the 32 healthy flies in place of the antihub-map group, the pre-specified criterion (Materials and methods) gave 1.31 (1.25 to 1.43), still short of the outcome "beyond replicate variation". The deficit under the antihub map was 8.5 percentage points against 23.9 under the hub map, a deficit ratio of 0.36 (0.12 to 0.55). At 27.1% the deficit ratio was 0.41 (0.22 to 0.50). The hub and shuffled maps were close at both of these levels ($p = 0.12$ at 11.5% and 0.09 at 27.1%).

At 48.3%, R was 1.45 (1.26 to 1.62). In the pre-specified analysis the lower bound lay just below 1.27, and it varies from one bootstrap to the next. The value quoted is its average over 300 repeated bootstraps (1.259, range 1.240 to 1.276), and 97% of them gave the same outcome, "unresolved".

**Comparison with the map comparison.** The map comparison, in which the maps remove equal numbers of neurons, was run at $s = 0.5$ and 1.0 only. As an approximation across designs, hub-map flies of the severity series interpolated to the antihub-map flies' own severities ($s = 0.28$ and 0.72) would have given antihub/hub ratios of about 2.0 and 4.8, against 1.20 and 1.52 at equal contacts lost.

How much matching reduced the difference between the antihub and hub maps depends on which map's severity is held fixed.

1. With the antihub-map flies fixed, the reduction was large. Antihub-map flies at 27.1% and shuffled-map flies at 48.3% of contacts lost had lost nearly the same fraction of neurons (17.1% and 17.5%) at nearly the same mean load (0.342 and 0.350), yet differed 2.32-fold (2.14 to 2.61) in whole-brain activity. The antihub/shuffled ratio at 27.1% was 1.45.
2. With the hub-map flies fixed, the reduction was modest. Antihub-map flies at 11.5% had lost about as many neurons as hub-map flies at 27.1% (6.7% and 6.0%; mean load 0.134 and 0.119). They were 1.73 (1.57 to 2.14) times as active as those hub-map flies, against 1.52 at equal contacts lost.

Computed within each design, the shuffled/hub ratio hardly changed with matching (1.45 at equal contacts lost against 1.53, 1.25 to 1.79, with equal numbers of neurons removed). The hub-map condition is the same in both designs, however, and the shuffled-map flies were less active at their matched severity (s = 0.73) than at s = 0.5 (35% against 47%, each relative to the healthy reference of its design; raw medians 6,199 against 8,111.5 spikes; Mann-Whitney $p = 0.001$). Against each of the four hub-map groups at s = 0.5 (S2 Fig and this design), the ratio was 1.53 to 1.95 with equal numbers of neurons removed and 1.17 to 1.49 at equal contacts lost, about a quarter lower.

The two hub-map groups at s = 0.5 in Fig 4B are independent groups of 32 flies. Their raw medians were 5,288 and 4,260.5 spikes (Mann-Whitney $p = 0.04$), a difference between groups of the kind shown in S2 Fig.

**Other readouts at equal contacts lost.** Values are for the hub, shuffled and antihub maps at 11.5% and 27.1%, and for the hub and shuffled maps at 48.3%. P values are Kruskal-Wallis tests across the maps present at each level.

1. Lateral horn rate: 91%, 92% and 98% at 11.5% ($p = 3 \times 10^{-7}$; hub against shuffled $p = 0.42$); 76%, 84% and 95% at 27.1% ($p = 3 \times 10^{-14}$); 44% against 64% at 48.3% ($p = 2 \times 10^{-5}$).
2. Kenyon cell rate: 88%, 93% and 98%; 73%, 84% and 97%; 37% against 59% ($p = 0.004$, $4 \times 10^{-5}$ and $1 \times 10^{-4}$).
3. MN9 drive: 84%, 97% and 95%; 66%, 82% and 95%; 32% against 50% ($p = 0.02$, $3 \times 10^{-5}$ and 0.02).

Post hoc, the pre-specified criterion was applied to each other readout with that readout's own replicate variation in place of 1.27 (Replicate groups of the default model, above). None was "beyond replicate variation" at any matched level. Lateral horn rate was "within replicate variation" at 11.5% and 27.1% (ratios 1.08 and 1.26, against its replicate variation of 1.35) and "unresolved" at 48.3% (1.44). Kenyon cell rate and MN9 drive were within replicate variation at 11.5% and unresolved at the two higher levels, Kenyon cell recruitment within it at the two lower levels and unresolved at 48.3%, and the PN rate, MBON rate, learning index and giant fibre rate within it at all three.

**Giant fibre rate at equal contacts lost.** Giant fibre rate was 99.6%, 97% and 98% of healthy at 11.5% (p = 0.18) and 83% under both the hub and the shuffled map at 48.3% (p = 0.86). At 27.1% it was 95%, 93% and 97% (p = 0.003), with the antihub map highest and the shuffled map, not the hub map, lowest. The hub and shuffled maps did not differ at any matched level (Mann-Whitney p = 0.09, 0.48 and 0.87). With equal numbers of neurons removed at s = 0.5 they had differed (78% against 90% of the healthy median; Mann-Whitney $p = 8 \times 10^{-4}$).

**MBON rate and learning index at equal contacts lost.**

1. MBON rate: 83% (73 to 89%), 87% (75 to 94%) and 83% (76 to 94%) of healthy at 11.5% (p = 0.79); 65% (55 to 78%), 65% (58 to 79%) and 67% (61 to 72%) at 27.1% (p = 0.91); 28% (24 to 42%) against 36% (30 to 51%) at 48.3% (p = 0.14).
2. Learning index: 76% (62 to 98%), 71% (63 to 87%) and 64% (54 to 83%) at 11.5% (p = 0.28); 51% (41 to 65%), 53% (38 to 68%) and 49% (41 to 61%) at 27.1% (p = 0.71); 30% (24 to 44%) against 37% (26 to 52%) at 48.3% (p = 0.60).

At 48.3% the intervals of the shuffled/hub ratio (1.30, 0.83 to 1.85, for MBON rate; 1.22, 0.75 to 1.8, for the learning index) include the whole-brain ratio (1.45), so a difference as large as that in whole-brain activity could have gone undetected. Replicate groups of one condition differ by up to 1.39-fold in MBON rate and 1.44-fold in the learning index among the three groups of S2 Fig, and by 1.56-fold in MBON rate when the hub-map group at 48.3% of this design is added, so these results do not establish equality.

With equal numbers of neurons removed at s = 0.5, only the antihub map had differed detectably in MBON rate. The antihub map removed 19.7% of contacts at that severity, and at the two matched levels on either side of it, MBON rate under the antihub map (83% and 67%) was close to that under the hub map (83% and 65%). The difference between maps in the learning index at s = 1.0 arose where the hub map removed 77% of contacts, beyond the matched range.

**Neurons and Kenyon cells removed at equal contacts lost.** Values are for 11.5%, 27.1% and 48.3% of contacts lost (Fig 4D).

1. Neurons removed: hub 2.4%, 6.0% and 11.9%; shuffled 3.5%, 8.7% and 17.5%; antihub 6.7% and 17.1%.
2. Kenyon cells removed (hub, shuffled, antihub): 2.9%, 3.8% and 6.7% at 11.5%; 7.4%, 9.9% and 17.2% at 27.1%; 14.7% and 19.6% (hub, shuffled) at 48.3%.

Antihub-map flies at 27.1% of contacts lost thus had 2.3 times as many Kenyon cells removed as hub-map flies, yet their MBON rate did not differ detectably (67% against 65%; Mann-Whitney $p = 0.67$). At 27.1% the rate per surviving Kenyon cell of the antihub-map flies was 116.8% of healthy (110.5 to 122.8%; Mann-Whitney $p = 3 \times 10^{-10}$ against healthy flies), against 79% for the hub-map flies.

**Threshold shift and loss of input at equal contacts lost.** The mean threshold shift over all neurons was 0.10, 0.14 and 0.27 mV for the hub, shuffled and antihub maps at 11.5%, 0.24, 0.35 and 0.68 mV at 27.1%, and 0.48 and 0.70 mV for the hub and shuffled maps at 48.3%. That the surviving Kenyon cells of antihub-map flies fire above the healthy rate fits an offsetting effect of their larger threshold shifts but does not isolate it, because these flies also kept their odour input and lost more Kenyon cells, which may reduce feedback inhibition of the surviving ones.

The PN rate was 98%, 99% and 101% of healthy at 11.5%, 92%, 93% and 99% at 27.1%, and 76.5% against 87% at 48.3% (Fig 4E). At 11.5% and 27.1% it did not differ detectably between maps (Kruskal-Wallis $p = 0.16$ and 0.17). Across flies it correlated with whole-brain activity (Spearman $\rho = 0.42$, 0.45 and 0.75 at the three levels; $p < 10^{-4}$). Among flies whose PN rate was at least 95% of healthy, the antihub map still exceeded the hub map at 11.5% (27 antihub-map against 19 hub-map flies; $p = 0.002$) and at 27.1% (22 against 11; $p = 3 \times 10^{-5}$). The hub/shuffled difference at 48.3% was not detectable in the 4 and 6 flies that remained (35% against 49% of healthy; $p = 0.07$). The PN rate averages all 686 projection neurons, stimulated or not, so this restriction corrects for loss of input only approximately.

With equal numbers of neurons removed at s = 0.5, the median PN rate was the same in the hub-map and shuffled-map groups compared (90% and 89%), but whole-brain activity differed 1.5-fold. The other three hub-map groups at s = 0.5 had lost more input (PN rate 76 to 78% of healthy in the severity series, the rule ablation and this design).

### Within single flies, the deficit due to neuron loss comes mostly from input and command neurons

**Flies and readouts.** In the within-fly analysis, in which a damaged fly is simulated again with part of its own damage undone, leaving out a readout for a fly when its deficit was smaller than 5% of the healthy value (Materials and methods) left 48 flies for whole-brain activity and lateral horn rate, 46 for Kenyon cell rate, 47 for MBON rate and MN9 drive, and 40 for giant fibre rate. In ten of the remaining fly-readout pairs the damaged fly exceeded its healthy copy, the same fly simulated without pathology (Kenyon cell rate in 2 flies, MBON rate in 4, MN9 drive in 4). These pairs were kept with a negative denominator, and all medians include them.

The damaged flies had lost a median of 11,856 neurons (7.1%) at s = 0.3 and 19,826 (11.9%) at s = 0.5. Their whole-brain activity was 41% (38 to 51%) and 27% (24 to 33%) of that of their healthy copies.

**Decomposition into neuron loss, synapse loss and threshold shift.** Thinned synapses are the synapses that surviving neurons had lost. The deficit due to synapse loss, the fraction of a fly's deficit recovered when all its thinned synapses are restored, was 64% of the whole-brain deficit at s = 0.3 and 60% at s = 0.5 (Mann-Whitney p = 0.44). By readout, the median deficit due to neuron loss, recovered when all removed neurons are restored, and that due to synapse loss were 36% and 67% for Kenyon cell rate, 46% and 69% for MBON rate, 57% and 44% for lateral horn rate, 84% and 16% for giant fibre rate, and 60% each for MN9 drive. The two summed to close to one for five of the six readouts (medians 0.94 to 1.16) and to 1.27 for MN9 drive, so within a fly the effects of the two loss rules nearly add. The effect of removing the threshold shift in these flies is given above (Within-fly effect of the threshold shift).

**Restoring one class at a time.** In the screen, the removed neurons of one class, or of one cell type, were restored on their own. At s = 0.3, where fewer flies had lost a stimulated projection neuron, classes other than the antennal lobe projection neurons (the PN class) recovered more. The central-brain intrinsic class, which holds the neurons of that superclass without a finer class label, recovered 10% of the whole-brain and 26% of the lateral horn deficit, against 8% and 16% for the PN class. The Kenyon cell class recovered 17% of the Kenyon cell deficit, against a median of 0 for the PN class.

Over both severities, the PN class recovered 7% of the MBON deficit (0 to 20%). This median includes the flies whose damaged MBON rate was above the healthy value and is sensitive to their inclusion. The PN class had the largest Banzhaf value, a class's average effect over many random combinations of restored classes (Supplementary methods, Class screen), for whole-brain activity and lateral horn and Kenyon cell rates (0.13, 0.35 and 0.31).

Restoring the PN class changed whole-brain activity in 36 flies (13 at s = 0.3 and 23 at s = 0.5). These are probably the flies that had lost at least one of the eight neurons that deliver odour A. Of 467 restorations of single PN cell types, only 4 changed any odour readout, and all 4 were odour A projection neuron types. With each neuron removed independently with probability 0.5 × s × its connectivity rank, 16.5 and 20.9 of the 24 flies at each severity are expected to lose at least one of the eight.

In the 36 flies, restoring the PN class alone recovered 16% of the whole-brain, 40% of the lateral horn, 33% of the Kenyon cell and 18% of the MBON deficit. As median per-fly ratios, this was 48%, 67%, 70% and 34% of what restoring all removed neurons recovered.

The descending-neuron class had the largest Banzhaf value for giant fibre rate (0.30). The MBON deficit due to neuron loss was divided among three classes, MBONs (15%, 13 to 18%), Kenyon cells (14%, 11 to 18%) and the PN class (Banzhaf values 0.19, 0.16 and 0.13). Over both severities, the central-brain intrinsic class recovered 22% of the lateral horn deficit, and MN9 drive depended most on it (30% of the MN9 deficit, 10 to 56%). The class is a catch-all, with a median of 3,354 removed neurons per fly, 21% of all removed.

The central-brain intrinsic class contains the anterior paired lateral (APL) neurons. The effect of restoring it in the five flies whose Kenyon cell and MBON rates were high relative to their own severity group (the group-relative criterion), and the one fly in which APL itself was tested, are described below (Loss of single neurons produces discrete circuit phenotypes, APL in the within-fly analysis).

The most effective single class in a fly recovered a median of 51% of what restoring all removed neurons recovered for whole-brain activity, 73% for lateral horn rate, 84% for Kenyon cell rate, 54% for MBON rate, 100% for MN9 drive and 90% for giant fibre rate. Class contributions added up. The effects of restoring single classes summed to 0.77 to 1.20 of the effect of restoring all removed neurons, and the additive model explained most of the variance across the random class combinations (median $R^2$ 0.84 to 0.997; Supplementary methods), except for MN9 drive (0.50).

**Flies with a giant fibre rate below 100 Hz.** The ten flies of the within-fly analysis with a giant fibre rate below 100 Hz (3 at s = 0.3 and 7 at s = 0.5) had damaged rates of 0 to 55 Hz. Restoring the descending-neuron class recovered 69 to 105% of their deficit, against a median of 19% in the 30 other flies with a giant fibre deficit.

In the two flies that had lost both giant fibres, restoring one of the two alone recovered 63% and 69% of the deficit, and the other 18% and 22%. In the other eight, the most effective class other than the descending neurons recovered a median of 5% of the deficit. The ten most effective tested types together recovered a median of 3.4% in these eight, 84% and 90% in the two flies in which the giant fibre type, DNp01, was tested, and a median of 23% in the 30 other flies with a deficit.

The 10 to 55 Hz rates of the eight fit the loss of the left giant fibre alone, the one on the side of the looming stimulus. All eight were below 100 Hz in their run of the severity series as well (13 to 55 Hz), in which 66 of the 67 such flies had lost it (Loss of single neurons produces discrete circuit phenotypes, below), so at least seven of the eight had lost it.

**Coverage of the screen of cell types.** Although an estimated 36 flies had lost a stimulated projection neuron, the odour A projection neuron types were tested only 4 times in 3 flies (Supplementary methods, Type selection). The tested types held 10% of the removed MBONs, 18% of the removed descending neurons, 26% of the removed neurons of the PN class and 24% of the removed central-brain intrinsic neurons, but 99.7% of the removed Kenyon cells.

For the MBON, descending-neuron and PN classes, the summed effect of the tested types was zero in the median fly. Restoring the whole classes recovered 15%, 37% and 33% of the MBON, giant fibre and lateral horn deficits. Only for Kenyon cells did the tested types add up to the class effect (12% against 13% of the Kenyon cell deficit).

Where a stimulated projection neuron type did enter the screen, it had the largest single-type effects on lateral horn and Kenyon cell rate. In one fly, two DM4_adPN neurons recovered 44% of the lateral horn and 91% of the Kenyon cell deficit. In another, restoring the two VA2_adPN neurons recovered as much as the whole PN class on every odour readout.

**Single cell types restored alone and together.** Per fly, a median of 15 of the 450 tested types changed whole-brain activity. The most effective tested type in a fly recovered a median of 3.8% of the lateral horn, 6.7% of the Kenyon cell, 10% of the MBON, 12.5% of the MN9 and 14.5% of the giant fibre deficit. It recovered more at $s = 0.3$ than at $s = 0.5$ for whole-brain activity (7.5% against 2.1%; Mann-Whitney $p = 0.002$) and for MN9 drive (40% against 8.6%; $p = 0.007$). The most effective single neurons of these types recovered less, a median of 2.5% of the whole-brain deficit.

Across the 40 flies with a giant fibre deficit, the most effective tested type, alone or tied, was LPLC2 in 14 and LC4 in 11 (one or both in 21; alone in 5 and 4). Ties occurred in 19 flies, because giant fibre rate moves in steps of one spike.

The ten most effective types together recovered 6.3% of the lateral horn, 13% of the Kenyon cell, 14% of the MBON, 6.3% of the MN9 and 18% of the giant fibre deficit. For whole-brain, lateral horn, Kenyon cell, MBON, MN9 and giant fibre readouts, this was 10%, 13%, 39%, 29%, 0% and 23% of what restoring all removed neurons recovered in the same fly (median per-fly ratios).

Effects combined sub-additively: restored together, the types recovered less than the sum of what each recovered alone. The ten most effective types together recovered 0.71, 0.78, 0.75, 0.58, 0.14 and 0.25 of the sum of their single effects for the same six readouts. The three most effective types together recovered less than the most effective single type in 20 of 48 flies for whole-brain activity and in 7 to 18 flies for the other five readouts.

Because the most effective types were chosen as the largest of 450 single effects, part of this shortfall may reflect selection of effects that do not reproduce. MN9 drive is also read at a resolution of 1 Hz against a median deficit of 9.5 Hz, so its fractions are coarse.

**Pre-specified conditions: wiring and robustness.** For the wiring condition, MN9 drive was the only readout for which every tested type of a fly gave the same fraction of the deficit restored, in 12 flies ($R^2$ = 1 by construction); its median leave-one-type-out $R^2$ was −0.016 over all 47 flies and −0.032 over the other 35.

Some types were consistently more important than the wiring predicted, with a median standardised residual of at least 2 in at least 10 flies, a floor added after the screen. These types recovered little:

1. whole-brain activity: LHPD2a1, KCg-m and KCab-m (median effects 0.6 to 1.1% of the deficit);
2. lateral horn rate: LHPD2a1, LHAV4a1_b and LHPV5d1 (2 to 6%);
3. Kenyon cell and MBON rates: Kenyon cell types KCab-m, KCa'b'-m and KCg-m, plus KCab-c for Kenyon cell rate (2 to 5%);
4. giant fibre rate: LPLC2, PVLP111 and LC4 (4 to 11%). LPLC2 and LC4 are the stimulated neurons that deliver looming, so their importance follows from the protocol.

No type met this criterion for MN9 drive.

High connectivity alone did not make a type important. The giant fibres, APL and six of the eight odour A projection neurons are among the most connected 0.4% of neurons, but of the 174 tested types whose lost neurons had a mean connectivity rank of at least 0.996, averaged over the flies in which the type was tested, 80 changed none of the six readouts in any fly and 128 recovered less than 5% of every readout's deficit in every fly in which they were tested.

For the robustness condition, the 54 types of the robustness test were the ten most effective for each readout (44 distinct types) and ten control types drawn at random from the other tested types. The required agreement with the unchanged model, the extended model with unchanged parameters (Materials and methods), was reached in 15 of 36 combinations of variant and readout: 11 of 24 for the parameter variants and 4 of 12 for the variants with new input noise, under which 1 and 3 of 6 readouts passed.

Six of the 15 passes rest on five or fewer types with a non-zero median: MN9 drive in five variants, and giant fibre rate with weights ×0.8. Most of the 54 types had a median effect of exactly zero under the unchanged model and all six variants (34 or 35 for whole-brain, lateral horn, Kenyon cell and MBON rates, and 47 for MN9 drive).

Ties in the top 10 were broken in a fixed type order. With that rule the top-10 overlap, the fraction of the ten highest-ranked types shared with the unchanged model, reached 0.6 in 34 of 36 combinations. Under random tie-breaking (1,000 draws), the mean overlap was 0.25 to 0.26 for MN9 drive and 0.37 to 0.47 for giant fibre rate, and 22 of 36 combinations reached 0.6.

The lateral horn ranking was the most stable: its top-10 overlap was 1.0 in all six variants, and its rank correlation reached 0.7 in four. In a post hoc analysis restricted to types with a non-zero median in either run, the rank correlations were 0.89 to 0.96 in every variant for lateral horn rate, 0.47 to 0.74 for whole-brain activity, 0.39 to 0.76 for Kenyon cell rate and 0.22 to 0.63 for MBON rate.

Single-type effects changed with the input noise. Across fly and type, the Spearman correlation of the fractions restored between the original realisation of the input noise and the two new ones was 0.050 and 0.043 for MN9 drive and between 0.089 and 0.570 for the other five readouts. Some effects held. Of the seven flies that had lost PVLP151, restoring it recovered 32 to 55% of the giant fibre deficit in three under each of the three realisations, and nothing in three.

**Full-protocol test: restoring a type alone and removing it alone.** Restored alone in the damaged fly, a type recovered at most 0.8% of the whole-brain, 1.8% of the lateral horn and 4.2% of the Kenyon cell and MBON deficits (medians over flies for each type). These odour readouts are identical to those of the shortened protocol and do not replicate them independently (Supplementary methods, Full-protocol test). GNG120 recovered 14% of the MN9 deficit (8%, 14% and 10% under the three realisations of the input noise in the shortened protocol, ranking first in each). PVLP151 recovered 21% of the giant fibre deficit (about 10% under each of those realisations).

Removing only the neurons of a type that the fly had lost from its healthy copy, one neuron for the median type and two for the median removal, reproduced at most 7.5% of the whole-brain, 3.7% of the lateral horn, 7.1% of the Kenyon cell, 10.5% of the MBON, 15% of the giant fibre (LPLC2) and 27% of the MN9 deficit (DNge106). For the learning index, 16 types reproduced more than 5% of the deficit, up to 25% (CRE080_c, per-fly values −0.53 to 0.55). Another 13 shifted it by more than 5% in the opposite direction, and 9 of the 10 control types had a median of zero.

Network state carries over between steps of the protocol. The learning index changed in 318 of the 323 removals that altered any readout before training, but in only 18 of the 160 that altered none. The learning index therefore picks up perturbations of earlier steps, and the effects of these removals on it need not identify pathways for learning.

For every readout except giant fibre rate, the largest effect of removing a type from a healthy copy exceeded the largest effect of restoring it in the damaged fly. Part of this asymmetry is built into the comparison. A restored neuron keeps the fly's threshold shift and its own thinned synapses and rejoins a network with many removed partners, whereas removal from the healthy copy takes intact neurons out of an intact network.

## The learning deficit mostly follows mushroom body output

**Learning index and MBON rate.** The learning index is a difference of valences, signed sums of MBON rates, measured before and after training, so its size is expected to scale with the overall level of MBON activity. Across the 384 flies of the severity series the two were correlated (Spearman $\rho = 0.73$; S3A Fig), and their medians declined together (S3B Fig; Table A). At low severity the relation held between severity groups more than between flies. The correlation was 0.35 over $s \leq 0.3$ (224 flies, healthy included) and 0.75 over $s \geq 0.4$ (160 flies). Within single severity groups, $\rho$ was −0.39 among the healthy flies ($p = 0.002$, $n = 64$), between −0.02 and 0.27 at $s = 0.05$ to 0.3 (all $p \geq 0.13$, $n = 32$ each) and between 0.41 and 0.70 at $s = 0.4$ to 1.0 (all $p \leq 0.02$). Controlling for severity, the partial rank correlation was 0.37 over all 384 flies ($p = 4 \times 10^{-14}$), 0.13 over $s \leq 0.3$ ($p = 0.047$) and 0.61 over $s \geq 0.4$ ($p = 6 \times 10^{-18}$).

The negative correlation among healthy flies reflects input noise in the naive presentation. The learning index subtracts the naive valence of odour A, which is measured in the same presentation as MBON rate ($\rho = 0.71$ between them), and with naive valence controlled it was unrelated to MBON rate in these flies (partial rank correlation 0.11, $p = 0.40$).

**The learning index divided by MBON rate.** Divided by each fly's own MBON rate, the learning index fell much less (S3B Fig; Table A). The ratio is defined for MBON rates of at least 0.5 Hz, which every fly kept up to $s = 0.3$; its healthy median, 26.4, is the reference value in Table A. MBON rate, a mean over MBONs of both signs, is a coarse measure of the gain that matters for learning.

**Table A. Severity series: mushroom body output, learning index, their ratio and the teaching signal.** Medians as % of the healthy median (64 flies). The ratio has 95% bootstrap intervals and two-sided Mann-Whitney p against the healthy flies, uncorrected (nine tests; $p < 0.05$ only at $s = 0.10$ and 0.6). The PAM pairing rate mainly counts the stimulated PAM neurons that survive, that is, the delivered teaching signal; within severity groups it did not track the learning index (Circuit readouts decline with severity, above).

| s | MBON rate | Learning index | Learning index / MBON rate (95% interval) | Flies with ratio defined | p vs healthy flies (uncorrected) | PAM pairing rate |
|---|---|---|---|---|---|---|
| 0.05 | 93.5% | 104.9% | 117.3% (96.8 to 138.0%) | 32 of 32 | 0.25 | 98.9% |
| 0.10 | 88.6% | 75.6% | 80.4% (61.5 to 106.7%) | 32 of 32 | 0.048 | 97.8% |
| 0.15 | 74.7% | 74.8% | 85.2% (67.3 to 113.8%) | 32 of 32 | 0.20 | 96.8% |
| 0.2 | 74.0% | 55.3% | 78.0% (47.8 to 114.7%) | 32 of 32 | 0.092 | 95.5% |
| 0.3 | 56.3% | 49.6% | 77.4% (59.0 to 107.5%) | 32 of 32 | 0.091 | 93.5% |
| 0.4 | 54.9% | 43.9% | 88.6% (65.1 to 107.4%) | 31 of 32 | 0.10 | 89.6% |
| 0.5 | 33.2% | 27.6% | 84.9% (62.3 to 105.9%) | 25 of 32 | 0.056 | 87.9% |
| 0.6 | 20.0% | 17.9% | 72.0% (25.4 to 110.4%) | 14 of 32 | 0.029 | 85.7% |
| 0.75 | 15.7% | 15.4% | 103.2% (63.1 to 132.9%) | 12 of 32 | 0.86 | 81.7% |
| 1.0 | 1.1% | 1.6% | undefined | 0 of 32 | – | 75.4% |

Pooled over s = 0.10 to 0.3, where the ratio is defined in every fly, it was 81.5% of healthy (67 to 99.8%; Mann-Whitney $p = 0.018$, $n = 128$). The upper bound of this interval ranged from 98.9 to 100.5% over repeated bootstraps of 4,000 resamples; the interval is quoted from 40,000 resamples, as is the next. The reduction depends on which groups are pooled: with s = 0.05 included the ratio was 89% (73 to 110%; $p = 0.11$, $n = 160$), and in no single severity group did the interval exclude 100% (Table A). Pooled over s = 0.10 to 0.6 the ratio was 82% (68 to 97%; $p = 0.006$, $n = 198$). Among damaged flies it did not decline further with severity ($\rho = -0.02$, $p = 0.84$ over s = 0.10 to 0.3), nor when tested over s = 0.05 to 0.3, 0.10 to 0.6 or all flies from s = 0.10 with a defined ratio (over the four tests, ρ from −0.14 to −0.02, p from 0.077 to 0.84, n = 128 to 210; Kruskal-Wallis $p = 0.92$ across the seven groups from s = 0.10 to 0.6). The trend over all 306 flies with a defined ratio, healthy flies included, is the twelfth test of the Holm family of severity trends ($\rho = -0.16$, raw and Holm $p = 0.005$); it is weak and consistent with a single step between $s \leq 0.05$ and $s \geq 0.10$. By s = 0.3, the fall in MBON rate thus accounts for most of the fall in the learning index. Severity nevertheless remained associated with the learning index after MBON rate was accounted for (partial rank correlation −0.28, $p = 2 \times 10^{-8}$, all 384 flies; −0.14, p = 0.036, s = 0.10 to 0.6, n = 224).

**Survivor selection from s = 0.4.** The ratio was undefined in 78 flies, all at s ≥ 0.4 (Table A), so from that severity it exists only in the flies that kept enough mushroom body output, and this selection (survivor selection) could hide a further decline. The flies in which it is defined kept higher MBON rates and learning indices than their groups (Table B), so the evidence that the reduction does not grow rests on s = 0.10 to 0.3, where every fly is included. The value at s = 0.75 (103.2%), which lies in the extrapolation, is the extreme case.

**Table B. Survivor selection of the learning index / MBON rate ratio (severity series).** Medians over all 32 flies of a group and over the flies with MBON rate at least 0.5 Hz.

| s | Flies with ratio defined | MBON rate, all flies (Hz) | MBON rate, flies with ratio (Hz) | Learning index, all flies (Hz) | Learning index, flies with ratio (Hz) |
|---|---|---|---|---|---|
| 0.4 | 31 of 32 | 1.26 | 1.28 | 27 | 28 |
| 0.5 | 25 of 32 | 0.76 | 0.90 | 17 | 18 |
| 0.6 | 14 of 32 | 0.46 | 0.66 | 11 | 16 |
| 0.75 | 12 of 32 | 0.36 | 0.87 | 9.5 | 21 |

**Rule ablation (S3C Fig).** The ratio is not a pure measure of plasticity. In the rule ablation at s = 0.5 it was at the healthy level under all three rules together (100%, 81 to 126%), but synapse loss alone raised it (141%, 116 to 166%) and neuron loss alone lowered it (73%, 50 to 99.6%), and the two differed (Mann-Whitney $p = 2 \times 10^{-6}$; Table C). Both loss rules without the threshold shift, with the ratio defined in a strongly selected subset (13 of 32 flies), differed from synapse loss alone ($p = 9 \times 10^{-4}$) but not detectably from neuron loss alone (p = 0.92) or from all three rules together (p = 0.10). The value under all three rules thus combines opposite effects of the two loss rules with those of the threshold shift and of survivor selection, and cannot be read as spared learning. It and the value of the severity series at the same severity (84.9%; Table A) come from independent groups of flies, and neither differs detectably from healthy flies.

**Table C. Learning index / MBON rate in the rule ablation.** 32 flies per condition; ratio as % of the healthy median of the severity series with 95% bootstrap intervals; two-sided Mann-Whitney p against the healthy flies; "undefined" means no fly kept an MBON rate of at least 0.5 Hz.

| Condition | s = 0.5: learning index / MBON rate (95% interval) | n defined | p vs healthy flies | s = 1.0: learning index / MBON rate (95% interval) | n defined |
|---|---|---|---|---|---|
| All three rules | 100.4% (81.3 to 125.7%) | 26 of 32 | 0.996 | undefined | 0 of 32 |
| Threshold shift only | 92.8% (71.6 to 110.9%) | 32 of 32 | 0.22 | 85.0% (69.6 to 103.2%) | 32 of 32 |
| Synapse loss only | 140.6% (115.7 to 166.3%) | 32 of 32 | $1 \times 10^{-4}$ | undefined | 0 of 32 |
| Neuron loss only | 73.3% (50.4 to 99.6%) | 31 of 32 | 0.002 | 70.0% (39.1 to 86.7%) | 19 of 32 |
| No threshold shift | 64.2% (26.8 to 113.0%) | 13 of 32 | 0.036 | undefined | 0 of 32 |

## Loss of single neurons produces discrete circuit phenotypes

**Motor phenotypes by severity.** Table D gives the counts behind Fig 6A to 6C and the graded decline of the remaining flies. Among damaged flies, giant fibre rate of flies at or above 100 Hz fell with severity (Spearman $\rho = -0.73$, $p = 6 \times 10^{-43}$, n = 253), and so did the MN9 drive of non-silent flies ($\rho = -0.79$, $p = 1 \times 10^{-56}$, n = 254).

**Table D. Giant fibre (GF) and MN9 readouts in the severity series.** 32 flies per severity (64 healthy). The healthy range of giant fibre rate was 186.7 to 201.7 Hz and that of MN9 drive 12 to 38 Hz. AUCs are against the healthy flies, using only flies at or above 100 Hz (giant fibre) or non-silent flies (MN9); an AUC of 1.00 is complete separation.

| s | GF below 100 Hz | GF from 100 Hz to below the lowest healthy value | GF at or above the lowest healthy value | Median GF rate, flies below 100 Hz (Hz) | Median GF rate, flies at or above 100 Hz (Hz) | AUC, flies at or above 100 Hz | MN9 silent | Median MN9 drive, non-silent flies (Hz) | AUC, non-silent MN9 flies |
|---|---|---|---|---|---|---|---|---|---|
| 0 (64 flies) | 0 | 0 | 64 | – | 195 | – | 0 | 20 | – |
| 0.05 | 1 | 5 | 26 | 47 | 193 | 0.59 | 0 | 19 | 0.60 |
| 0.10 | 3 | 13 | 16 | 47 | 188 | 0.73 | 4 | 17 | 0.63 |
| 0.15 | 3 | 11 | 18 | 47 | 188 | 0.74 | 4 | 16 | 0.73 |
| 0.2 | 2 | 19 | 11 | 48 | 183 | 0.77 | 5 | 14 | 0.82 |
| 0.3 | 5 | 18 | 9 | 48 | 177 | 0.92 | 6 | 13 | 0.94 |
| 0.4 | 5 | 22 | 5 | 37 | 175 | 0.97 | 9 | 10 | 0.99 |
| 0.5 | 8 | 24 | 0 | 28 | 153 | 1.00 | 6 | 9 | 0.98 |

| s | GF below 100 Hz | GF from 100 Hz to below the lowest healthy value | GF at or above the lowest healthy value | Median GF rate, flies below 100 Hz (Hz) | Median GF rate, flies at or above 100 Hz (Hz) | AUC, flies at or above 100 Hz | MN9 silent | Median MN9 drive, non-silent flies (Hz) | AUC, non-silent MN9 flies |
|---|---|---|---|---|---|---|---|---|---|
| 0.6 | 9 | 22 | 1 | 25 | 157 | 0.997 | 4 | 6 | 0.999 |
| 0.75 | 15 | 17 | 0 | 13 | 138 | 1.00 | 9 | 4 | 1.00 |
| 1.0 | 16 | 16 | 0 | 0 | 118 | 1.00 | 19 | 2 | 1.00 |

**Loss of the giant fibres.** Of the 320 damaged flies, 45 had lost only the left giant fibre, the one on the side of the looming stimulus, 21 both, 48 only the right and 206 neither. All 66 that lost the left giant fibre fell below 100 Hz, against 1 of the 48 that lost only the right one and none of the 206 that lost neither. Of the 67 flies below 100 Hz, 22 were silent; the 45 that still fired had rates of 5 to 98 Hz and a median first-spike latency of 10 ms (interquartile range 9 to 13 ms), against 3 ms in damaged flies at or above 100 Hz. A removed neuron never fires, so the 21 flies that lost both giant fibres account for 21 of the 22 silent flies, and at least 44 of the 45 that lost only the left one still fired, consistent with the unstimulated right giant fibre continuing to respond. Only two damaged flies had rates between 60 and 100 Hz (66.7 Hz at s = 0.4 and 98.3 Hz at s = 1.0).

**Arena check of flies with a low giant fibre rate.** In the arena cohort, the separate set of flies that ran the circuit protocol and were then placed in the virtual arena, 16 flies had a giant fibre rate below 100 Hz in the circuit protocol and 15 of them met at least one looming threat in the arena. Thirteen of these still took off, with as many take-offs as threats met (circuit giant fibre rates 17 to 72 Hz); the other two, both at s = 0.5 and with silent giant fibres, never took off, on three threats each. Two differences limit this check: the arena delivers looming on whichever side faces the threat, not on the left side only as the circuit protocol does, and it counts a take-off with or without a threat, so take-off counts are not escape success.

**MN9.** Of the 66 damaged flies with a silent MN9, 54 had lost one MN9 neuron, 7 both and 5 neither; of the 89 flies that lost exactly one, 35 still responded. The two MN9 neurons have connectivity ranks at the 96.6th and 58.6th percentiles, so their chances of removal differ. The median MN9 drive of non-silent flies fell from 20 Hz in healthy flies to 9 Hz (45% of healthy) at s = 0.5 (Table D). None of the 18 flies of the arena cohort with a silent MN9 fed in the arena, where feeding requires an MN9 rate above 10 Hz.

**High-activity flies.** The healthy-referenced criterion, which defines the high-activity flies, requires Kenyon cell and MBON rates both above the healthy maxima (0.896 and 2.82 Hz), and the group-relative criterion both above 1.5 times the median of the fly's own severity group (Fig 6F). No healthy fly met the corresponding criterion relative to the healthy medians (0.813 and 2.30 Hz). From s = 0.5 every fly that met the group-relative criterion had a Kenyon cell rate below the healthy maximum (Table E), and one of them, at s = 0.5, an MBON rate above the healthy maximum (3.19 Hz).

**Table E. High-activity flies in the severity series (of 32 per severity) and the expectation for loss of at least one of two neurons with nearly the largest load, $32 \times [1 - (1 - 0.5 \times s)^2]$.** Binomial P is the probability of the observed group-relative count or fewer with n = 32 and $p = 1 - (1 - 0.5 \times s)^2$; from s = 0.5 the comparison is not informative, because a criterion relative to the group median cannot flag a phenotype shared by a large part of the group.

| s | Healthy-referenced criterion | Group-relative criterion | Median Kenyon cell rate of flies meeting the group-relative criterion (× healthy median) | Expected, loss of at least one of two neurons with nearly the largest load | Binomial P (group-relative count or fewer) |
|---|---|---|---|---|---|
| 0.05 | 1 | 1 | 2.22 | 1.58 | 0.527 |
| 0.10 | 3 | 3 | 2.06 | 3.12 | 0.619 |
| 0.15 | 6 | 6 | 1.71 | 4.62 | 0.830 |
| 0.2 | 2 | 2 | 1.62 | 6.08 | 0.042 |
| 0.3 | 4 | 4 | 1.30 | 8.88 | 0.035 |
| 0.4 | 6 | 7 | 1.17 | 11.52 | 0.066 |
| 0.5 | 0 | 3 | 0.87 | 14.0 | not informative |
| 0.6 | 0 | 3 | 0.53 | 16.32 | not informative |
| 0.75 | 0 | 5 | 0.42 | 19.5 | not informative |
| 1.0 | 0 | 7 | 0.13 | 24.0 | not informative |

Summed over s = 0.05 to 0.15 the count was 10 by either criterion against 9.32 expected (about 9.3), so high-activity flies were as frequent as flies expected to lose at least one of the two neurons; over s = 0.2 to 0.4 it was 13 by the group-relative criterion and 12 by the healthy-referenced one against 26.48 (about 26.5), about half as frequent. The expectation holds for any pair of neurons with nearly the largest load, such as the two giant fibres, so the agreement at low severity is consistent with the APL neurons (connectivity rank 0.99999) but does not identify them, and the shortfall from s = 0.2 shows that not every predicted loss produced the phenotype.

**High Kenyon cell rates in the rule ablation (Fig 6E).** At s = 0.5 the threshold shift alone moved most flies slightly above the healthy maximum Kenyon cell rate (20 of 32; group median 113% of the healthy median), and the nine flies above the healthy maximum under neuron loss alone are the flies that met the group-relative criterion (Table F).

**Table F. Kenyon cell rate in the rule ablation (32 flies per condition).** Healthy maximum 0.896 Hz, healthy median 0.813 Hz. Columns 3 to 6 are numbers of flies.

| Condition | s | Kenyon cell rate above healthy maximum | above 1.2 × healthy maximum | above 1.5 × healthy median | Group-relative criterion met | Highest Kenyon cell rate (Hz) | Highest / healthy maximum |
|---|---|---|---|---|---|---|---|
| All three rules | 0.5 | 0 | 0 | 0 | 6 | 0.861 | 0.96 |
| Threshold shift only | 0.5 | 20 | 1 | 0 | 0 | 1.116 | 1.25 |
| Neuron loss only | 0.5 | 9 | 8 | 3 | 9 | 1.417 | 1.58 |
| Synapse loss only | 0.5 | 0 | 0 | 0 | 0 | 0.508 | 0.57 |
| No threshold shift | 0.5 | 0 | 0 | 0 | 4 | 0.594 | 0.66 |
| All three rules | 1.0 | 0 | 0 | 0 | 13 | 0.152 | 0.17 |
| Threshold shift only | 1.0 | 31 | 12 | 0 | 0 | 1.215 | 1.36 |
| Neuron loss only | 1.0 | 0 | 0 | 0 | 10 | 0.875 | 0.98 |
| Synapse loss only | 1.0 | 0 | 0 | 0 | 0 | 0.128 | 0.14 |
| No threshold shift | 1.0 | 0 | 0 | 0 | 6 | 0.080 | 0.09 |

**APL in the within-fly analysis.** The within-fly analysis covered s = 0.3 and 0.5 only, so no fly at s ≤ 0.2 was tested. Five of its 48 flies met the group-relative criterion in the severity series, three at s = 0.3 and two at s = 0.5. The three at s = 0.3 were also above the healthy maxima (Kenyon cell rates 0.98, 0.94 and 1.13 Hz; MBON rates 3.55, 3.03 and 4.87 Hz). The high activity of the five flies was relative to the group, not necessarily to healthy flies. In their runs of the within-fly analysis, two of them had a Kenyon cell rate below that of their healthy copy, and a third was within 2% of it. In all five, restoring the central-brain intrinsic class, which contains APL (2,454 to 4,293 neurons per fly), lowered Kenyon cell rate by 0.24 to 0.50 Hz, to 46 to 74% of the healthy copy, and MBON rate by 1.28 to 2.74 Hz, to 28 to 62%, whereas no other class lowered them by more than 0.010 and 0.021 Hz. The class is too large for this to show that APL is responsible.

In one of the two flies at s = 0.5 both APL neurons had been removed and APL was among the tested types. In the severity series this fly's Kenyon cell and MBON rates were 1.9 and 3.5 times the medians of its severity group, although below the healthy maxima; under the shortened protocol of the within-fly analysis they were 0.68 and 2.48 Hz. Restoring the two APL neurons alone gave 0.37 and 0.67 Hz, restoring the whole class 0.38 and 0.66 Hz, and restoring either APL neuron alone intermediate values (Kenyon cell rate 0.50 or 0.55 Hz, MBON rate 1.41 or 1.79 Hz). The medians of the 24 flies of the within-fly analysis at s = 0.5 were 0.34 and 0.71 Hz, and the rates of the fly's healthy copy 0.81 and 2.33 Hz. No other of the 450 types tested in this fly lowered MBON rate by more than 0.021 Hz.

In the other four flies APL was not among the tested types. The screen chose types by the number of neurons they had lost, and in these flies every type chosen had lost at least two neurons, or three in one fly. The random control types came from classes other than APL's, so a single lost APL neuron could not have been tested, and in that one fly neither could the loss of both.

## Separation from healthy flies follows the change relative to trial-to-trial spread

**First severity at which the AUC criterion was reached (Fig 7B).** Separation of a severity group from the healthy flies is measured as the AUC. Table G gives, for each readout, the first severity at which the data reached an AUC of 0.80 and the distribution of that severity over bootstrap resamples.

**Table G. First severity with AUC ≥ 0.80, severity series.**

| Readout | Observed first s (AUC ≥ 0.80) | Bootstrap median | Bootstrap 2.5 to 97.5 percentile | Resamples never reaching 0.80 | Resamples reaching it by s = 0.2 | AUC one step before observed |
|---|---|---|---|---|---|---|
| Whole-brain activity | 0.10 | 0.10 | 0.05 to 0.15 | 0% | 100% | 0.753 |
| Lateral horn rate | 0.10 | 0.10 | 0.05 to 0.15 | 0% | 100% | 0.779 |

| Readout | Observed first s (AUC ≥ 0.80) | Bootstrap median | Bootstrap 2.5 to 97.5 percentile | Resamples never reaching 0.80 | Resamples reaching it by s = 0.2 | AUC one step before observed |
|---|---|---|---|---|---|---|
| MBON rate | 0.2 | 0.15 (not stable) | 0.10 to 0.3 | 0% | 96% | 0.790 |
| Kenyon cell rate | 0.2 | 0.2 | 0.10 to 0.2 | 0% | 99% | 0.784 |
| Kenyon cell recruitment | 0.2 | 0.2 | 0.05 to 0.3 | 0% | 96% | 0.754 |
| MN9 drive | 0.2 | 0.2 | 0.15 to 0.3 | 0% | 88% | 0.767 |
| Naive valence | 0.3 | 0.2 | 0.15 to 0.3 | 0% | 64% | 0.798 |
| Giant fibre rate | 0.3 | 0.2 | 0.10 to 0.3 | 0% | 70% | 0.785 |
| Learning index | 0.4 | 0.3 | 0.15 to 0.5 | 0% | 38% | 0.793 |
| Projection neuron rate | 0.4 | 0.4 | 0.3 to 0.5 | 0% | 2% | 0.653 |
| Learning index / MBON rate | never | 0.6 | 0.2 to 0.6 | 89% | 0% | - |
| Giant fibre latency | 0.75 | 0.75 | 0.4 to 1.0 | 0.2% | 0% | 0.768 |

The bootstrap median of MBON rate was not stable across repeated bootstraps (0.15 or 0.2; Supplementary methods, Statistics). For the learning index divided by MBON rate the bootstrap median and range refer to the 11% of resamples that reached the criterion.

**Paired comparisons with the learning index.** Within the same flies, whole-brain activity minus learning index AUC was 0.19 (0.03 to 0.31) at s = 0.05, 0.16 (0.02 to 0.31) at 0.10, 0.28 (0.15 to 0.41) at 0.15, 0.22 (0.12 to 0.34) at 0.2 and 0.21 (0.11 to 0.31) at 0.3 (95% paired-bootstrap intervals). Lateral horn rate minus learning index was 0.22 (0.06 to 0.34), 0.17 (0.002 to 0.32), 0.29 (0.18 to 0.42), 0.20 (0.09 to 0.32) and 0.21 (0.11 to 0.31) at the same severities; the sign of its lower bound at s = 0.10 was not stable across repeated bootstraps. MN9 drive and giant fibre rate separated better than the learning index only at s = 0.3, and Kenyon cell and MBON rates at none. Part of this ordering is built into the readouts, because the learning index is computed from MBON rates and the motor readouts lie downstream of central activity.

**Test-retest.** Table H compares the two runs of the flies simulated twice with the same damage and different input noise (Supplementary methods, Simulated cohorts, Flies simulated twice). It gives the Spearman correlation between the runs and, in brackets, the within-pair SD, the square root of half the mean squared difference between the runs, divided by the SD over both runs. A healthy pair is two runs of the same intact network, so a ρ near 0 and a ratio near 1 follow by construction. At s = 0.5, flies with pan-neuronal pathology did differ as individuals: 16 rerun with new input noise largely kept their whole-brain activity (ρ = 0.96 between runs, against −0.16 for 16 healthy flies run twice) and, less closely, their learning index (0.71). The flies with Kenyon-cell-only pathology were not reproducible for most readouts. Only the projection neuron rate was correlated between healthy reruns ($p = 0.003$ uncorrected, 0.03 after Bonferroni correction over the ten readouts), which is attributed to chance.

**Table H. Test-retest Spearman ρ (within-pair SD / total SD).**

| Readout | Healthy network, 16 pairs of runs: ρ (within-pair / total SD) | s = 0.5, pan-neuronal, 16 flies | s = 0.5, Kenyon-cell-only, 8 flies |
|---|---|---|---|
| Whole-brain activity | −0.16 (1.07) | 0.96 (0.16) | −0.10 (1.00) |
| Lateral horn rate | 0.28 (0.72) | 0.93 (0.13) | −0.10 (1.09) |
| Projection neuron rate | 0.69 (0.59) | 0.90 (0.10) | −0.51 (1.15) |
| Kenyon cell rate | 0.19 (0.84) | 0.93 (0.19) | −0.24 (1.07) |
| Kenyon cell recruitment | 0.04 (0.95) | 0.89 (0.22) | 0.52 (0.71) |
| MBON rate | −0.18 (1.10) | 0.90 (0.20) | 0.12 (0.81) |
| Naive valence | 0.36 (0.87) | 0.76 (0.40) | −0.19 (1.06) |
| Learning index | −0.16 (1.07) | 0.71 (0.40) | 0.05 (1.11) |
| MN9 drive | 0.00 (0.93) | 0.94 (0.19) | 0.21 (0.75) |
| Giant fibre rate | 0.28 (0.96) | 1.00 (0.03) | 0.52 (0.82) |

**Spread, change and separation (Fig 7C).** Table I gives the trial-to-trial spread of each readout, the robust coefficient of variation (CV) of the 64 healthy flies, with its change and AUC at three severities.

**Table I. Trial-to-trial spread among healthy flies, change of the group median relative to the healthy median and AUC, severity series.**

| Readout | Healthy robust CV (%) | Rank of CV (1 = lowest of 10) | Median change at s = 0.05 / 0.10 / 0.2 (%) | AUC at s = 0.05 / 0.10 / 0.2 |
|---|---|---|---|---|
| Whole-brain activity | 9.56 | 7 | −9.56 / −16.56 / −43.42 | 0.753 / 0.829 / 1.000 |
| Lateral horn rate | 4.00 | 3 | −3.89 / −5.63 / −18.06 | 0.779 / 0.830 / 0.980 |
| Projection neuron rate | 2.84 | 2 | 0.00 / +1.29 / −3.92 | 0.515 / 0.523 / 0.658 |
| Kenyon cell rate | 5.34 | 4 | −4.74 / −7.77 / −18.66 | 0.645 / 0.680 / 0.915 |
| Kenyon cell recruitment | 5.56 | 5 | −4.75 / −4.83 / −12.11 | 0.686 / 0.639 / 0.880 |
| MBON rate | 9.31 | 6 | −6.50 / −11.43 / −26.01 | 0.652 / 0.720 / 0.878 |
| Naive valence | 11.44 | 8 | −11.11 / −14.81 / −26.54 | 0.661 / 0.648 / 0.798 |
| Learning index | 32.85 | 10 | +4.88 / −24.39 / −44.72 | 0.559 / 0.664 / 0.776 |
| MN9 drive | 30.58 | 9 | −5.00 / −15.00 / −30.00 | 0.604 / 0.677 / 0.848 |
| Giant fibre rate | 1.90 | 1 | −1.28 / −4.27 / −6.84 | 0.602 / 0.755 / 0.785 |

Across the eight graded readouts, the rank correlation of AUC with the absolute median change divided by the spread, with the absolute change alone and with the inverse spread alone was 0.86 ($p = 0.007$), 0.26 ($p = 0.53$) and −0.02 ($p = 0.96$) at $s = 0.05$; 0.74 ($p = 0.037$), 0.38 ($p = 0.35$) and −0.07 ($p = 0.87$) at $s = 0.10$; and 0.90 ($p = 0.002$), 0.10 ($p = 0.82$) and 0.10 ($p = 0.82$) at $s = 0.2$. These p values use the t approximation; the exact permutation values for the change divided by the spread were 0.011, 0.046 and 0.005. With MN9 drive and giant fibre rate included, $\rho$ was 0.73 at $s = 0.10$ ($p = 0.016$, exact 0.020; $n = 10$). Projection neuron rate, the least variable graded readout, changed by only 1.3% at $s = 0.10$ (AUC 0.52), as expected for a measure of delivered odour input. Separation also depends on how many flies of a group a change affects: giant fibre rate had the largest change relative to its spread, yet only the third-highest AUC at $s = 0.10$ (0.76; Table I), because half of the flies, 16 of 32, stayed within the healthy range; its AUC without the 3 flies below 100 Hz was 0.73.

**High activity and the mushroom body AUCs (grey curves in Fig 7A).** Table J gives the AUCs with and without the flies whose Kenyon cell and MBON rates were high relative to their own severity group. MBON rate and Kenyon cell recruitment reached an AUC of 0.80 at $s = 0.2$ but fell below it again at $s = 0.4$, where 7 of 32 flies had such high activity; without these flies, both readouts and Kenyon cell rate had AUCs of 0.928 or more from $s = 0.15$ to 0.5.

**Table J. AUC of the mushroom body readouts with all 32 flies and without the flies that met the group-relative criterion.**

| Readout | s = 0.15 | s = 0.2 | s = 0.3 | s = 0.4 | s = 0.5 |
|---|---|---|---|---|---|
| Flies meeting the group-relative criterion (of 32) | 6 | 2 | 4 | 7 | 3 |
| Kenyon cell rate: all flies / without these flies | 0.784 / 0.97 | 0.915 / 0.98 | 0.873 / 1.00 | 0.811 / 1.00 | 0.976 / 1.00 |
| Kenyon cell recruitment: all flies / without these flies | 0.754 / 0.93 | 0.880 / 0.94 | 0.855 / 0.98 | 0.792 / 0.98 | 0.951 / 1.00 |
| MBON rate: all flies / without these flies | 0.790 / 0.97 | 0.878 / 0.94 | 0.870 / 0.99 | 0.789 / 0.98 | 0.911 / 0.99 |

Because these flies were selected on Kenyon cell and MBON rates, removing them raises the AUCs of those two readouts by construction; the comparison measures how much separation these few flies cost, most at s = 0.15 and 0.4, where they were most numerous.

## Circuit readouts against virtual-arena behaviour in the same flies

**The arena cohort.** Circuit readouts were compared with virtual-arena behaviour in the same flies by simulating a separate set of 176 flies, the arena cohort (32 healthy and 24 at each of six severities from 0.05 to 0.5). Each fly ran the circuit protocol with two steps added before training: a state reset, which returns the dynamic state to rest and leaves the weights unchanged, and a naive-valence probe, odour presentations that give the reference valence for steering in the arena. It then spent 120 s in a virtual arena where it could feed, take off from looming threats and approach either odour (Materials and methods). In order, each fly ran: steps 1 to 3 of the circuit protocol; the state reset; the naive-valence probe; training as in step 4, once; the memory test of step 5; and the arena, after a second state reset. Its learning index pairs the valences of step 1, measured before the reset, with those of step 5; the severity series has no reset or probe, which limits comparison of the learning index between the two sets of flies. Jumps per threat is undefined for the 5 flies that met no threat, and the odour preference index for the 6 that entered neither plume.

**Separation by circuit readouts in the arena cohort (S4 Fig).** The result for whole-brain activity and lateral horn rate was reproduced: both again reached an AUC of 0.80 at a lower severity than any other readout, s = 0.10 (bootstrap range 0.05 to 0.15). The rest of the order was not. The learning index reached it at s = 0.15 (bootstrap range 0.05 to 0.5), having reached it by that severity in 64% of resamples, against 0.4 in the severity series. MBON rate, Kenyon cell rate and MN9 drive also reached it at s = 0.15, in about half of the resamples by that severity; giant fibre rate reached it at s = 0.2, naive valence at 0.3, and Kenyon cell recruitment and PN rate at 0.5; giant fibre latency and the learning index divided by MBON rate did not reach it by s = 0.5. Every circuit readout except the learning index divided by MBON rate worsened with severity (Spearman rank correlation, Holm-corrected $p < 10^{-3}$; for the ratio, Holm-corrected $p = 0.52$).

Within the same flies, whole-brain activity was taken to separate better than the learning index (Materials and methods) only at s = 0.15 to 0.3: whole-brain activity minus learning index AUC was −0.05 (−0.24 to 0.16) at s = 0.05, 0.14 (−0.01 to 0.31) at 0.10, 0.17 (0.05 to 0.31) at 0.15, 0.27 (0.14 to 0.42) at 0.2, 0.20 (0.08 to 0.33) at 0.3 and 0.04 (0.00 to 0.09) at 0.5, where the AUCs were 1.00 and 0.96; lateral horn rate behaved alike. The learning-index AUC was not monotone in the arena cohort (0.81 at s = 0.15, 0.73 at 0.2, 0.80 at 0.3).

The bootstrap ranges of the two sets of flies overlap, and learning-index medians did not differ systematically between the arena cohort and the severity series: 67 against 61.5 in healthy flies (Mann-Whitney $p = 0.13$), 52.5 against 64.5 at s = 0.05 ($p = 0.04$, uncorrected) and p between 0.11 and 0.89 at the other severities. The learning index declined with severity to a similar extent ($\rho = -0.52$ in the arena cohort, −0.50 in the severity series restricted to the same severities), and its healthy robust coefficient of variation was 24% in the arena cohort and 33% in the severity series. The learning index is therefore unstable between the two sets of flies, as its large trial-to-trial spread predicts.

**Separation by arena readouts (S4 Fig).** Two arena readouts declined with severity, time feeding ($\rho = -0.51$, Holm-corrected $p = 2 \times 10^{-12}$, n = 176) and jumps per threat ($\rho = -0.49$, Holm-corrected $p = 5 \times 10^{-11}$, n = 171), and both separated damaged from healthy flies only at higher severity than whole-brain activity and lateral horn rate did. Time feeding first reached an AUC of 0.80 at s = 0.5 (bootstrap range 0.15 to 0.5; never in 8% of resamples), and so did jumps per threat (AUC 0.79 at s = 0.2 and 0.73 at 0.3; never in 41% of resamples). Whole-brain activity was taken to separate better than every arena readout at every severity from 0.10 to 0.5, and lateral horn rate from 0.15; the learning index at none. The criterion for a circuit readout separating better than the arena readouts is given in Supplementary methods, Statistics; the closest arena readout is the one with the lowest lower bound of the paired AUC difference. At the first severity at which whole-brain activity met this criterion, its AUC exceeded that of the closest arena readout, jumps per threat, by 0.17 (0.01 to 0.31). The learning index had a higher point AUC than every arena readout at five of the six severities, but its difference from the closest one was not detectable at any (at s = 0.5, 0.09, −0.01 to 0.21). At s = 0.05 no readout of either kind reached an AUC of 0.80 (highest 0.72, lateral horn rate), and no circuit readout was taken to separate better than any arena readout. Over all pairs of 12 circuit and 4 arena readouts at the six severities, the circuit readout was taken to separate better in 95 of 288 pairs and the arena readout in 1.

**Arena odour readouts.** The two arena odour readouts, time near the trained odour and the odour preference index, showed no trend with severity (Holm-corrected p = 0.77): their Spearman correlations with severity were −0.04 and −0.07, respectively, and their AUCs lay between 0.50 and 0.63 at every severity. This does not show that memory is preserved, because these readouts failed as a measure of memory in this model: healthy flies did not prefer the trained odour (mean preference index +0.11, Wilcoxon signed-rank p = 0.45, n = 32), and their preference did not follow their circuit learning index ($\rho = -0.21$, p = 0.25).

**Trial-to-trial spread of the arena readouts.** Across the 32 healthy flies of the arena cohort the arena readouts varied far more than their circuit counterparts: the trial-to-trial spread was 102% for time feeding, 119% for time near the trained odour and 23% for jumps per threat (n = 31), against 11% for whole-brain activity, 4.9% for lateral horn rate, 16% for MN9 drive and 1.4% for giant fibre rate. Six healthy flies never fed although their MN9 rates (17 to 32 Hz) exceeded the 10-Hz feeding threshold, so their time feeding of zero reflects whether the walk brought them onto food, not their brain. Jumps per threat is quantised by the few threats met, a median of 3 in healthy flies (range 0 to 7).

**Circuit and arena readouts of the same fly.** MN9 drive and time feeding were correlated across all 176 flies ($\rho = 0.49$, $p = 3 \times 10^{-12}$), mostly through their shared dependence on severity: the partial rank correlation controlling for severity was 0.22 (p = 0.003), and among healthy flies there was no relation ($\rho = -0.08$, p = 0.65). Giant fibre rate and jumps per threat behaved alike ($\rho = 0.42$, $p = 9 \times 10^{-9}$, n = 171; partial 0.16, p = 0.04).

**Jumps per threat.** The decline of jumps per threat with severity came from the loss of take-offs outside a threat. At most one take-off can fall within a threat met (Supplementary methods, Virtual arena), so every take-off beyond the number of threats met is a take-off outside a threat (Table K).

**Table K. Jumps per threat in the arena cohort.** Columns 2 to 6 refer to the flies that met at least one threat; the last two columns refer to all flies. Take-offs beyond threats met are a lower bound on take-offs outside the threats met.

| s | Flies that met a threat | More than one take-off per threat | Exactly one | Fewer than one | Median jumps per threat | Flies (all) with take-offs beyond threats met | Take-offs beyond threats met / all take-offs |
|---|---|---|---|---|---|---|---|
| 0 (healthy) | 31 | 18 | 13 | 0 | 1.17 | 18 of 32 | 26 / 137 |
| 0.05 | 24 | 11 | 13 | 0 | 1.00 | 11 of 24 | 16 / 97 |
| 0.10 | 23 | 4 | 19 | 0 | 1.00 | 4 of 24 | 12 / 91 |
| 0.15 | 24 | 2 | 22 | 0 | 1.00 | 2 of 24 | 2 / 72 |
| 0.2 | 24 | 0 | 24 | 0 | 1.00 | 0 of 24 | 0 / 89 |
| 0.3 | 22 | 2 | 20 | 0 | 1.00 | 2 of 24 | 5 / 70 |
| 0.5 | 23 | 0 | 21 | 2 | 1.00 | 0 of 24 | 0 / 83 |

Only 2 of the 171 flies that met a threat took off fewer times than the threats they met, and with values capped at 1 the trend with severity shrank to $\rho = -0.16$ ($p = 0.03$). The uncapped trend (uncorrected $p = 8 \times 10^{-12}$) was matched by that of an indicator of more than one take-off per threat ($\rho = -0.49$, $p = 8 \times 10^{-12}$). The number of threats met did not change with severity ($\rho = 0.00$, $p = 0.98$, $n = 176$).

## Supplementary methods

### Connectome

**Source tables.** MaleCNS v1.0 was released by FlyEM (HHMI Janelia Research Campus) with the University of Cambridge, the MRC Laboratory of Molecular Biology and Google Research. Three of its released tables [1] were used: body annotations and connectome weights, both at a minimum synapse confidence of 0.5, and body neurotransmitter predictions. The criterion for retaining neurons, every reconstructed cell with a superclass annotation, follows that of the MaleCNS network of the drosophila-brain-mlx project (https://github.com/Kisame76/drosophila-brain-mlx). A neuron's side is its annotated soma side or, where that is neither left nor right, its root side.

**Rows removed.** The raw weight table has 151,856,684 connection rows, of which 126,273,746 (83% of rows, not of synapses) were dropped because an endpoint had no superclass annotation. Of the 166,700 neurons, 103,720 were predicted cholinergic, 51,371 GABAergic or glutamatergic, and 11,609 carried another prediction or none; the 1,113,526 outgoing rows of this last group were also dropped. No minimum number of contacts per connection was applied.

**FlyWire.** FlyWire v630 (127,400 neurons) and v783 [2, 3] had their connections made excitatory or inhibitory by the convention of the base model, the published whole-brain model [4], GABA and glutamate negative and every other transmitter positive, not by the convention used for MaleCNS.

## Neuron model and simulation engine

The membrane and synaptic variables are advanced at 0.1 ms by their exact exponential solutions. In an isolated neuron at rest, one Poisson input event triggers a spike on the next step (0.98 spikes per event at 150 Hz). Within a step the engine, the simulator written for this study, follows the update order of the reference Brian2 implementation [5]: state update, threshold, synaptic delivery and Poisson input, reset. As in the base model [4], a spike resets the membrane potential and sets the synaptic variable to zero, and neither is integrated during the refractory period; the adaptation current of the extended model is not reset and decays throughout. The base model declares the synaptic variable with Brian2's `(unless refractory)` clause. Two treatments of synaptic input that arrives during the refractory period, discarded or accumulated, were compared with the reference runs (below); every simulation of the paper discards it.

## Engine validation

**Reference runs.** The engine was compared with two kinds of reference: Brian2 2.10.1 running the unmodified published code [6] on the same machine, at the 150-Hz default of that code and at 200 and 100 Hz, and the two activation-only example outputs distributed with the code, generated at 200 and 100 Hz (below). Each reference is a single realisation of the input noise, the random timing of the input spikes, over 30 trials. The engine was run with three realisations at each rate; the per-neuron comparisons and each panel of S1 Fig show one realisation, and ranges are over the three.

**Refractory input discarded.** Against the local Brian2 run at 150 Hz, the Pearson correlation was 0.9997 across the 438 neurons active in either run and the slope 1.004 (S1A Fig). Over the three realisations the engine produced 13,544 to 13,707 spikes per trial, against 13,539 in Brian2, and drove MN9 at 82.8 to 84.1 Hz, against 82.3 Hz, differences of the same order as the variation between its realisations.

**Refractory input accumulated.** Against the same Brian2 run, the correlation fell to 0.958 and the slope rose to 1.147 (S1C Fig), with 16,681 to 16,784 spikes per trial and MN9 at 105.4 to 105.8 Hz. It follows that under Brian2 2.10.1 the `(unless refractory)` clause discards input that arrives during the refractory period.

**Distributed example outputs.** The example output distributed with the code was generated with 200-Hz input, not the 150-Hz default: its 21 stimulated neurons fire at 191.3 to 202.1 Hz (about 191 to 202 Hz), against 143.1 to 151.0 Hz in the local Brian2 run at the default, and the notebook distributed with the code states that neurons are excited at 200 Hz by default. This output was therefore compared with runs at 200 Hz (S1B and S1D Fig). Brian2 2.10.1 at 200 Hz gave 16,997 spikes per trial, against 17,052 in the distributed output, and MN9 at 93.6 Hz, against 93.3 Hz (Pearson 0.9998, slope 0.997, 459 neurons). The engine with refractory input discarded gave 16,919 to 17,012 spikes per trial and MN9 at 91.9 to 93.7 Hz over the three realisations (plotted realisation: Pearson 0.9998 across 452 neurons, slope 0.997). With refractory input accumulated it gave 20,687 to 20,757 spikes per trial and MN9 at 119.1 to 121.2 Hz (Pearson 0.967, slope 1.123).

A second distributed output, generated at 100 Hz, gave the same pattern: MN9 fired at 67.0 Hz in the distributed output, 65.7 Hz in Brian2 2.10.1 and 68.1 to 68.6 Hz in the engine with refractory input discarded (Pearson 0.9996 with the distributed output, slope 1.000, 422 neurons), against 79.9 to 81.9 Hz with refractory input accumulated (Pearson 0.957). The treatment of refractory input inferred above therefore also holds for the outputs generated by the authors of the code, whose environment file lists Brian2 2.5.1.

**Remaining differences.** At 200 Hz the engine's spike totals lay 0.2 to 0.8% below the distributed output in all three realisations, and at 100 Hz its MN9 rate lay 1.0 to 1.6 Hz above the distributed output and 2.4 to 2.9 Hz above the local Brian2 run. The largest of these differences is about two standard errors of the difference between two 30-trial means. With a single realisation per reference, a systematic offset of this size is not excluded.

**MaleCNS gustatory check.** The base model, with the published weight, no adaptation and a recurrent antennal lobe, was run with right-side input only, at 150 Hz for 30 trials of 1 s. Without a stimulus no neuron fired. The 11 putative sugar neurons of type LB3c drove the MN9 annotated as left at 28.3 ± 5.9 Hz (mean ± standard deviation, SD, over trials) and the other MN9 at 0.2 Hz, with 5,167 neurons active; the 17 LB3c and LB3b neurons drove them at 47.5 ± 8.4 Hz and 0.6 Hz, with 5,244 active. The 19 bitter neurons LB1a to LB1d left both MN9 silent while activating 15,033 neurons, including 3,352 Kenyon cells. This replicates the published drive of MN9 by sugar-sensing neurons [4] only qualitatively: comparable sugar stimuli activated far more neurons on MaleCNS than on FlyWire (about 5,200 against 432), and the published suppression of sugar-evoked MN9 firing by bitter input [4] was not tested.

## Model extensions: adaptation and a feedforward antennal lobe

**Requirements.** The three requirements set before the model was changed (Materials and methods) were, in full: activity ends after each stimulus, with about zero neurons active 500 to 700 ms after odour, sugar and looming; sugar still drives MN9 strongly and looming the giant fibres, for which no numerical limit was set; and one odour recruits roughly 5 to 20% of Kenyon cells, the required range, which covers the responses reported in experiments (about 5 to 10% on average [7, 8] and below 20% across odour concentrations [7]).

**Measure of persistent activity.** The stimuli of the tests of persistent activity, firing that continues after the stimulus has ended (Materials and methods), differ from those of the circuit protocol: the ORNs of DM1, DM4 and VA2 on the left for odour, LB3c and LB3b on the right for sugar and LPLC2 on the right for looming, all on the base model on MaleCNS with a recurrent antennal lobe. The odour left 8,696 of 166,700 neurons, including 2,281 of 4,064 Kenyon cells, still firing 500 to 700 ms after a 300-ms stimulus at 150 Hz (neurons firing in at least one of four runs with different input noise; Table L). On FlyWire v783 only the ORN stimulus was tested; 3,360 of 5,177 Kenyon cells were active 500 to 700 ms after its offset, and 8,162 to 8,225 neurons in every 100-ms bin after it ended, so persistent activity was not confined to MaleCNS.

**Size of the odour response.** Six unilateral ORN stimuli (DM1, DM4 and VA2 together, DA2, or V, on either side) each activated 11,933 to 12,413 neurons during the stimulus (10 runs). In an exploratory intensity sweep, 35 DM1 ORNs at 20 Hz activated as many projection neurons as 179 ORNs of three glomeruli at 100 Hz (547 to 562 over all 24 conditions of stimulus size, rate and adaptation). Every condition of the sweep includes DM1 and only the numbers of active neurons were recorded, so these tests show that ORN drive recruited a similarly large population whatever the glomerulus or intensity, not that the population was the same.

**Adaptation sweep.** The sweep covered $\tau_a$ of 0.3 and 1 s with $\Delta a$ of 0.5, 1, 2 and 4 mV on the network with ORN input and a recurrent antennal lobe (Table L). Persistent activity after sugar and looming ended in four settings, of which the adopted one, $\tau_a$ = 1 s with $\Delta a$ = 1 mV, kept the most MN9 drive; only $\tau_a$ = 1 s with 4 mV also ended persistent activity after the odour, and it was rejected because it cut MN9 drive by 85%. At no setting did the Kenyon cell response fall within the required range (45.5 to 57.7% of Kenyon cells responding across the eight settings; Table L).

**Table L. Spike-frequency adaptation sweep.** Base model on MaleCNS with olfactory receptor neuron input and a recurrent antennal lobe; stimuli as described under Measure of persistent activity. Persistent neurons are those active 500 to 700 ms after offset; Kenyon cell responses, MN9 and giant fibre rates are measured during the 300-ms

stimulus (four runs, union over runs for the fraction responding). Sugar was applied on the right and looming to the right LPLC2, so the giant fibre on the stimulated side is the right one.

| $\tau_a$ (s) | Δa (mV) | Neurons persistent after odour | Neurons persistent after sugar | Neurons persistent after looming | Kenyon cells responding (%) | Kenyon cell rate (Hz) | Responding MN9 (Hz) | Giant fibre, stimulated side (Hz) | Giant fibre, other side (Hz) |
|---|---|---|---|---|---|---|---|---|---|
| none | 0 | 8,696 | 933 | 92 | 61.5 | 39.28 | 95.0 | 265.0 | 140.8 |
| 0.3 | 0.5 | 8,662 | 416 | 253 | 57.6 | 17.16 | 75.8 | 240.0 | 131.7 |
| 0.3 | 1 | 7,984 | 0 | 79 | 55.8 | 10.92 | 54.2 | 228.3 | 125.0 |
| 0.3 | 2 | 8,096 | 34 | 0 | 49.7 | 6.37 | 39.2 | 208.3 | 104.2 |
| 0.3 | 4 | 6,771 | 0 | 0 | 45.5 | 3.72 | 16.7 | 182.5 | 75.8 |
| 1 | 0.5 | 9,080 | 0 | 31 | 57.7 | 15.05 | 63.3 | 238.3 | 129.2 |
| 1 | 1 (adopted) | 7,768 | 0 | 0 | 55.5 | 9.26 | 55.0 | 220.8 | 119.2 |
| 1 | 2 | 1,025 | 0 | 0 | 51.5 | 5.40 | 32.5 | 202.5 | 100.0 |
| 1 | 4 | 0 | 0 | 0 | 50.5 | 3.16 | 14.2 | 169.2 | 60.0 |

**Exploratory alternatives without adaptation.** Three further changes also left persistent activity after the odour in place: removing synapses among Kenyon cells (7,594 neurons persistent; 30.5% of Kenyon cells recruited), removing input to sensory neurons (8,211 persistent, 28% of them Kenyon cells), and lowering the unitary weight to 0.157 mV (4,004 persistent, with 19.5% of Kenyon cells recruited). The control rows of the synapse and weight tests match the adaptation sweep exactly, so their protocol is taken to be that of the sweep; the sensory-input test and the intensity sweep above have no control row, and for them the recurrent antennal lobe, the 500- to 700-ms window and, for the sensory-input test, the stimuli of the sweep are assumed.

**Feedforward antennal lobe calibration.** PNs were driven at 100 Hz for 300 ms in eight runs. Truncated to whole neurons, a mean of 1 neuron per run was active 500 to 700 ms after odour A offset with the adopted adaptation and 11 without (0 and 0 for odour B). With adaptation odour A recruited 5.6% of Kenyon cells and odour B 1.9% (4.2% and 1.5% without adaptation), so odour B fell below the 5% lower bound of the required range (Materials and methods), and the two Kenyon cell sets overlapped with a Jaccard index of 0.012.

**Healthy Kenyon cell responses in the extended model.** In the 64 healthy flies of the severity series, odour A recruited a median of 16.6% of Kenyon cells (range 13.6 to 18.7%) and odour B 8.5% (7.0 to 9.6%), fewer than odour A in every fly (median ratio B/A 0.51). The two sets overlapped with a median Jaccard index of 0.071, against 0.059 expected for independent sets of the same sizes (median per-fly ratio of observed to expected 1.21; above expectation in 63 of 64 flies).

**Sensitivity to adaptation in the extended model.** The adaptation increment was halved in the healthy copies of 12 flies under the shortened protocol of the within-fly analysis, and the changes were measured during the stimuli. Halving it changed MN9 drive by a factor of 1.46, whole-brain activity by 1.30 and the fraction of Kenyon cells recruited by 0.79, but Kenyon cell and giant fibre rates by only 1.05, so the readouts depend unequally on the adaptation setting. That adaptation returns the network to rest after each stimulus is inferred from the sweep in the recurrent network (Table L) and from the feedforward calibration; persistent activity after sugar and looming was not retested in the feedforward network.

## Synaptic weight

**Synapse-count ratio.** The analysis matched cell types between MaleCNS v1.0 and FlyWire v783 by their type annotations and kept the 6,808 types whose mean number of input synapses per neuron was at least 50 in both datasets. MaleCNS type means were taken over the neurons of the brain, that is every neuron outside the ventral nerve cord and ascending superclasses, and MaleCNS inputs were counted only from the 166,700 retained neurons. The median per-type ratio, MaleCNS over FlyWire, was 1.755 (interquartile range 1.408 to 2.259), which at face value implies a unitary weight of 0.157 mV, 0.57 times the published value. This lower weight lies outside the range of 0.8 to 1.2 times the published weight covered by the robustness test of the within-fly analysis (Materials and methods).

**Tests of 0.157 mV.** In a calibration run on the base model on MaleCNS, right-side sugar stimulation (LB3c and LB3b at 150 Hz for 300 ms; 16 runs) drove one MN9 at 24.4 Hz while the other stayed silent, with about 150 neurons active in each 100-ms bin during the stimulus and none from 100 ms after offset. Odour-evoked activity (left DM1, DM4 and VA2 ORNs; 4 runs) persisted at 0.157 mV: 3,806 to 3,966 neurons were active in every 100-ms bin up to 700 ms after offset, against 8,132 to 8,357 at 0.275 mV. The exploratory test of the lower weight (Model extensions, Exploratory alternatives without adaptation) was a separate run of both weights under the protocol of the adaptation sweep and gives the like-for-like MN9 comparison: 22.5 Hz at 0.157 mV against 95.0 Hz at 0.275 mV (ratio 0.24).

## Pathology

**Rules.** The neurons exempt from pathology are those of the eight superclasses whose name contains "sensory". Synapse loss is a binomial draw on the outgoing contact counts of every neuron with a load, removed neurons included, which matters only when a removed neuron is restored in the within-fly analysis. Of the two observations behind this rule (Table 1), the loss of boutons at the larval neuromuscular junction came without a significant change in the number of active zones [9].

**Stimulated neurons.** The 34 sugar receptor neurons are sensory and have no load. The odour PNs, the left-side LPLC2 (94 neurons) and LC4 (71 neurons) and the 316 PAM neurons of 15 types can be damaged. The connectivity ranks of the odour A PNs are 0.9994 and 0.9993 (DM1_lPN, left and right), 0.9980 and 0.9982 (DM4_adPN), 0.9988 and 0.9990 (VA2_adPN), and 0.4570 and 0.7218 (DM4_vPN). The odour B PNs rank 0.956 to 0.997 (mean 0.985), the left LPLC2 0.816 to 0.932 (mean 0.88), the left LC4 0.854 to 0.954 (mean 0.91) and the PAM neurons 0.054 to 0.877 (mean 0.499). Under the hub map a set of stimulated neurons is expected to lose 0.5 × s × its mean connectivity rank (Table M).

**Table M. Expected removal of stimulated neurons under the hub map.**

| Population | s = 0.5 | s = 1.0 |
|---|---|---|
| Odour-A PNs (8) | 22.4% | 44.8% |
| Odour-B PNs (9) | 24.6% | 49.3% |
| Left LPLC2 (94) | 22.0% | 44.0% |
| Left LC4 (71) | 22.8% | 45.5% |
| PAM neurons (316) | 12.5% | 25.0% |
| PAM, mean over the 15 types | 12.1% | 24.3% |

**Connectivity rank.** The connectivity rank is rank/N in (0, 1], ties ordered by body identifier. The number of synaptic contacts is taken in the network of excitatory and inhibitory connections before the antennal lobe was made feedforward, so outputs of the 11,609 neurons whose predicted transmitter gives no sign are absent from their own totals and from their partners' inputs, and contacts within the antennal lobe are included although the simulation removes them. In the antihub map the least connected non-sensory neuron receives the largest load; because the exempt sensory neurons lie towards the poorly connected end, the load under the antihub map is not exactly one minus the rank. Kenyon cells are the 4,064 neurons whose cell type begins with "KC". Among the neurons at which responses are measured, the two giant fibres (cell type DNp01) rank 0.998 and 0.996 and the two MN9 neurons 0.966 and 0.586. In FlyWire, highly connected neurons are disproportionately GABAergic [10], so the hub map probably damages inhibitory neurons more; the contribution of this bias was not tested.

## Anchoring to data and assumptions

The measured Kenyon cell loss of about 17% at 40 days was obtained in flies expressing amyloid-β42 in the mushroom body alone (201Y-GAL4 driver) [11]: it is 16.5% from the published means, 110.3 against 132.1 Kenyon cells labelled with green fluorescent protein per confocal section. The larger loss with pan-neuronal expression, about 30% at 40 days, was read from the graph in Figure 7A of that study. Any removal probability above about 0.29 would reach 17% Kenyon cell loss at some $s \leq 1.0$, so the measurement does not constrain the value 0.5. The input gain of about 1.85 that corresponds to the measured 46% fall in rheobase, the smallest current that makes a neuron fire, is 1/0.54, a conversion and not a measurement; the model's largest threshold shift, which narrows the gap between rest and threshold from 7 mV to 5 mV, corresponds to a given input being about 1.4 times as effective.

## Circuit protocol

**Stimuli.** Odour A drives 4 PNs per side and odour B 5 on the left and 4 on the right. Looming is delivered through LPLC2 and LC4, the two cell types from which the giant fibres receive nearly all their direct optic lobe input [12]. Every set of neurons stimulated at some point has no refractory period throughout the protocol; in the arena cohort, the flies placed in the virtual arena after the circuit protocol, this includes the right-side LPLC2 and LC4 neurons, which the arena can drive. In the healthy flies of the severity series the naive valences of the two odours have medians of 81 and 18 Hz, so odour B evoked a 4.5 times smaller naive valence than odour A.

**State carried over.** No state variable is reset between steps, so each step starts with residual adaptation: 37% of the current present at the end of the previous stimulus after a 1-s gap and 14% after a 2-s gap.

**Learning.** Compartments were assigned from the MaleCNS connectome before the outputs of dopaminergic neurons were dropped: an MBON type belongs to a dopaminergic type's compartment if it receives at least 10% of that type's synapses onto MBONs. The compartment map was built from PAM and protocerebral posterior lateral 1 (PPL1) types, but only the 15 PAM types (4 to 50 neurons each) were tested for activity, so the PPL1 compartments (16 neurons, 8 types) never change. The change of the Kenyon cell to MBON synapses was applied once, after the whole training step, so odour B during training ran on unchanged weights.

**Readouts.** In the 64 healthy flies of the severity series, training changes valence by a median of +66 Hz for odour A against +3 Hz for odour B, so the learning index is almost entirely the change for the trained odour. MN9 drive and giant fibre rate are mean rates of the two MN9 neurons and of the two giant fibres over the 0.5 s of sugar and the 0.3 s of looming, so their smallest possible change, one spike, is 1 Hz and 1.67 Hz. Looming on one side drives mainly the giant fibre on that side: right-side LPLC2 stimulation at the adopted setting of the adaptation sweep gave 221 against 119 Hz (Table L).

## Virtual arena

Table N gives the parameters of the virtual arena (Materials and methods). The PNs of both sides receive the same drive, which carries no direction: the model body supplies the direction of the odour gradient, and the brain only the sign and strength of steering. A take-off needs no threat, and because take-offs are at least 1 s apart and a threat lasts 0.6 s, at most one falls within a threat; a value of jumps per threat above 1 therefore includes take-offs with no threat.

**Table N. Parameters of the virtual arena.** c, odour concentration; σ, plume standard deviation; θ, heading; x, fly position; $x_A$ and $x_B$, odour source positions; SD, standard deviation.

| Component | Parameter | Value |
|---|---|---|
| Arena | Radius and wall | 50 mm; a fly that passes 48.5 mm from the centre is placed back on that circle and turned towards the centre, with Gaussian noise of SD 0.6 rad |
| Arena | Start | Position uniform within 40 mm of the centre, random heading, hunger 1 |
| Coupling | Time steps | Body step 10 ms; 100 brain steps of 0.1 ms per body step |
| Coupling | Input cap | Every input (Poisson) rate capped at 300 Hz, including the drive to the uniglomerular PNs and to LPLC2 and LC4 |
| Food | Patches | Two discs of radius 8 mm centred at (0, 27) and (0, −27) mm |
| Food | Sugar drive | 150 Hz × hunger to LB3c and LB3b neurons of both sides, only on a patch and on the ground |
| Food | Hunger | Starts at 1; falls by 0.25 per s of feeding and recovers by 0.03 per s otherwise, within 0 to 1 |
| Food | Feeding | On a patch, on the ground, not taking off, and MN9 rate (mean of the two MN9 over the last 100 ms) above 10 Hz |
| Odour | Sources | Odour A at (−21, 0) mm and odour B at (21, 0) mm, 42 mm apart |
| Odour | Concentration | $c = \exp(-d^2/2\sigma^2)$ with σ = 14 mm, d the distance to the source |
| Odour | Drive | 150 Hz × c to the uniglomerular PNs of that odour on both sides |
| Odour | Inside a plume | $c > \exp(-0.5)$ (about 0.61), that is within 14 mm (1σ) of the source |
| Looming | Threats | 10 in the 120-s run: the first at 6 s, then one every 10 s plus a uniform 0 to 4 s, until 2 s before the end (onsets 6.0 to 112.55 s, intervals 10.06 to 13.86 s); centres uniform within 18 mm of the arena centre; drawn once and shared by all flies |
| Looming | Shadow | Radius grows linearly from 2 to 24 mm over 0.6 s |
| Looming | Drive | 150 Hz per threat to LPLC2 and LC4 on the side of the fly facing the shadow centre, only while the fly is within the shadow radius + 6 mm and on the ground |
| Looming | Threat encountered | Fly within the shadow radius + 6 mm at any body step of the expansion, on the ground or not |
| Escape | Take-off | At least 2 giant fibre spikes, summed over both DNp01, in the last 30 ms; fly on the ground; at least 1 s since the previous take-off; no threat required |

| Component | Parameter | Value |
|---|---|---|
| Escape | Take-off movement | 22 mm in 0.35 s, directed away from the shadow centre (along the current heading if no shadow is present), with Gaussian noise of SD 0.5 rad |
| Walking | Speed and heading | 9 mm/s; heading diffusion 2.2 rad/√s; no walking while feeding or airborne |
| Steering | Equation | $d\theta/dt = 6$ rad/s × $\tanh(V_{learned}$ / 30 Hz) × $\sin(\varphi - \theta)$, only while walking; φ is the direction of the vector $c_A (x_A - x) + c_B (x_B - x)$ towards the two sources |
| Steering | Valence relative to naive valence | $V_{learned}$ = valence over the last 300 ms minus the fly's naive valence at the current $c_A$ and $c_B$ |
| Steering | Naive valence (naive-valence probe) | Odour A, then odour B, at 0.25, 0.5 and 1.0 × 150 Hz, 1 s each with 0.5-s gaps, after a state reset and before training; interpolated linearly in c from valence 0 at c = 0 through the three levels, and summed over the two odours |
| Dopamine | Teaching signal | None in the arena; plasticity only in the training step before it |
| Run | Duration and state | 120 s; membrane, synaptic and adaptation state reset at the start, weights (including the trained change) kept |

## Simulated cohorts

**Batches.** The severity series, the rule ablation, the map comparison and Kenyon-cell-only pathology were each run in batches of 48 flies, as were the flies of the contact-matched comparison, with conditions permuted across batch positions so that batch effects cannot coincide with condition. The arena cohort was simulated in one batch with severities permuted across positions.

**Shared input noise.** Input spikes were drawn for a whole batch at once, and the first four designs used the same random-number seed for the input, so flies that sat in the same position of the same batch in two designs received identical input spikes. Among the three replicate groups of the default model this concerns 5 pairs of flies at s = 0.5 and 7 at s = 1.0, and 3 to 10 of the 32 flies of each rule-ablation, map and Kenyon-cell-only group sat in the position of a healthy fly of the severity series. In the last batch of the rule-ablation and Kenyon-cell-only designs, which held fewer than 48 flies, the input spikes were identical only during the first odour presentation. Conditions were permuted into positions, so the sharing adds no bias, but the groups concerned are not fully independent in input noise.

**Flies simulated twice.** A second run of 40 flies, with the same damage but different input noise, was used only for the test-retest comparison: 16 healthy flies of the severity series and 8 flies each of the rule ablation (all three rules), the map comparison (hub map) and Kenyon-cell-only pathology at s = 0.5.

**Healthy groups.** The healthy flies of the contact-matched comparison had a median whole-brain activity of 17,814.5 spikes against 17,273 in the healthy flies of the severity series, and a median learning index of 76.0 against 61.5. The two groups ran the same protocol and differ only by input noise, yet the healthy flies of the contact-matched comparison had 1.24 times the learning index and 1.03 times the whole-brain activity of those of the severity series (Mann-Whitney p = 0.022 and 0.042, uncorrected; 0.19 and 0.34 after Holm correction over the nine readouts of Fig 2 other than naive valence). The healthy flies of the arena cohort had a median learning index of 67.0.

## Maps matched for synaptic contacts lost

**Contacts lost against severity, and replicate variation.** The mean fraction of contacts lost in the three flies per map whose damage was drawn without simulating them was interpolated linearly between severities. The replicate variation of every other readout, the counterpart of the 1.27 of whole-brain activity, is given under Supplementary results (Circuit readouts decline with severity, Replicate groups of the default model).

**Outcomes as worded before the run.** The three outcomes of the pre-specified criterion (Materials and methods) were written before the run as "burden explains the map effect", "placement matters beyond burden" and "unresolved"; burden is the fraction of contacts lost and placement is where the damage falls. The main text calls the first two "within replicate variation" and "beyond replicate variation".

**Summaries and post hoc analyses.** Besides the pre-specified criterion, the analysis written before the run summarised seven readouts (whole-brain activity, lateral horn, Kenyon cell and MBON rates, learning index, MN9 drive and giant fibre rate) as medians with bootstrap intervals and Kruskal-Wallis tests across maps at each matched level, without a pre-specified interpretation, together with each group's median contacts lost, neurons removed and Kenyon cells removed. Everything else was added after the run, including: pairwise Mann-Whitney tests between maps; the criterion applied to other readouts with each readout's own replicate variation at s = 0.5 in place of 1.27; the per-fly mean load over all neurons, sensory neurons included at zero, and the corresponding mean threshold shift (2 mV × mean load); projection neuron rate as a measure of delivered odour input; Kenyon cell rate per surviving Kenyon cell (Kenyon cell rate divided by the fraction of Kenyon cells not removed); and deficit ratios, (100 − % of healthy) of one map over that of another.

## Within-fly analysis: undoing parts of a fly's own damage

**Versions of a fly.** A restored neuron returns with its own thinned synapses, the outgoing contacts it had lost under the synapse-loss rule: each was lost with probability at most 0.18 at s = 0.3 and 0.30 at s = 0.5.

**Shortened protocol.** The protocol is odour A for 1 s, a 1-s gap, sugar for 0.5 s, a 1-s gap and looming for 0.3 s (3.8 s). It yields every circuit readout of Table 2 except the learning index, the learning index divided by MBON rate and the PAM pairing rate. Its MN9 and giant fibre readouts are measured earlier in the input stream and after a different activity history than in the full protocol, which has odour B and a gap before them.

**Readouts above the healthy value.** The pairs with a damaged value above the healthy value in the screen, in which one class or one cell type at a time was restored, are counted in Supplementary results. In the robustness test and the full-protocol test, further such pairs were kept in two flies: MBON rate, Kenyon cell rate and the learning index in one, and MN9 drive in the other. For a fly whose damaged value was above the healthy value, a positive fraction means a move down towards the healthy value. A fraction above 1 means the readout moved past the healthy value. Small denominators give values far outside 0 to 1: in one fly the MBON rate was 2.48 Hz damaged against 2.33 Hz healthy, so restoring its two lost APL neurons gave a fraction of 11.7. MN9 drive and giant fibre rate move in steps (Circuit protocol, Readouts), so small deficits give coarse fractions; 16 of the 48 flies had a giant fibre deficit of only 5 to 20% of healthy.

**Classes and cell types.** A fly had a median of 23.5 damaged classes, classes that had lost at least one neuron. Cell types come from the type annotation, with untyped neurons grouped by class.

**Class screen.** Every damaged class was restored singly (1,123 simulations). In 256 further versions per fly, each class was included independently with probability 0.5 and all included classes were restored together. The ridge regression of the fraction of the deficit restored on the inclusion indicators used a penalty of $10^{-3} \times 256$ with the intercept not penalised; its coefficients are the classes' Banzhaf values, and its in-sample $R^2$ measures additivity. Class medians are reported for classes damaged in at least 10 flies.

**Type selection.** The union of the three most effective classes for each of the six readouts gave a median of 8 chosen classes per fly (range 4 to 12). The 400 selected types and the 50 control types were restored one at a time (21,600 simulations). In 47 of 48 flies every tested type within the chosen classes had lost at least two neurons; types that lost a single neuron were otherwise tested only as random controls (1,325 tests). The 450 tested types per fly held a median of 18% of the fly's removed neurons (range 10 to 63%) and 17% of their synapses, out of a median of 4,982 damaged types. Types with few neurons were therefore rarely tested: DNp01 in 2 flies, APL in 1, the odour A PN types DM1_lPN in 1, DM4_adPN in 2 and VA2_adPN in 1, and DM4_vPN in none.

**Single neurons and combinations.** Within the five most effective tested types for each readout (a median of 18 types per fly, range 12 to 24), up to 20 neurons per type, those with the most input and output synapses, were restored singly (8,060 simulations). The 1, 3, 10 and 30 most effective tested types for each readout were restored together (1,104 simulations). Together with five simulations of each fly as a whole, namely damaged, healthy and with each of the three kinds of damage undone, these runs comprised 44,415 simulations.

**Derived quantities.** For an effect relative to all removed neurons, the fraction of the deficit restored by a class or a set of types was divided, fly by fly, by that of all removed neurons; flies in which the divisor was 0.1 or less were left out. A class was compared with its tested types by setting the effect of restoring the class against the sum of the effects of restoring each of its tested types singly in the same fly. The descending-neuron class is a superclass, and the most effective tested type is the maximum over the fly's 450 tested types.

**Wiring condition.** For each fly and readout with at least 10 tested types, the fractions of the deficit restored were regressed on five measures of the wiring of each type's lost neurons: their number, synapse count, direct contacts and two-step contacts onto the neurons of the readout, each as the logarithm of one plus the value, and their mean connectivity rank. For whole-brain activity the readout neurons are all neurons, and two-step contacts pass through any intermediate neuron. The leave-one-type-out $R^2$ is computed within each fly, and the standardised residual is the leave-one-out residual divided by the standard deviation of that fly's residuals. No measure encodes which neurons the protocol stimulates, so a stimulated type can exceed the prediction.

**Robustness condition.** Types were ranked for each readout by their median fraction of the deficit restored pooled over all 48 flies, among types with a finite fraction in at least 3 flies. For MN9 drive only six types had a positive median fraction, and the other four of its ten most effective were taken from the 2,331 types tied at zero. Each of the 54 types was restored in those of the 12 flies in which it had lost neurons (3 to 12 flies per type, median 9; 483 type-fly pairs). The unchanged model, the extended model with unchanged parameters, and each of the six variants also ran the damaged fly and the healthy copy of each fly (3,549 simulations in all). A post hoc analysis separated the 24 parameter-variant combinations from the 12 input-noise combinations, flagged passes resting on at most five types with a non-zero median in either run, recomputed the correlation over types with a non-zero median in either run and the top-10 overlap under random tie-breaking (1,000 draws), and correlated per-fly type effects between the unchanged model and each variant with new input noise.

**Full-protocol test.** The restore-alone and remove-alone runs comprised 483 + 483 runs of single types and 24 runs of damaged flies and healthy copies (990 simulations). In a remove-alone run only the neurons of one type that the fly had lost are removed from its healthy copy, and the fraction of the deficit reproduced is $1 - f$, with $f$, the fraction of the deficit restored (Materials and methods), computed for that copy. The restore-alone values of the seven odour A readouts repeat those of the unchanged model in the robustness test, which had the same input and first step; new are the learning index, MN9 drive and giant fibre rate, measured later in the protocol, and the remove-alone runs. In total the within-fly analysis comprised 48,954 simulations.

## Statistics

**Bootstrap.** Unless stated otherwise, intervals use 4,000 bootstrap resamples with replacement (2.5th and 97.5th percentiles). A few intervals of the severity series and the rule ablation, whose bounds were not stable across repeated bootstraps at this number, are from 40,000 resamples. In the within-fly analysis, damaged readouts were bootstrapped with each fly paired to its own healthy copy, and intervals of effects relative to all removed neurons come from resampling the per-fly values, like those of median fractions of the deficit restored. Intervals drawn in Figs 3 and 4 and in S3 Fig come from separate resamples and can differ from those quoted in the text by Monte Carlo error, by at most 0.8 percentage points in Fig 4; in Fig 3C the hub-map MBON upper bound at $s = 0.5$ is drawn 5 points above the 59% quoted.

**Smallest detectable difference.** The smallest detectable difference, quoted where a comparison found none, is a normal approximation for 80% power at a two-sided $\alpha$ of 0.05 with the pooled per-fly standard deviation.

**Replicate groups.** The three replicate groups of the default model were compared in 22 Kruskal-Wallis tests (11 readouts × 2 severities), of which 14 are shown in S2 Fig and three were below 0.05 (Supplementary results, Replicate groups of the default model).

**Partial rank correlations.** The variables were converted to ranks, the linear effect of the rank of the control variable was removed from both, and the Pearson correlation of the residuals is given with its p value.

**Separation from healthy flies.** Per-severity AUC intervals use 2,000 bootstrap resamples of flies within each group. For the first severity at which the AUC criterion was reached, flies were resampled within the healthy group and within each severity group. When the cumulative fraction of resamples that had reached it was within 0.1 of one half at some severity, the bootstrap was repeated 10 further times; a bootstrap median that changed between these repetitions is described as not stable across repeated bootstraps, and the observed severity is quoted. The point estimate of a paired AUC difference is the full-sample difference.

**Chance level of the direction-free AUC.** Because max(AUC, 1 − AUC) is at least 0.5, its expected value without any group difference exceeds 0.5. Computed exactly from the null distribution of the Mann-Whitney statistic without ties, it has a mean of 0.550 (95th percentile 0.623) for 32 against 64 flies, as in the severity series, and 0.563 (95th percentile 0.654) for 24 against 32 flies, as in the arena cohort.

**Arena cohort.** Trends with severity were tested as in the severity series, with Holm correction across sixteen readouts: the twelve tested in the severity series and the four arena readouts (Table 2). A circuit readout was taken to separate better than the arena readouts at a severity when the 95% interval of its paired AUC difference against each of the four lay above zero, without multiplicity adjustment. Healthy flies' preference for the trained odour was tested with a Wilcoxon signed-rank test against zero.

## Figure details

**Fig 4D.** The per-fly interquartile ranges, at most 0.17 percentage points for all neurons and 0.9 for Kenyon cells, are not drawn.

**Fig 5.** In Fig 5A, values drawn at the axis edge are triangles, or crosses for the pairs whose damaged value lies above the healthy value. In Fig 5B, the open circles are the median for the PN class in the 36 flies (35 for Kenyon cell rate) in which restoring the class changed whole-brain activity. In Fig 5C, DNp01 was the most effective tested type (printed as best tested type) in both flies in which it was tested. In Fig 5D a point below the k = 1 value means that the median over flies of what the combined types recovered was smaller than the median of what each fly's most effective single type recovered. In Fig 5E, most of the 54 types have a median of zero for MN9 drive and giant fibre rate, so the printed overlaps for these readouts reflect the fixed tie order; the overlaps under random tie-breaking are given in Supplementary results.